\documentclass[twocolumn,times]{aastex63}
\usepackage{multirow,bigdelim,amsmath,soul,longtable,comment,booktabs}
\usepackage{xcolor}
\usepackage{colortbl}
\usepackage{mathtools}
\usepackage{savesym}
\savesymbol{tablenum}
\usepackage{siunitx}
\usepackage[T1]{fontenc}
\usepackage{enumitem}
\DeclarePairedDelimiterXPP\BigOSI[2]%
  {\mathcal{O}}{(}{)}{}%
  {\SI{#1}{#2}}

\begin{document}
\title{Eight Years of Light from ASASSN-15oi: Towards Understanding the Late-time Evolution of TDEs}

\correspondingauthor{A. Hajela}
\email{aprajita.hajela@gmail.com}

\author[0000-0003-2349-101X]{A. Hajela}
\affiliation{DARK, Niels Bohr Institute, University of Copenhagen, Jagtvej 155, 2200 Copenhagen, Denmark}
\author{K. ~D. Alexander}
\affiliation{Department of Astronomy and Steward Observatory, University of Arizona, 933 North Cherry Avenue, Tucson, AZ 85721-0065, USA}
\author[0000-0003-4768-7586]{R. Margutti}
\affil{Department of Astronomy, University of California, Berkeley, CA 94720-3411, USA}
\affil{Department of Physics, University of California, Berkeley, CA 94720-7300, USA}
\author[0000-0002-7706-5668]{R. Chornock}
\affil{Department of Astronomy, University of California, Berkeley, CA 94720-3411, USA}
\author{M. Bietenholz}
\affiliation{SARAO/Hartebeesthoek Radio Astronomy Observatory, PO Box 443, Krugersdorp, 1740, South Africa}
\author[0000-0003-0528-202X]{C. T. Christy}
\affiliation{Department of Astronomy and Steward Observatory, University of Arizona, 933 North Cherry Avenue, Tucson, AZ 85721-0065, USA}
\author[0000-0002-3019-4577]{M. Stroh}
\affiliation{Center for Interdisciplinary Exploration and Research in Astrophysics (CIERA), Northwestern University, Evanston, IL 60208, USA}
\author[0000-0003-0794-5982]{G. Terreran}
\affiliation{Las Cumbres Observatory, 6740 Cortona Drive, Suite 102, Goleta, CA 93117-5575, USA}
\affiliation{Department of Physics, University of California, Santa Barbara, CA 93106-9530, USA}
\author{R. Saxton}
\affiliation{Telespazio UK for ESA, ESAC, Apartado 78, 28691 Villanueva de la Ca\~{n}ada, Madrid, Spain}
\author{S. Komossa}
\affiliation{Max-Planck Institut für Radioastronomie, Auf dem Hügel 69, 53121 Bonn, Germany}
\author[0000-0002-7735-5796]{J. S. Bright}
\affiliation{Department of Astronomy, University of California, Berkeley, CA 94720-3411, USA}
\affiliation{Astrophysics, Department of Physics, University of Oxford, Keble Road, Oxford OX1 3RH, UK}
\author[0000-0003-2558-3102]{E. Ramirez-Ruiz}
\affiliation{Department of Astronomy and Astrophysics, University of California, Santa Cruz, CA 95064, USA}
\author[0000-0001-5126-6237]{D.~L. Coppejans}
\affiliation{Department of Physics, University of Warwick, Gibbet Hill Road, CV4 7AL Coventry, United Kingdom}
\author{J.~K. Leung}
\affiliation{David A. Dunlap Department of Astronomy and Astrophysics, University of Toronto, 50 St. George Street, Toronto, ON M5S 3H4, Canada}
\affiliation{Dunlap Institute for Astronomy and Astrophysics, University of Toronto, 50 St. George Street, Toronto, ON M5S 3H4, Canada}
\affiliation{Racah Institute of Physics, The Hebrew University of Jerusalem, Jerusalem 91904, Israel}
\author{Y. Cendes}
\affiliation{Department of Physics, University of Oregon, Eugene, OR 97403, USA}
\author{E. Wiston}
\affil{Department of Astronomy, University of California, Berkeley, CA 94720-3411, USA}
\author[0000-0003-1792-2338]{T.~Laskar}
\affiliation{Department of Physics \& Astronomy, University of Utah, Salt Lake City, UT 84112, USA}
\affiliation{Department of Astrophysics/IMAPP, Radboud University, P.O. Box 9010, 6500 GL, Nijmegen, The Netherlands}
\author[0000-0002-5936-1156]{A. Horesh}
\affiliation{Racah Institute of Physics, The Hebrew University of Jerusalem, Jerusalem 91904, Israel}
\author[0000-0001-9915-8147]{G. ~Schroeder}
\affiliation{Center for Interdisciplinary Exploration and Research in Astrophysics (CIERA),  Northwestern University, Evanston, IL 60208, USA}
\affiliation{Department of Physics and Astronomy, Northwestern University, 2145 Sheridan Road, Evanston, IL 60208-3112, USA}
\author[0000-0002-8070-5400]{Nayana A.~J.}
\affil{Department of Astronomy, University of California, Berkeley, CA 94720-3411, USA}
\author[0000-0002-7721-8660]{M.~H. Wieringa}
\affil{CSIRO Space and Astronomy, Australia Telescope National Facility, PO Box 76, 1710, Epping, NSW, Australia}
\author{N. Velez}
\affiliation{Department of Physics, University of Oregon, Eugene, OR 97403, USA}
\author{E.~Berger}
\affiliation{Center for Astrophysics | Harvard \& Smithsonian, Cambridge, MA 02138, USA}
\author{P.~K. Blanchard}
\affiliation{Center for Interdisciplinary Exploration and Research in Astrophysics (CIERA), Northwestern University, Evanston, IL 60208, USA}
\author{T. Eftekhari}
\affiliation{Center for Interdisciplinary Exploration and Research in Astrophysics (CIERA), Northwestern University, Evanston, IL 60208, USA}
\affiliation{Department of Physics and Astronomy, Northwestern University, 2145 Sheridan Road, Evanston, IL 60208-3112, USA}
\author[0000-0001-6395-6702]{S. Gomez}
\affiliation{Space Telescope Science Institute, 3700 San Martin Dr, Baltimore, MD 21218, USA}
\author[0000-0002-2555-3192]{M. Nicholl}
\affiliation{Astrophysics Research Centre, School of Mathematics and Physics, Queens University Belfast, Belfast BT7 1NN, UK}
\author{H. Sears}
\affiliation{Center for Interdisciplinary Exploration and Research in Astrophysics (CIERA), Northwestern University, Evanston, IL 60208, USA}
\affiliation{Department of Physics and Astronomy, Northwestern University, 2145 Sheridan Road, Evanston, IL 60208-3112, USA}
\author[0000-0003-1152-518X]{B.~A. Zauderer}
\affiliation{National Science Foundation, 2415 Eisenhower Avenue, Alexandria, Virginia 22314, USA}

\begin{abstract}
We present the results from an extensive follow-up campaign of the Tidal Disruption Event (TDE) ASASSN-15oi spanning $\delta t \sim 10 - 3000$\,d, offering an unprecedented window into the multiwavelength properties of a TDE during its first $\approx 8$ years of evolution. ASASSN-15oi is one of the few TDEs with strong detections at X-ray, optical/UV, \emph{and} radio wavelengths and featured two delayed radio flares at $\delta t \sim 180$\,d and $\delta t \sim 1400$\,d. 
Our observations at $> 1400$\,d reveal an absence of thermal X-rays, a late-time variability in the non-thermal X-ray emission, and sharp declines in the non-thermal X-ray and radio emission at $\delta t \sim 2800$\,d and $\sim 3000$\,d, respectively. The UV emission shows no significant evolution at $>400$\,d and remains above the pre-TDE level. We show that a cooling envelope model can explain the thermal emission consistently across all epochs. We also find that a scenario involving episodic ejection of material due to stream-stream collisions is conducive to explaining the first radio flare. Given the peculiar spectral and temporal evolution of the late-time emission, however, constraining the origins of the second radio flare and the non-thermal X-rays remains challenging. Our study underscores the critical role of long-term, multiwavelength follow-up.
\end{abstract}


\section{Introduction} \label{Sec:intro}

A star straying too close to a supermassive black hole (SMBH) of mass $\lesssim 10^{8}\,M_{\odot}$
in the center of a galaxy is torn apart by the strong tidal forces of the black hole. This triggers an energetic and short-lived transient known as a TDE (\citealt{Hills75,Rees88,Evans89,Guillochon2013}). Through emitting radiation across the electromagnetic spectrum, TDEs provide an exclusive window to study previously dormant SMBHs, accretion physics on human-accessible timescales, initiation and cessation of the launched outflows, and the make up of the nuclear environment of the host galaxies \citep[e.g.,][]{Bonnerot21rev,Dai21_rev,Gezari21rev}. 
Early TDE datasets typically had limited temporal and spectral coverage, but as the quality and quantity of data have improved,  observations have uncovered a broad diversity at every wavelength of the electromagnetic spectrum. In this context, the TDE ASASSN-15oi stands out for its uniquely detailed dataset, with X-ray, optical, UV, and radio detections spanning $\sim3000$ days. We briefly review the current landscape of TDE observations before focusing on ASASSN-15oi as a valuable test case of TDE emission models.  

The first TDE candidates were discovered as X-ray transients, and observations over the past three decades have revealed diverse X-ray behaviors among TDEs detected at different wavelengths.
Reviews of X-ray selected TDEs can be found in \citealt{Komossa2002_rev, Auchettl2017,Saxton2021_err} and \citealt{Komossa2023_rev}, while \citealt{Guolo2023} and \citealt{Yao2023} discuss X-ray properties of optically-discovered TDEs. ``Soft'' X-rays, characterized by a power-law spectrum with photon index $\Gamma_{\rm X} > 3$ or a blackbody spectrum with temperature $kT_{\rm BB} \sim 0.05$\,keV, emerges from regions at a distance $\sim r_{\rm{g}} \approx 10^{11} - 10^{12}$\,cm from the SMBH, where $r_{\rm{g}}$ is the gravitational radius around a $\sim 10^{6}\,M_{\odot}$ non-rotating SMBH. This implies that soft X-rays are emitted from the material actively accreting onto the SMBH \citep{Dai2018_unified}. ``Hard'' X-rays ($\Gamma_{\rm X} \lesssim 2$ power-law) can result either from synchrotron radiation or by Compton up-scattering of thermal disc photons by electrons in a corona or in an outflow. TDEs detected at X-ray energies show either, both, or neither of these components. ASASSN-15oi is one of the few TDEs to show both a ``hard'' and ``soft'' X-ray component (this work, also \citealt{Gezari17,Holoien18}).

The physical origin of the optical and UV emission in TDEs is less certain. This emission arises at radii $\sim 10^{15}$ cm $>> r_{\rm g}$. One potential scenario involves the reprocessing of disc X-rays by material at these extended radii \citep{Guillochon2014,Roth2016,Dai2018_unified,Thomsen2022}. Such material could be disc winds \citep[e.g.,][]{Miller2015}, or outflows launched during the circularization process \citep{Metzger2016_tde} or during stream-stream collisions \citep[called collision induced outflows or CIO in][]{LuBonnerot2020_CIO}. This scenario predicts that the post-optical-peak luminosity should trace the mass fallback rate with a characteristic $t^{-5/3}$ decay, as indeed observed in many TDEs \citep[e.g.,][]{Mockler19,Gezari21rev}. Another mechanism proposed by \citet{Piran15} suggests that the optical luminosity is powered directly by the stream-stream collisions, with the gravitationally bound stellar debris stream returning to the SMBH and self-intersecting.  Recently, a third model (inspired by \citealt{Loeb97}) was revived and re-analysed by \citet{Metzger2022}, where the optical emission results from the radiative cooling of a quasi-spherical, pressure-supported envelope formed as a result of the stellar disruption. This \emph{cooling envelope} model predicts the optical/UV luminosity to decline as $t^{-3/2}$, co-incidentally resembling the $t^{-5/3}$ decay from the re-processing scenario. 

Radio emission from TDEs is primarily powered by synchrotron radiation from outflows interacting with the surrounding medium at larger distances ($\gtrsim 10^{16}$\,cm) from the SMBH \citep[][and references therein]{Alexander20}. Among TDEs that have radio detections, a few show evidence of powerful on-axis relativistic jets. \textit{Swift}\,J1644+57 is the canonical example of this subset \citep[e.g.,][]{Bloom2011,Burrows2011, Levan11, Zauderer11, Berger12, Eftekhari18, Cendes21_J1644}. 
Another classic radio TDE,  ASASSN-14li,
showed evidence for only a non-relativistic outflow instead (\citealt{Alexander16,Krolik16}, but see also \citealt{vanVelzen16}). While some TDEs show radio emission similar to ASASSN-14li's, others remain undetected in radio observations, even up to several hundred days post-disruption \citep{Alexander20}. 

Early radio campaigns typically ceased observations within a few months if no emission was detected, however, recently a large fraction of TDEs ($\sim 40\%$) were found to exhibit significant radio emission $\sim$ months -- years post-discovery \citep{Cendes2023_latetimeTDEs, ASKAP_lateTDEs}. Some examples include iPTF16fnl \citep{Horesh_16fnl_2021}, AT\,2018hyz \citep{Cendes_18hyz_2022, Sfaradi_18hyz_2024}, and IGR\,J12580+0134 \citep{Perlman_igr_2022}. Among this subset, ASASSN-15oi was the first TDE to show a late-time radio brightening at $\delta t \sim 180$\,d, and about 3.5 years later, it exhibited a second brightening \footnote{Some TDEs with early radio emission have also shown delayed re-brightenings: AT\,2019azh re-brightened at $\sim 200$\,d \citep{Sfaradi_19azh}, similar to the timescale of the first radio flare in ASASSN-15oi, but unlike ASASSN-15oi it also showed prompt radio emission starting just 21 d post-discovery. AT\,2020vwl re-brightened after $\sim 1000$\,d \citep{Goodwin_ATEL_202vwlrebrightening}, similar to the timescale of the second radio flare  of ASASSN-15oi.  Radio emission modeling of AT\,2020vwl in \cite{Goodwin_2020vwl_2023} suggests a prompt outflow launch as well.} in the VLA Sky Survey (VLASS) data, reported by \cite{Horesh21}. Our team independently identified this bright radio source in VLASS data, and began a multi-wavelength follow-up campaign that covers the time period $\delta t \approx 1471 - 2970$\,d, which we report here.

ASASSN-15oi was optically discovered on 2015 August 14 (\citealt{Brimacombe_15oidisc}; we measure $\delta t$ in this paper as the time elapsed since discovery) at a distance of $\sim 216$\,Mpc \citep{Holoien16, Gezari17} by the All-Sky Automated Survey for SuperNovae (ASAS-SN; \citealt{Shappee14}). Observations of ASASSN-15oi at optical, UV, and X-ray wavelengths at $\delta t \lesssim 600$\,d were previously reported in \citet{Holoien16, Gezari17, Holoien18}, and \citet{Hinkle21}. The temporal evolution of the optical and UV luminosity at $\delta t < 100$\,d initially showed an exponential decay.   The optical  emission subsequently faded to the level of 
the host galaxy at $\gtrsim 300$\,d \citep{Holoien16,Gezari17, Holoien18} while the emission at UV wavelengths remained above the pre-TDE host galaxy level. 
The X-ray emission from ASASSN-15oi, on the other hand, was initially weaker ($\sim 10^{42}$\,erg/s) compared to those of other X-ray bright TDEs in the literature \citep[$\sim 10^{43}$\,erg/s,][]{Guolo2023}, but it later brightened by a factor of $\sim 6$ between $\delta t\sim 80$ and $\sim 230$ d. Previous studies identified two contributing components to the X-ray emission: a blackbody component responsible for the delayed brightening of the X-ray emission; and a power-law component reported as remaining nearly constant, leading to its association with a pre-existing Active Galactic Nucleus (AGN) \citep{Gezari17}. We note that the archival mid-infrared colors of ASASSN-15oi's host galaxy, with $W1-W2 \sim 0.06$, from the Widefield Infrared Survey Explorer (\emph{WISE}; \citealt{Wright2010}) are inconsistent with the presence of a strong AGN activity \citep{Assef2013,Assef2018}, as was also reported in \citet{Holoien18}.

Our aim in this study is to investigate the emission from ASASSN-15oi using the complete multi-wavelength dataset spanning 8 years since its discovery. Our dataset includes previously published data ($\delta t \lesssim 600$\,d), other publicly available observations ($\delta t \gtrsim 600$\,d), as well as new observations that we acquired following the onset of the second radio flare at $\delta t \approx 1471 - 2970$\,d. We re-analyse the previously reported observations in a consistent manner with the new data, and report our inferences on the origin of the emission from ASASSN-15oi at the different stages. The paper is organized as follows: in \S\ref{Sec:DataAnalysis}, we outline the details of the multi-wavelength observations and our data reduction. In \S\ref{Sec:RadioModeling}, we fit the radio spectral energy distributions (SEDs) with different models under the umbrella assumption that the radio observations are powered by synchrotron emission. In \S\ref{Sec:Discussion} we divide the emission from ASASSN-15oi into thermal and non-thermal components, discuss the possible origins of the emission at different wavelengths, and investigate any potential inter-correlations. We provide a summary and conclusions in \S\ref{Sec:Conclusions}.

\section{Observations and Data Analysis} \label{Sec:DataAnalysis}
In this section, we present new data spanning the time period $\delta t \approx 600 - 3000$\,d, acquired by a variety of telescopes: \textit{Swift}-UVOT, \textit{Swift}-XRT\footnote{For \textit{Swift} data, we provide a self-consistent flux calibration of the entire dataset that accounts for the spectral evolution of the source. },  X-ray Multi-Mirror Mission (\textit{XMM-Newton}), Karl G. Jansky Very Large Array (VLA), Very Long Baseline Array (VLBA), Australia Telescope Compact Array (ATCA), MeerKAT telescope and Atacama Large Millimeter/submillimeter Array (ALMA). Furthermore to perform a systematic analysis, we homogeneously reduce all the observations over the time period of $\delta t = 8 - 3000$\,d. 

\subsection{UV: \textit{Swift}-UVOT} \label{SubSec:DataUVOT}
\begin{figure*}
    \centering
    \includegraphics[scale=0.7]{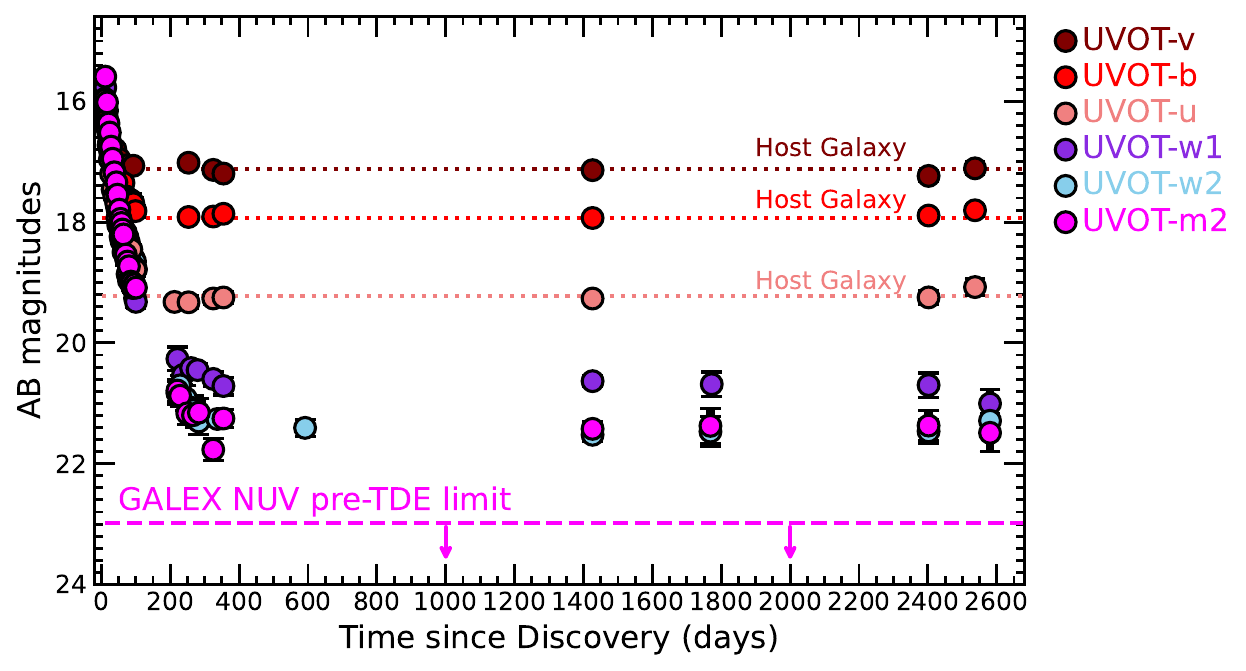}
    \caption{Long-term evolution of ASASSN-15oi as captured by \textit{Swift}-UVOT. Filled circles mark the total emission (i.e., TDE + host galaxy) at the location of ASASSN-15oi. While at $\delta t\gtrsim 300\,$d the optical emission is dominated by the host galaxy light (horizontal dotted lines as determined by \citealt{Hinkle21}),  the UV emission at the location of ASASSN-15oi remains bright and does not relax to the pre-TDE UV flux levels (horizontal dashed line). Magnitudes have been corrected for Galactic extinction.}
    \label{Fig:UVOT}
\end{figure*}

\textit{Swift}-UVOT \citep{Gehrels04,Roming05} 
observed ASASSN-15oi from 2015 August 23 until 2022 October 19 ($\delta t=8 - 2580$\,d). We analyse all available \textit{Swift}-UVOT photometric data following the prescriptions of \cite{Brown09}, with the updated calibration files \citep{Poole08,Breeveld10}. All photometry has been extracted using a 5\arcsec\, radius aperture and a 36\arcsec\, region free of sources for the background. When possible, we merged observations to achieve a minimum source detection significance of $\approx 10\,\sigma$. Finally, we correct for Galactic extinction assuming the \cite{Fitzpatrick99}  reddening law  and $R_{\rm{V}}=3.1$.\footnote{\cite{Hinkle21} adopt a \cite{Cardelli89} reddening law. Here we adopt the \cite{Fitzpatrick99} reddening law following the findings of \cite{Schlafly11}. For the $R_{\rm{V}}=3.1$ curves the difference in the $A_{\lambda}$ values of \textit{Swift}-UVOT filters are $\le0.015$ mag. } The resulting extinction corrections are 
$A_{v}=0.185$ mag,  
$A_{b}=0.245$ mag,
$A_{u}=0.294$ mag,
$A_{w1}=0.377$ mag,
$A_{w2}=0.551$ mag,
$A_{m2 }=0.547$ mag.
The \textit{Swift}-UVOT photometry is presented in Table \ref{tab:uvotphot} and the long-term UV evolution of ASASSN-15oi is shown in Figure \ref{Fig:UVOT}. For observations at $\delta t<500\,$d we find excellent agreement with the photometry presented in \cite{Hinkle21}. We use the UV/optical bolometric luminosities (for epochs where optical emission was above the host galaxy level) from \cite{Hinkle21} for our analysis.

In stark contrast to the X-ray and radio emission that show dramatic temporal variability on short time scales (\S\ref{SubSec:DataXRT}, \S\ref{SubSec:DataXMM}, \S\ref{SubSec:DataVLA}), the late-time 
\textit{Swift}-UVOT photometry at $\delta t>600$\,d is consistent with constant flux (Figure \ref{Fig:UVOT}).  We find no significant evidence for fading at $\delta t\approx 500-2600\,$d in any of the  \textit{Swift}-UVOT filters. However, while at optical wavelengths  (i.e., \textit{Swift}-UVOT  $u$, $b$, and $v$ filters) the flux is consistent with the pre-TDE host galaxy level as determined by the updated host galaxy modeling of \cite{Hinkle21}, at UV wavelengths (i.e., \textit{Swift}-UVOT  $uvw1$, $uvw2$, and $uvm2$ filters) we confirm the presence of an excess of emission that was reported by \cite{Holoien18} using data at earlier epochs $\delta t\approx 250-600\,$d. As of $\delta t=2600\,$d,  with $m_{w1}\approx 21$\,mag,  $m_{w2}\approx 22$\,mag and $m_{m2}\approx 22$\,mag (observed mags, AB system), ASASSN-15oi is $\approx 0.7$\,mag,  $\approx 1.5$\,mag and  $\approx 0.9$\,mag brighter, respectively, than the best-fitting pre-TDE host galaxy model of \cite{Hinkle21}. From a completely observational perspective that is independent from the host galaxy light modeling, pre-TDE  \emph{Galaxy Evolution Explorer} (\emph{GALEX}) observations constrain the UV emission of the host galaxy to be $m_{NUV}>22.98$\,mag \citep{Holoien18,Hinkle21}. For the 
blackbody spectrum with $T\sim 2\times 10^4\,\rm{K}$ that best fits the late-time UVOT data, the \emph{GALEX} NUV to  \textit{Swift}-UVOT $uvm2$ filter correction term is $\delta mag\approx 0.05\,$ mag, which implies that at $\delta t \approx 2600\,$d, the UV emission is $\gtrsim 2.3$ times brighter than in the pre-TDE era. Similarly persistent UV excesses of emission have been found in  ASASSN-14li \citep{Brown17}, as well as in other TDEs \citep{vanVelzen2019, Mummery2024_galscaling}. 
We note that since optical discovery, ASASSN-15oi has radiated $\sim 1.2 \times 10^{51}\,\rm{erg}$ at UV and optical wavelengths.
Finally, we note the presence of an unrelated optical/UV source at $\approx 10\arcsec$ from ASASSN-15oi (S1 hereafter), that is also bright in \emph{Swift}-XRT observations (see \S\ref{SubSec:DataXRT}).

\subsection{X-rays: \textit{Swift}-XRT} \label{SubSec:DataXRT}
\begin{figure*}
    \centering
    \includegraphics[scale=0.7]{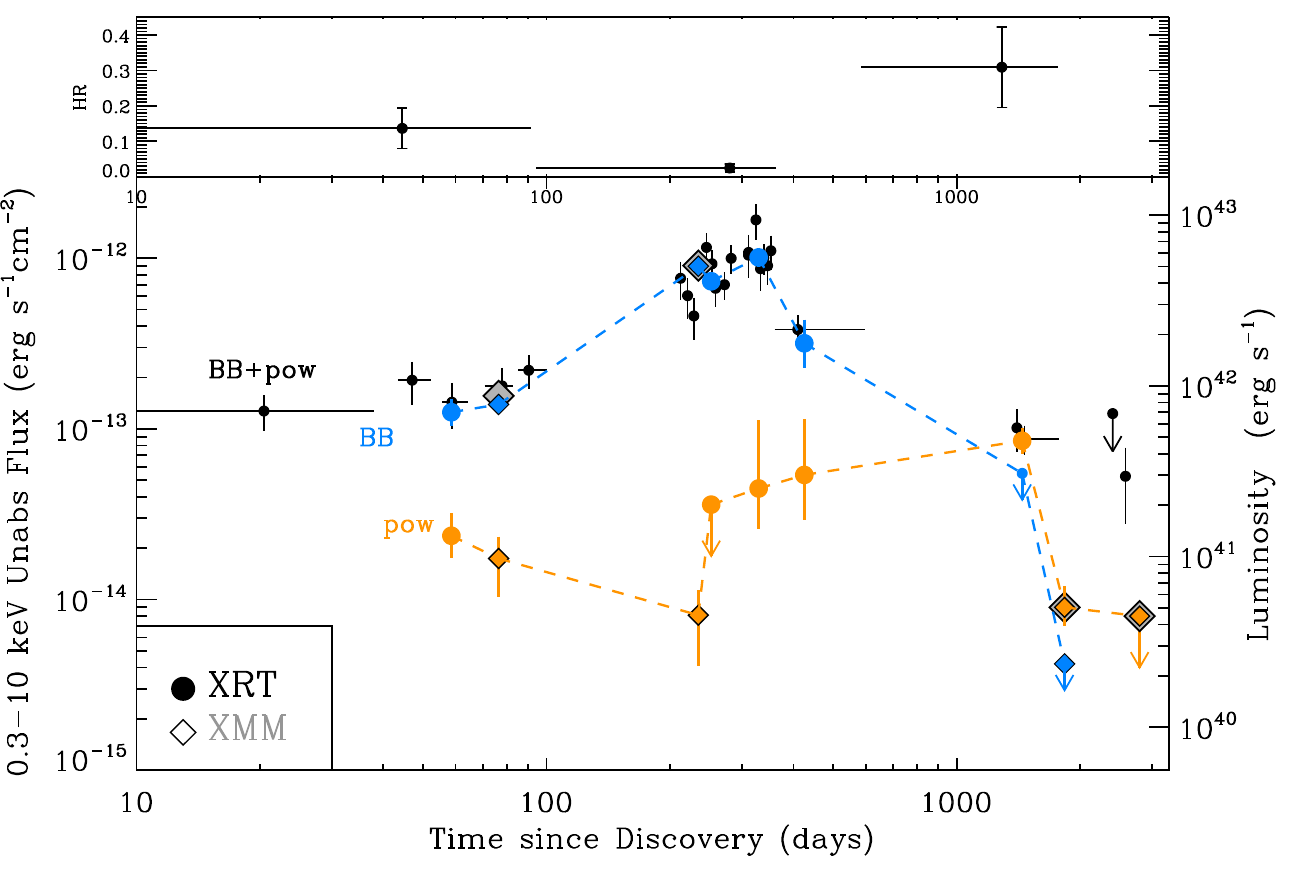}
    \caption{Long-term evolution of ASASSN-15oi as captured by \textit{Swift}-XRT (circles) and \textit{XMM-Newton} (diamonds) in 0.3-10 keV. The data have been corrected for Galactic absorption. The rise of the X-ray emission at $\delta t\le 350\,$d is dominated by a blackbody component (blue), while a power-law spectral component dominates the emission at $\delta t \ge 1400\,$d. At $> 1400\,$d, the thermal X-rays become non-detectable and the non-thermal X-rays show variability, showing a sharp decline in the latest observation at $\delta t= 2794\,$d.   \emph{Upper Panel:} Temporal evolution of the Hardness Ratio (HR) here defined as the ratio of the \textit{Swift}-XRT  counts in the 0.3--1.5 keV to the 1.5--10. keV energy bands. The very soft emission around the time of the light-curve peak is followed by hardening as the blackbody component subsides.}
    \label{Fig:XrayLC}
\end{figure*}

We analyse \textit{Swift}-XRT data \citep{Gehrels04,Burrows05} using online tools\footnote{The Swift-XRT data products generator, \url{https://www.swift.ac.uk/user_objects/docs.php}} \citep{Evans09} and custom \texttt{IDL} scripts following the prescriptions in  \cite{Margutti13}. \textit{Swift}-XRT observed  ASASSN-15oi starting on 2015 August 23 ($\delta t = 8$\,d).
Observations acquired under IDs 00033999 and 00095141 extend to 2020 June 21 
($\delta t = 1773 $\,d). They showed a progressive brightening of the source until $\delta t \approx 350$\,d, followed by rapid fading by a factor $\sim 10$ in luminosity by  $\delta t \approx 600$\,d, as can be seen in Figure \ref{Fig:XrayLC} (see also Fig.\ 3 from \citealt{Gezari17} and Fig.\ 6 from \citealt{Holoien18}). Additional XRT observations of ASASSN-15oi acquired at $\approx 1400-1460$\,d showed fainter but persistent X-ray emission. A later-time epoch of \textit{Swift}-XRT data was acquired on 2022 March 11  
($\delta t = 2401 $\,d) under ID 00096018\@. No X-ray source is detected at the location of the transient and we infer a $3\sigma$ upper limit on the count-rate of $4.3\times 10^{-3}\rm{c\,s^{-1}}$ (exposure time of 7.2\,ks, 0.3--10 keV). Finally,  ASASSN-15oi was observed with \textit{Swift}-XRT between 2022 August and October ($\delta t \approx 2540-2620$\,d).  We found marginal evidence (at the level of $\approx 3.5\,\sigma$, Gaussian equivalent) for X-ray emission at the location of ASASSN-15oi with count-rate $(1.3 \pm 0.6)\times 10^{-3}\rm{c\,s^{-1}}$ (exposure time of 7.0\,ks, 0.3 -- 10 keV).

We extracted seven spectra (Table \ref{tab:xraydata}) 
by grouping observations close in time. Following \cite{Holoien16,Holoien18,Gezari17}, we fitted the spectra in the 0.3 -- 10 keV energy range  with an absorbed two-component model featuring a (non-thermal) power-law and a (thermal) blackbody (BB), i.e., \texttt{tbabs*(cflux*pow+cflux*bbody)} within \texttt{XSPEC} v12.12.1. The Galactic neutral hydrogen column density in the direction of the transient is $N_{\rm{H,gal}} = 5.6 \times 10^{20}\,\rm{cm^{-2}}$ \citep{Kalberla05}. We found no evidence for intrinsic absorption and we thus assume $N_{\rm H,int}=0\,\rm{cm^{-2}}$. The temporal evolution of the relative flux of the thermal and non-thermal spectral components in ASASSN-15oi implies a time-varying count-to-flux conversion factor. Following \citet{Margutti13}, we use the results from the time-resolved spectral analysis to perform a self-consistent flux calibration of the count-rate light-curve. Specifically, we derive a count-to-flux conversion as a function of time since discovery by linearly interpolating in the log space the count-to-flux conversion factors derived from the seven spectra.  For each epoch, the best-fitting blackbody temperature ($kT_{\rm BB}$), blackbody radius $R_{\rm BB}$, power-law index $\Gamma_{\rm X}$, unabsorbed fluxes corresponding to respective components, and total absorbed fluxes are reported in Table \ref{tab:xraydata}. The X-ray light-curve is shown in Figure \ref{Fig:XrayLC} and the unfolded spectra are shown in Figure \ref{Fig:XMMspecevol}.

\subsection{X-rays: \textit{XMM-Newton}} \label{SubSec:DataXMM}
\begin{figure*}
    \centering
    \includegraphics[scale=0.25]{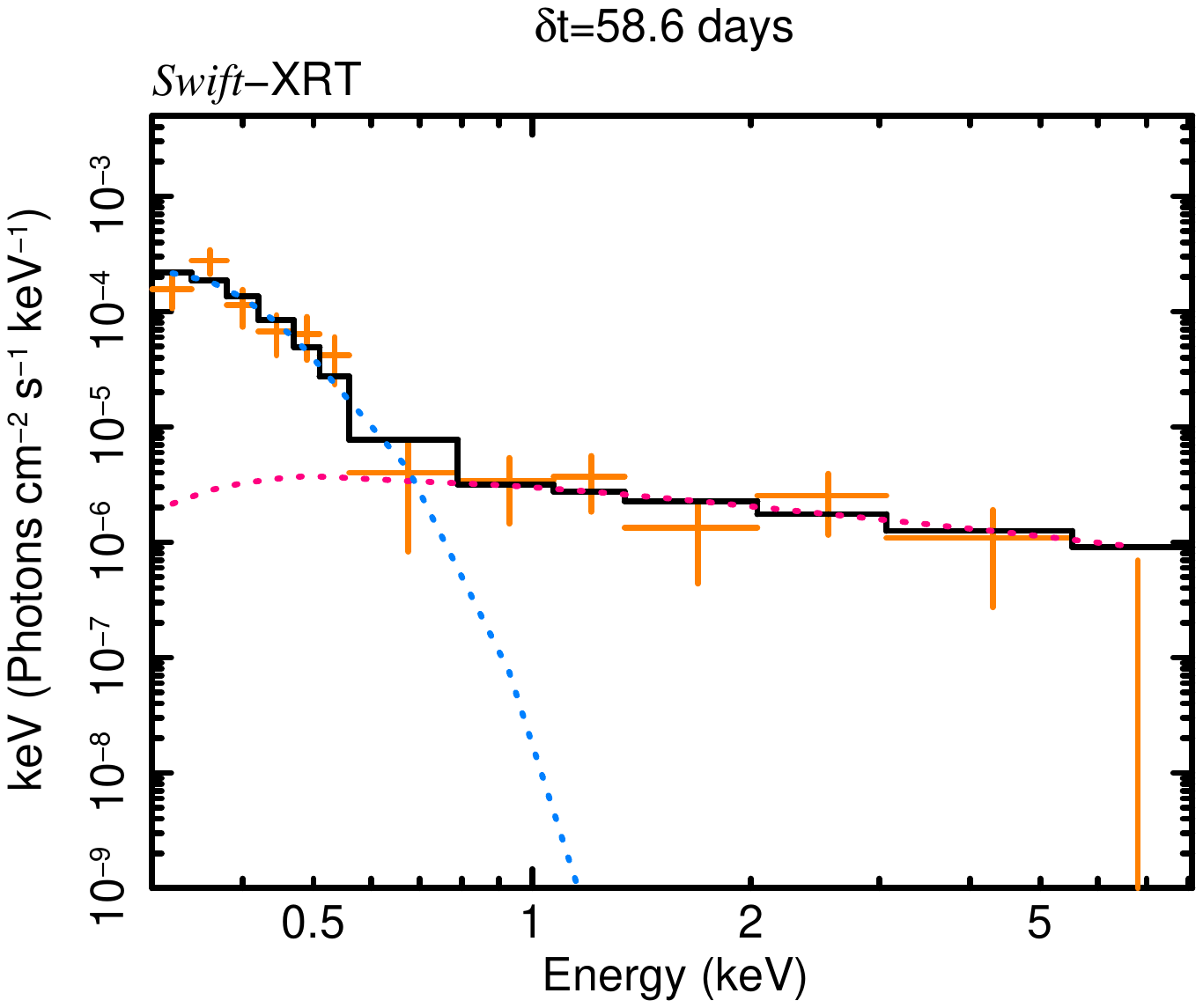}
    \includegraphics[scale=0.25]{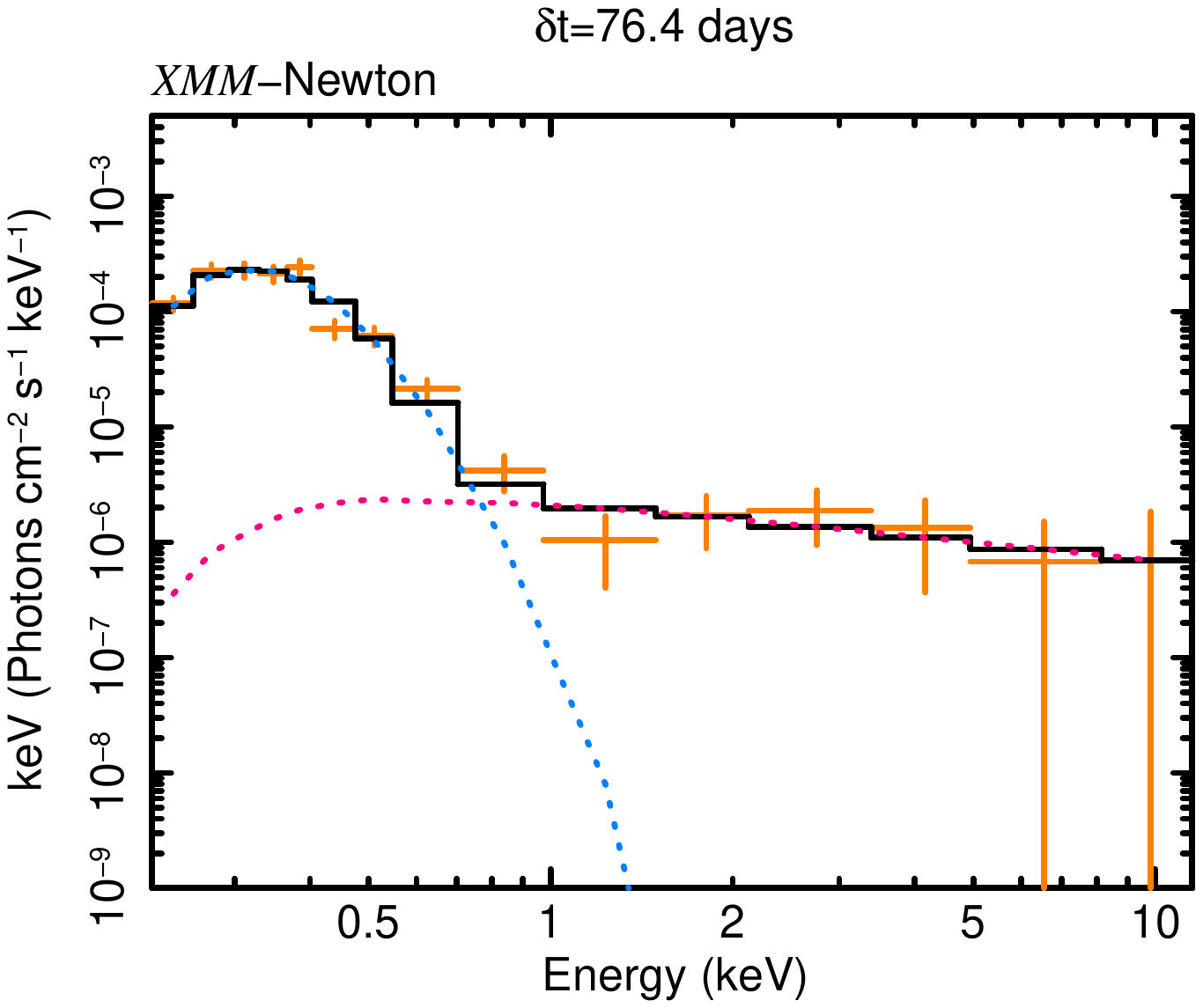}
    \includegraphics[scale=0.25]{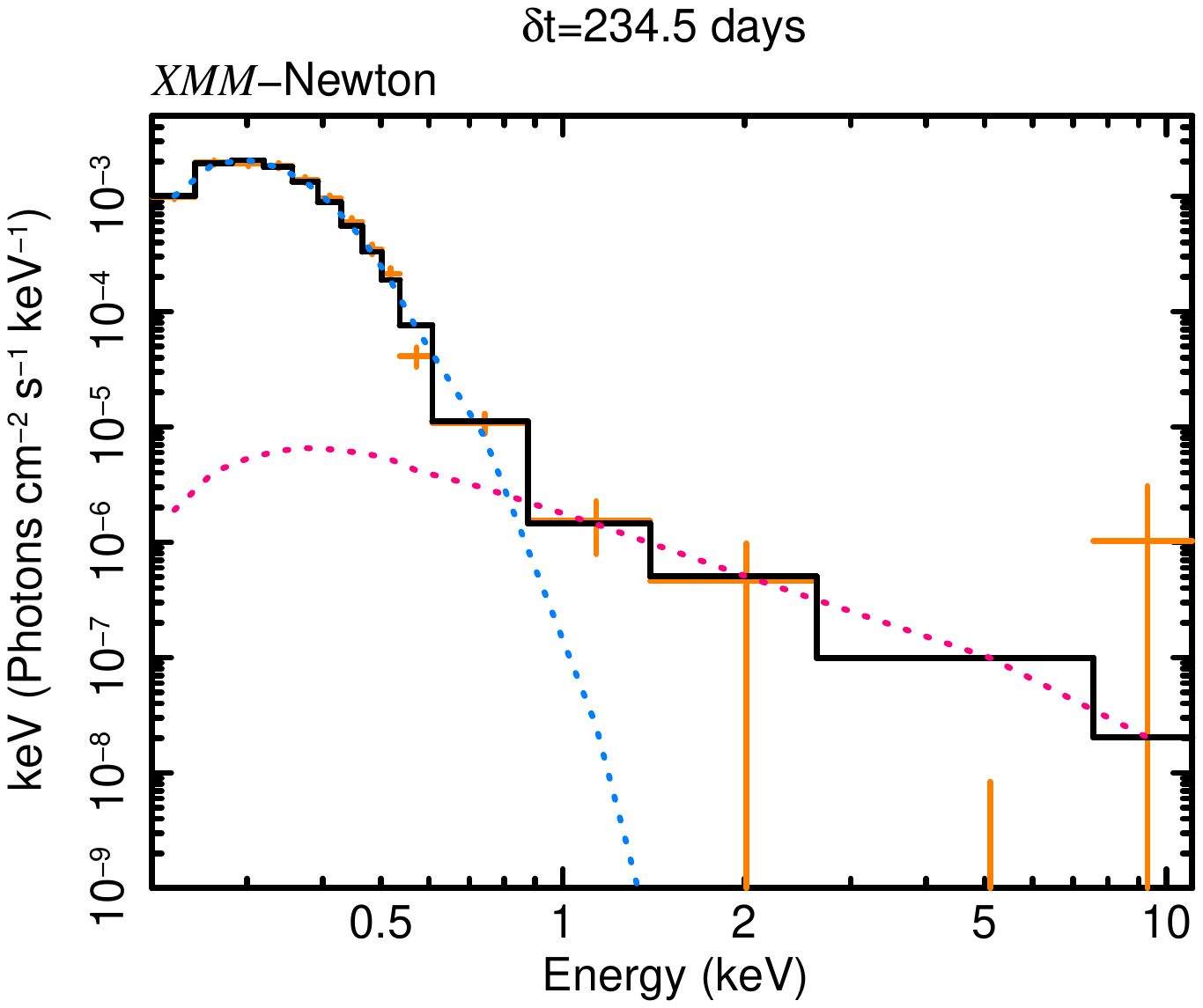}
    \includegraphics[scale=0.25]{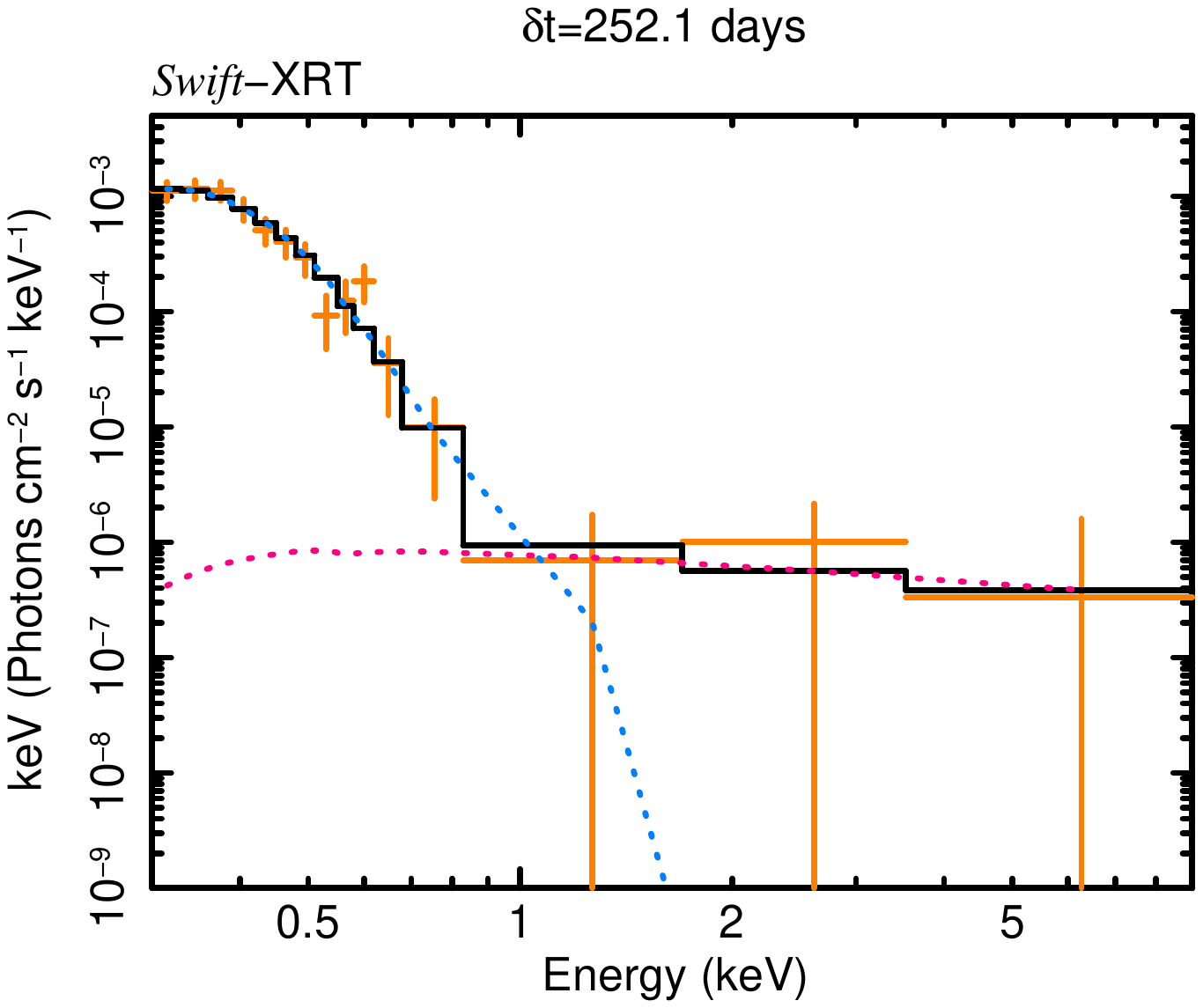}
    \includegraphics[scale=0.25]{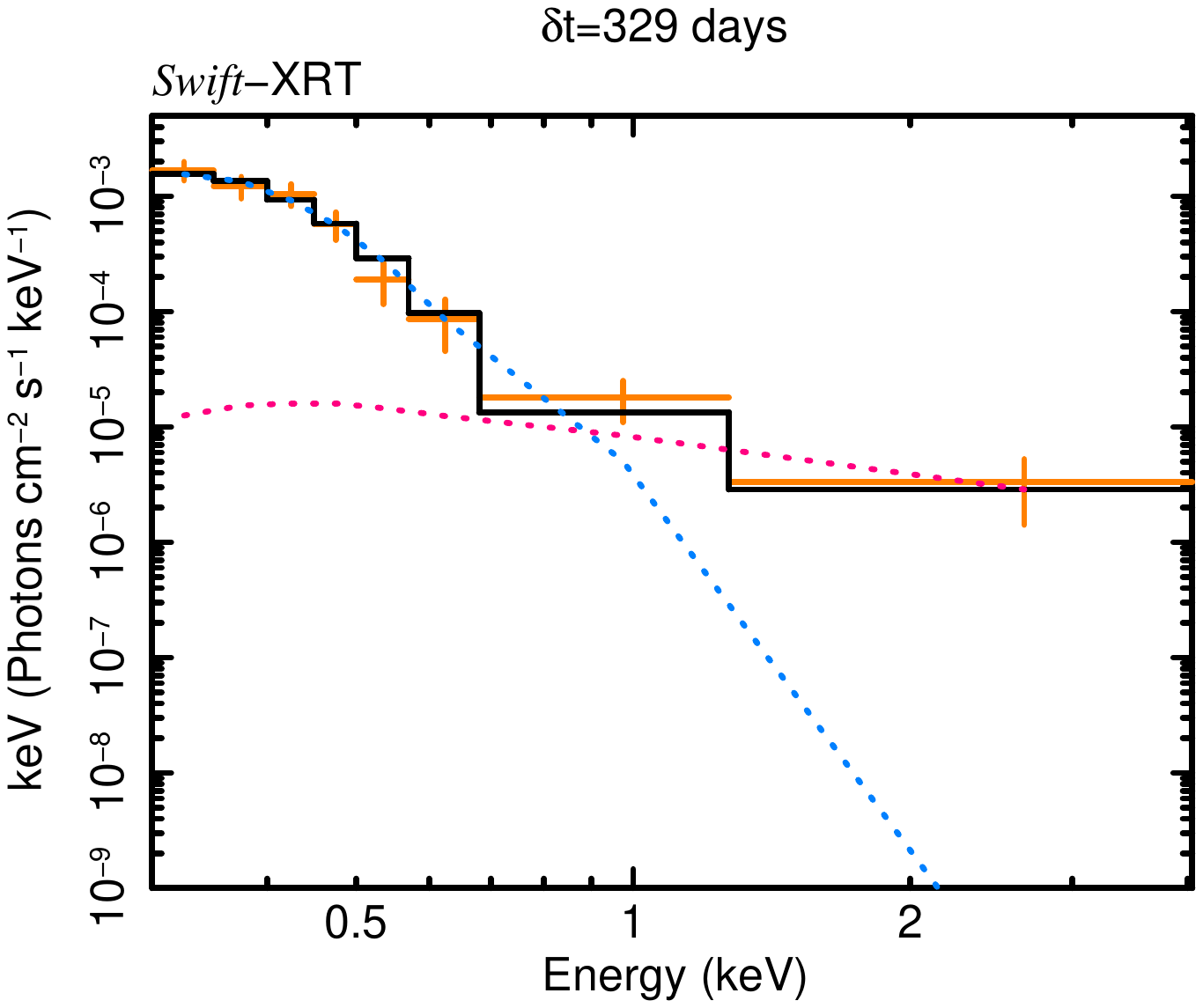}
    \includegraphics[scale=0.25]{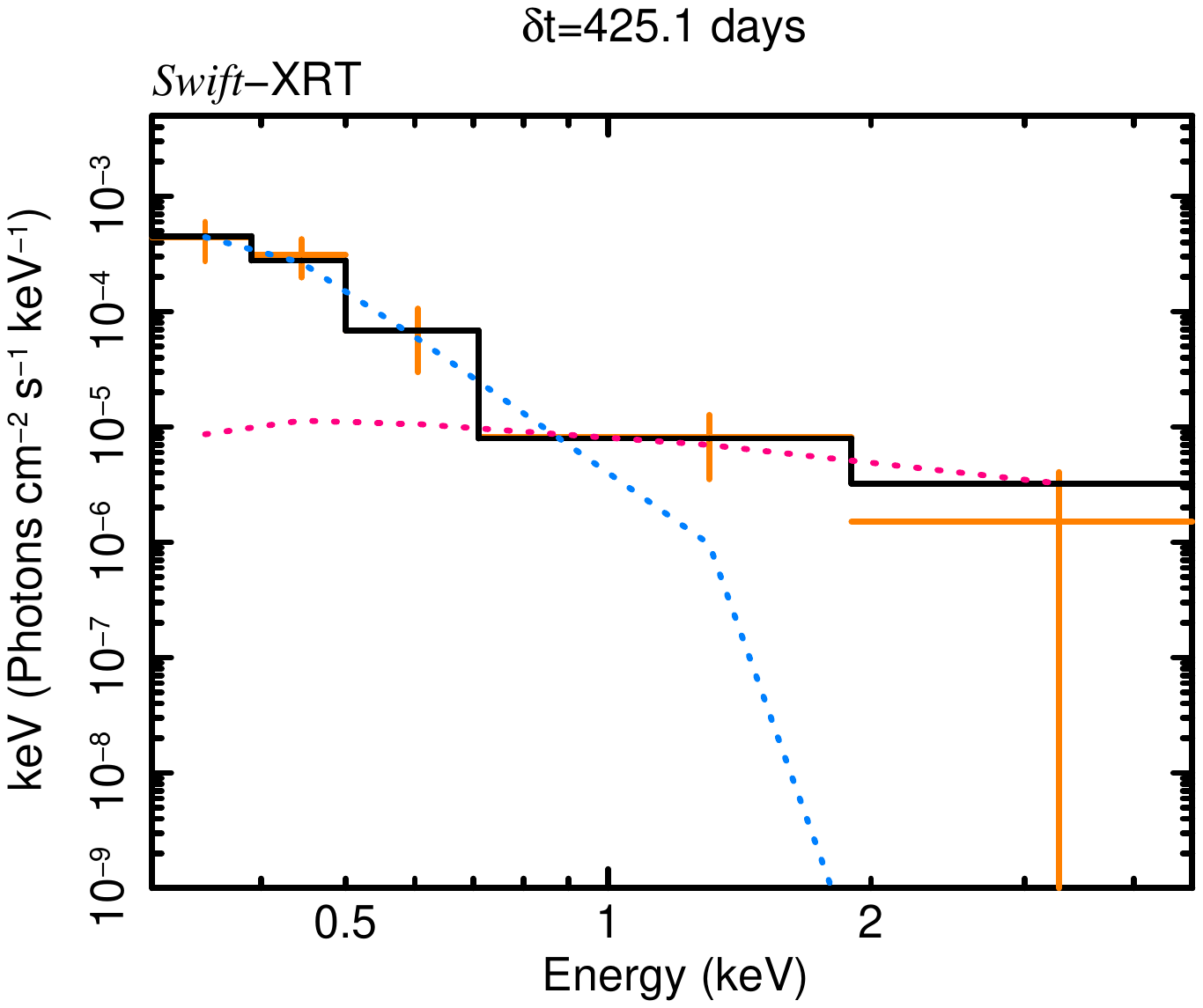}
    \includegraphics[scale=0.25]{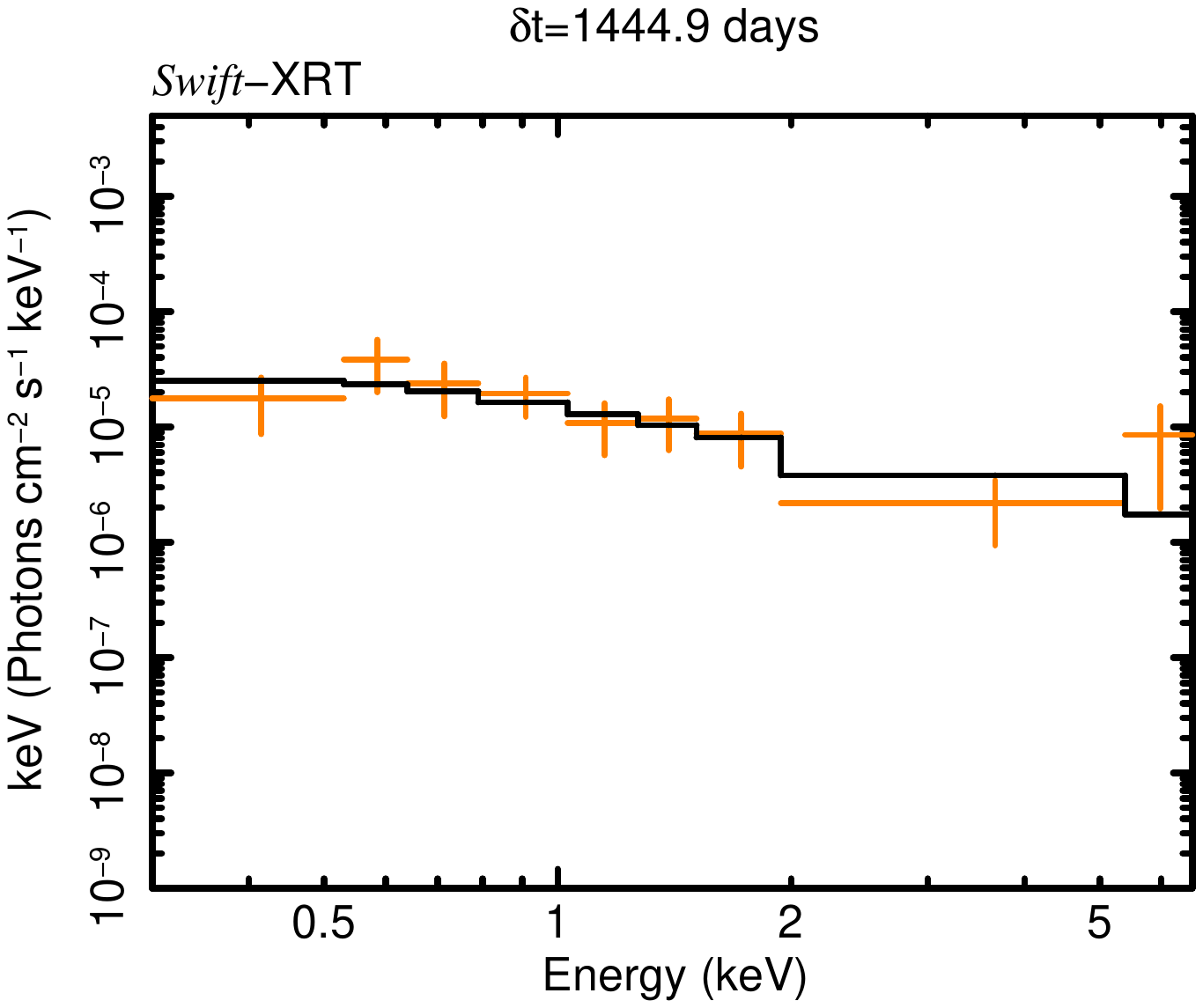}
    \includegraphics[scale=0.25]{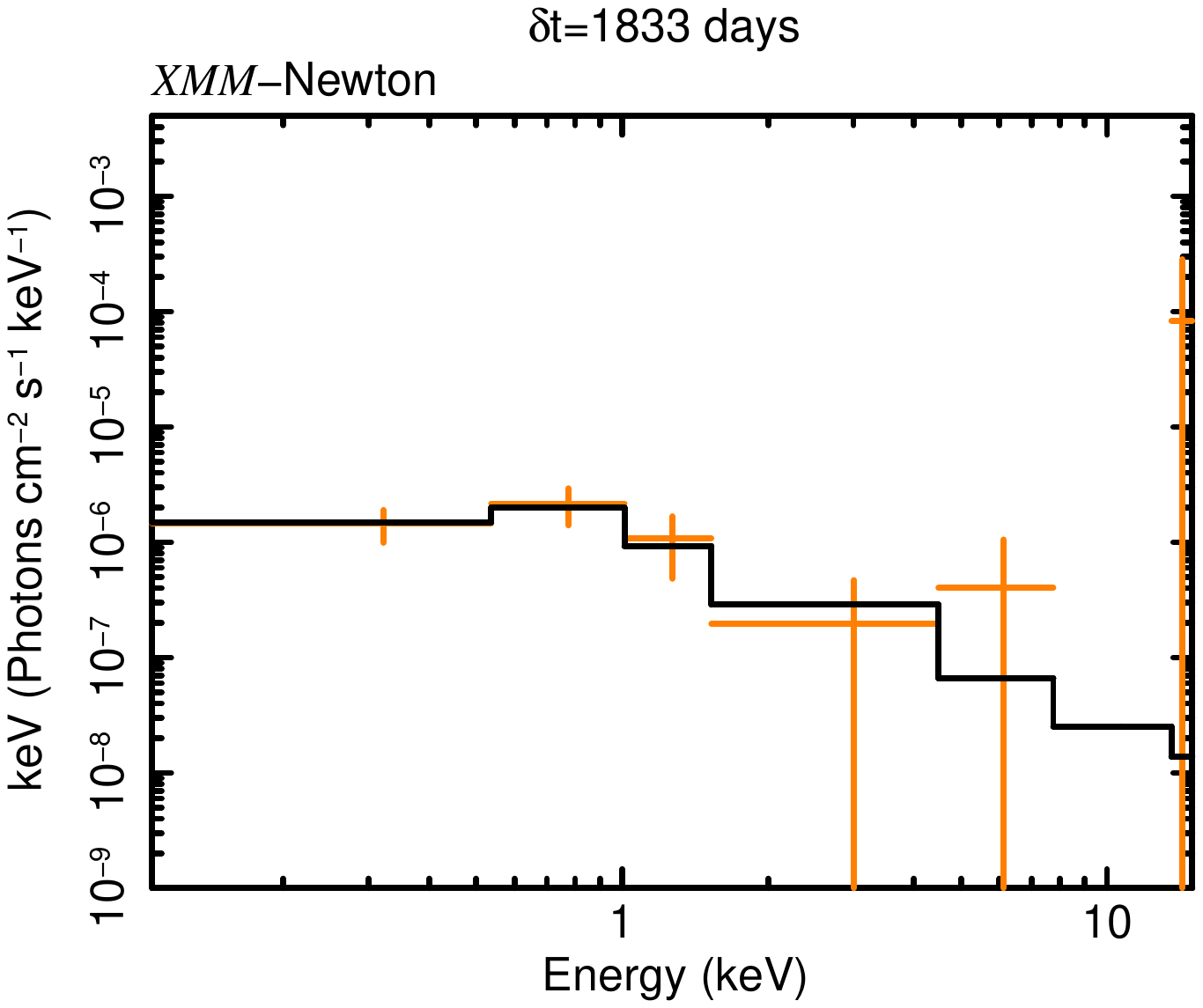}
    \caption{Unfolded X-ray spectra of ASASSN-15oi ranging from $\delta t \approx 76.4$\,d -- $1833$\,d in the 0.2 -- 12 (0.3 -- 10) keV energy range as observed by \textit{XMM-}Newton (\textit{Swift-}XRT). We plot spectra here where we have a high signal-to-noise. Blue (pink) dashed line: blackbody (power-law) component. For all the plots the y-axis covers the same dynamical range. At late times the blackbody component that was responsible for the source brightening until $\delta t\approx 350$\,d becomes undetectable and the spectrum consists of an absorbed simple power-law. }
    \label{Fig:XMMspecevol}
\end{figure*}

Following the prominent re-brightening of ASASSN-15oi at radio frequencies at $\delta t \sim 1400$\,d (\citealt{Horesh21}) we acquired deep X-ray observations with \textit{XMM-Newton} (\citealt{Jansen2001}) on 2020 October 08 ($\delta t = 1830$\,d, exposure times $\approx\,23.6, 23.8, \rm{and}\, 12.5$\,ks for MOS1, MOS2, and pn cameras, respectively; obsID 0872390301; PI Hajela). We acquired additional observations on 2023 April 08 ($\delta t = 2794$\,d, exposure times $\approx\,21.9, 21.7, \rm{and}\, 17.5$\,ks for MOS1, MOS2, and pn cameras, respectively; obsID 0903330101; PI Hajela). The results from the analysis of the earlier two \textit{XMM} observations (taken at $\delta t = 76.4$ and $\delta t = 234.5$\,d; PI Gezari) were reported in \citet{Gezari17}  and \citet{Holoien18}. 

All \textit{XMM} observations were performed with the European Photon Imaging Camera (EPIC) detector in full-frame mode with the thin filter. We processed all the observations using standard routines in the Scientific Analysis System (SASversion 1.3) software package and the corresponding calibration files. Filtering out time intervals with high background flaring activity results in the net exposure times\footnote{We note that the slightly different net exposure times in \cite{Gezari17} for the first two observations likely result from different filtering criteria.} reported in Table \ref{tab:xraydata}.
An X-ray source is clearly detected in three \textit{XMM} observations at the position of the radio/optical transient. However, there is no evidence of significant emission in the last epoch at $\delta t = 2794$\,d. We extracted a spectrum from a circular region of 15\arcsec\ (300 pixels) radius centered at the source position, and used a nearby 75\arcsec\, source-free background region. We perform our spectral analysis using the data from all three EPIC cameras\footnote{\cite{Gezari17} used only observations from EPIC-pn camera. Since pn detector has the largest number of counts, the results will remain unaffected.} (MOS1, MOS2 and pn). 

 We fit the data in the 0.2 - 12\,keV energy range with the same two-component model that we used for \emph{Swift}-XRT data: \texttt{tbabs*ztbabs*(cflux*pow+cflux*bbody)} within \textit{XSPEC} (v.12.12.1).  We do not find evidence for significant intrinsic absorption from a joint-fit of the the XMM epochs at $\sim 80 - 1400$\,d ($N_{\rm{{H,int}}} < 1.4 \times 10^{20}\,\rm{cm^{-2}}$), therefore we proceed with $N_{\rm{H,int}} = 0$ as before. We used $N_{\rm{H,gal}} = 5.6 \times 10^{20}\,\rm{cm^{-2}}$  and $N_{\rm{H,int}} = 0\,\rm{cm^{-2}}$, and report the best-fitting parameters and fluxes in Table \ref{tab:xraydata}, plot the light-curve in Figure \ref{Fig:XrayLC}, and the unfolded spectra in Figure \ref{Fig:XMMspecevol}.

Our results for the first two epochs (at $\delta t = 76.4$\,d and $234.5$\,d) are broadly in agreement with those of \cite{Gezari17} with an exception of the power-law X-ray flux that they report as constant between epochs, but we identify as varying.
In our new observation at $\delta t=1833$\,d, we find that a blackbody component is not required and the data are well explained by a non-thermal power-law spectrum only (Figure \ref{Fig:XMMspecevol}). This result is consistent with our findings from the \emph{Swift}-XRT spectrum at $\delta t \approx 1400\,$d, which was also dominated by the power-law spectral component (\S\ref{SubSec:DataXRT}).  Our latest \emph{XMM} observation at $= 2794$\,d shows a sharp flux decline as the emission becomes non-detectable.

Taken together, the \emph{Swift}-XRT and \textit{XMM-Newton} campaigns lead to the following observational results\footnote{Pre-TDE \emph{ROSAT} observations of the host galaxy 2MASX\,J20390918--3045201 provide no useful constraint: \cite{Holoien18} report a pre-TDE X-ray flux limit of $F_x<1.9\times 10^{-12}\,\rm{erg\,cm^{-2}s^{-1}}$ (0.1--2.4 keV), which is larger than the measured X-ray fluxes for ASASSN-15oi.}. (i) Since optical discovery, this transient has radiated $\sim 10^{51}\,\rm{erg}$ in X-rays. (ii) While S1 (\S\ref{SubSec:DataUVOT}) falls within the \emph{XMM} and \emph{Swift}-XRT PSF, we found no evidence for a shift of the centroid of the source of X-ray emission between the second (blackbody dominated) and the third  (power-law dominated) \emph{XMM} epochs. We thus exclude with confidence the possibility that the X-ray power-law is associated with S1 and unrelated to the TDE. (iii) For $\delta t\lesssim 425\,$d the soft X-ray spectrum consists of a mixture of a thermal blackbody component with $kT_{\rm BB}\sim0.05\,$keV and a non-thermal power-law component with $\Gamma_{\rm X}\sim 2$, i.e., $F_{\nu}\propto \nu^{-1}$ (Figure \ref{Fig:XMMspecevol}). (iv) The brightening of X-ray emission by a factor $\approx 6$ in the time interval $\delta t\approx 10-330\,$d is due to the flux increase of the blackbody component (Figure \ref{Fig:XrayLC}). During this time the $T_{\rm BB}$ shows no evolution, whereas the effective $R_{\rm BB}$ increases with time from $\sim 2.3\times10^{11}\,\rm{cm}$ to $\sim 1.3\times10^{12}\,\rm{cm}$.  Results (iii)+(iv) confirm the findings of \cite{Holoien16,Holoien18,Gezari17}.
(v)  Differently from \cite{Holoien18}, we find that the blackbody component rapidly fades between $330 - 425$\,d  and becomes undetectable by the time of the following \textit{Swift}-XRT observation at $\delta t\approx 1400\,$d. (vi) The non-thermal power-law component in the spectra is consistently present and dominates at $>1400$\,d. The power-law flux initially decreases until around $\approx 230$\,d when the blackbody flux increases. It then brightens rapidly until $\approx 330$\,d, followed by a gradual increase until $\approx 1400$\,d. At $\gtrsim 1400$\,d, significant variability is observed with a dip in the observed flux at $\approx 1800$\,d between the brighter $\approx 1400$\,d and $\approx 2580$\,d observations.

\subsection{Radio: VLA} \label{SubSec:DataVLA}

\cite{Horesh21} started monitoring ASASSN-15oi with the VLA at $\delta t = 8$\,d. They reported the first radio detection at $\delta t = 182$\,d, showing a significant brightening of the source compared to earlier upper limits (Figure \ref{fig:radioLCcomparisonAH}).  The first radio brightening was followed by a second, even more dramatic radio re-brightening at $\delta t>1400$\,d.  ASASSN-15oi's second radio flare was discovered in VLASS observations \citep{Lacy20}. It was observed as part of regular survey operations on 2019 July 1 ($\delta t = 1417$\,d), revealing that the flux density at 3 GHz had increased by a factor of $\sim3000$ since the 2017
observations of 
\cite{Horesh21}. To further explore the nature of this second radio flare, we obtained multi-frequency observations with the VLA under the DDT program 20A-492 (PI: Alexander). The data were taken when the array was in the C configuration at $\delta t = 1741$\,d. 
To monitor the spectral and temporal evolution, we obtained subsequent broadband observations with the VLA under the programs 21A-303 (PI Hajela; $\delta t = 2129$\,d) in the B configuration, and 23A-241 (PI Cendes; $\delta t = 2970$\,d) in the A configuration. We observed ASASSN-15oi in a standard phase referencing mode for 18-20 minutes at the mean frequencies of 1.5\,GHz (L-band), 3\,GHz (S-band), 6\,GHz (C-band), and 10\,GHz (X-band). We used 3-bit samplers for C and X bands and 8-bit samplers for L and S bands. 

For the observations acquired between $\delta t = 1741 - 2129$\,d, we use the VLA calibration pipeline packaged with \textsc{CASA} v.6.2.1.7, with 3C\,48 = J0137+3309 as the flux calibrator, and ICRF J210101.6$-$293327 (J2101$-$2933) as the complex gain calibrator. For $\delta t = 2970$\,d, we used 3C\,286 = J1331+3030 as the flux calibrator and the new VLA calibration pipeline packaged with \textsc{CASA} v.6.5.4.9. After manually inspecting the data, we further flagged antennas with bad solutions as well as additional weak radio-frequency interference and then re-ran the pipeline. To densely sample the SED, we divide the dataset in every observing band further into sub-bands, and image each sub-band individually using the \texttt{CLEAN} algorithm with Briggs weighting and a robust factor of $0.5$ in \textsc{CASA}. 
Where necessary, we also use the automated self-calibration pipeline.
We note the presence of a bright source $\sim 0.3\,\arcdeg$ away in the 1.5- and 3-GHz bands, however its residual sidelobes are below the image background rms near ASASSN-15oi. ASASSN-15oi is detected as a very bright radio source at all frequencies. We measure the flux density of ASASSN-15oi using the {\tt imtool} package within {\tt pwkit} \citep{Williams17} and report the results in Table \ref{tab:radiodata}. Since the flux density scale calibration has an accuracy of 3\% -- 5\% for our observing bands, we add a conservative 5\% systematic uncertainty in quadrature to our uncertainties. We also include this uncertainty in the early-time observations reduced by \cite{Horesh21}.

\subsection{Radio: VLBI} \label{SubSec:DataVLBI}

We observed ASASSN-15oi with the VLBA of
the National Radio Astronomy Observatory (NRAO), using all antennas
except Brewster, under the program VLBA/22A-382 (PI: Hajela).  The total observing time was $5.5$\,h, and the midpoint of the observations was of 2022 February 22.75 (UT; MJD = 59632.75; $\delta t = 2384$\,d). We recorded a total bandwidth of 512 MHz, centered on $8.30$\,GHz, in  two polarizations, using a total bitrate of 4096 Mbps. 
The VLBI data were correlated with NRAO’s VLBA processor,
and the analysis carried out with NRAO’s \textsc{AIPS} (Astronomical Image Processing System; \citealt{Greisen_aips}).  The initial flux density calibration was done through measurements of the system temperature at each telescope, and then improved through self-calibration of the reference source.  A correction was made for the dispersive delay due to the ionosphere using the \textsc{AIPS} task \texttt{TECOR}, although the effect at our frequency is not large.  We phase-referenced the
observations of ASASSN-15oi to PMN J2036$-$2830, using a
$\sim 3.7$ min cycle, of which $\sim 2.5$~min were on ASASSN-15oi.

\begin{figure}[th]
    \centering
    \includegraphics[scale=0.45]{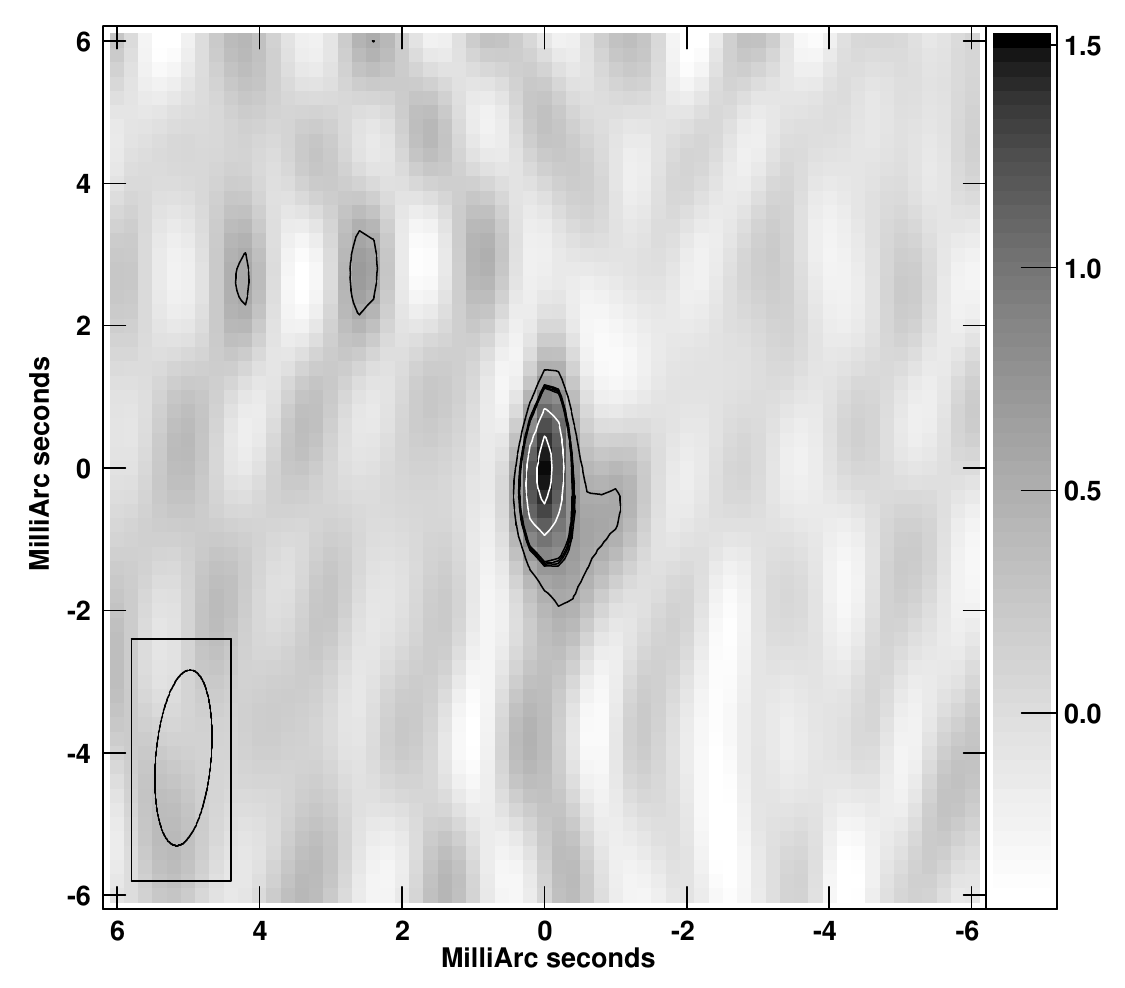}
    \caption{A VLBI image of ASASSN-15oi on 2022 February \,22 ($\delta t=2384\,$d), at $8.3$~GHz.
    The peak brightness was $1.5$ mJy~beam$^{-1}$, and the image
    background rms brightness was $140$ $\mu$Jy~beam$^{-1}$.  The
    greyscale is labelled in mJy\,beam$^{-1}$.  The
    contours are drawn at $-35$, $35$, {\bf $50$} (emphasized), 70 and 90\% of the peak brightness.  The FWHM resolution is shown at lower left, and was $2.47 \times 0.78$\,mas at p.a.\ $-5$\arcdeg.  North is up and east to the left.}
    \label{fig:vlbi}
\end{figure}

\begin{figure}
    \centering
    \hspace*{-0.2in}
    \includegraphics[scale=0.29]{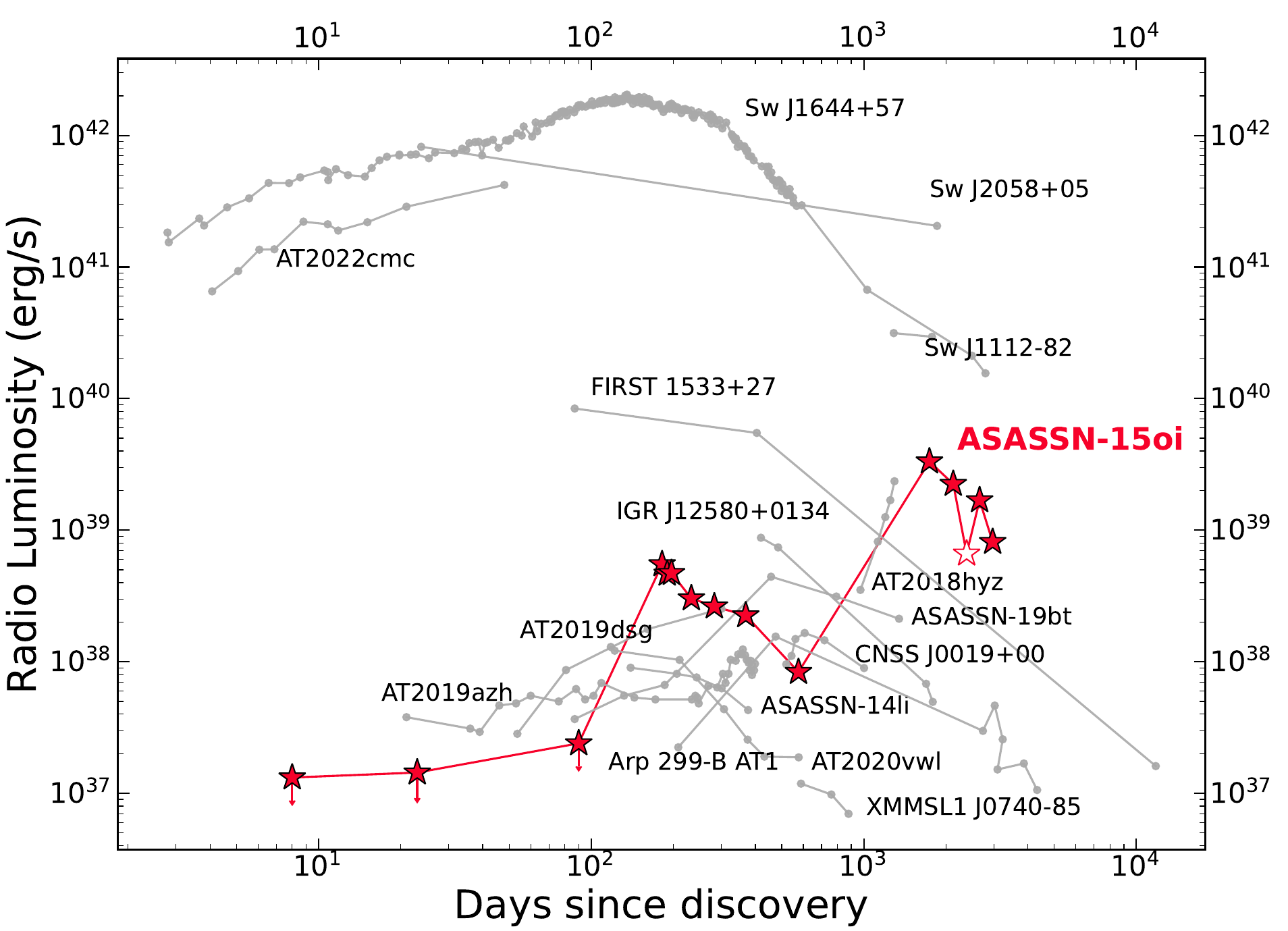}
    \caption{Temporal evolution of the radio luminosity $\nu L_{\nu}$  of ASASSN-15oi ($7-8.5$\,GHz; red stars)
    in the context of selected radio-bright TDEs. The filled red stars show VLA and ATCA measurements, while the VLBA one is shown as an open red star. Other TDEs are: \emph{Sw}\,J1644+57 ($15.5$\,GHz compiled from \citealt{Zauderer11,Berger12,Zauderer13,Eftekhari18,Cendes21_J1644}), AT\,2022cmc ($11.5$\,GHz from \citealt{Andreoni_cmc_2022}) among the jetted-TDEs; AT\,2018hyz ($5-8$\,GHz from \citealt{Cendes_18hyz_2022}), AT\,2019azh ($15.5$\,GHz from \citealt{Sfaradi_19azh}), ASASSN-19bt ($5.5$\,GHz from \citealt{Christy2024}), AT\,2020vwl ($9$\,GHz from \citealt{Goodwin_2020vwl_2023}). Luminosity of all other TDEs have been adopted from \cite{Alexander20} in $3-9$\,GHz frequency range. At $\delta t\approx 1800$\,d, ASASSN-15oi is the most luminous known radio TDE among those not reported to be associated with relativistic jets.}  
    \label{fig:radioLCcomparisonAH}
\end{figure}

We show the VLBI image of ASASSN-15oi, made using the \texttt{CLEAN} algorithm, in Fig.\ \ref{fig:vlbi}, 
The total \texttt{CLEAN}ed flux density is $1.2$ mJy, and the image background rms is $140$\,$\mu$Jy\,beam$^{-1}$. For an ideal, unresolved source, the peak brightness should also be $1.2$\,mJy, however, the peak  brightness in the image is somewhat larger $1.5$ mJy\,beam$^{-1}$. 
So, the relatively low-dynamic-range (11:1\@) image allows only an approximate determination of the flux density. 
The angular size of ASASSN-15oi is smaller than our FWHM resolution of 
$2.47 \times 0.78$\,mas at position angle (p.a.) $-5$\arcdeg. 

For marginally- or unresolved sources, such as ASASSN-15oi, the best 
values or limits for the flux density and source size come from fitting 
models directly to the visibility data \citep[e.g.,][]{Bietenholz+2010}, rather than from imaging.  Fitting a circular Gaussian to the visibilities  by least squares (\textsc{AIPS} task \texttt{OMFIT}), we find that the best fit  has a FWHM size of $0\pm1$\,mas and a flux density of 
$1.4\pm0.3$ mJy, with the caveat that the fitted size is 
positively correlated with the fitted flux density. At the distance of ASASSN-15oi, the FWHM translates to a $3\sigma$ physical source size of $\sim 9 \times 10^{18}$\,cm.

\begin{figure}
    \centering
    \includegraphics[scale=0.3]{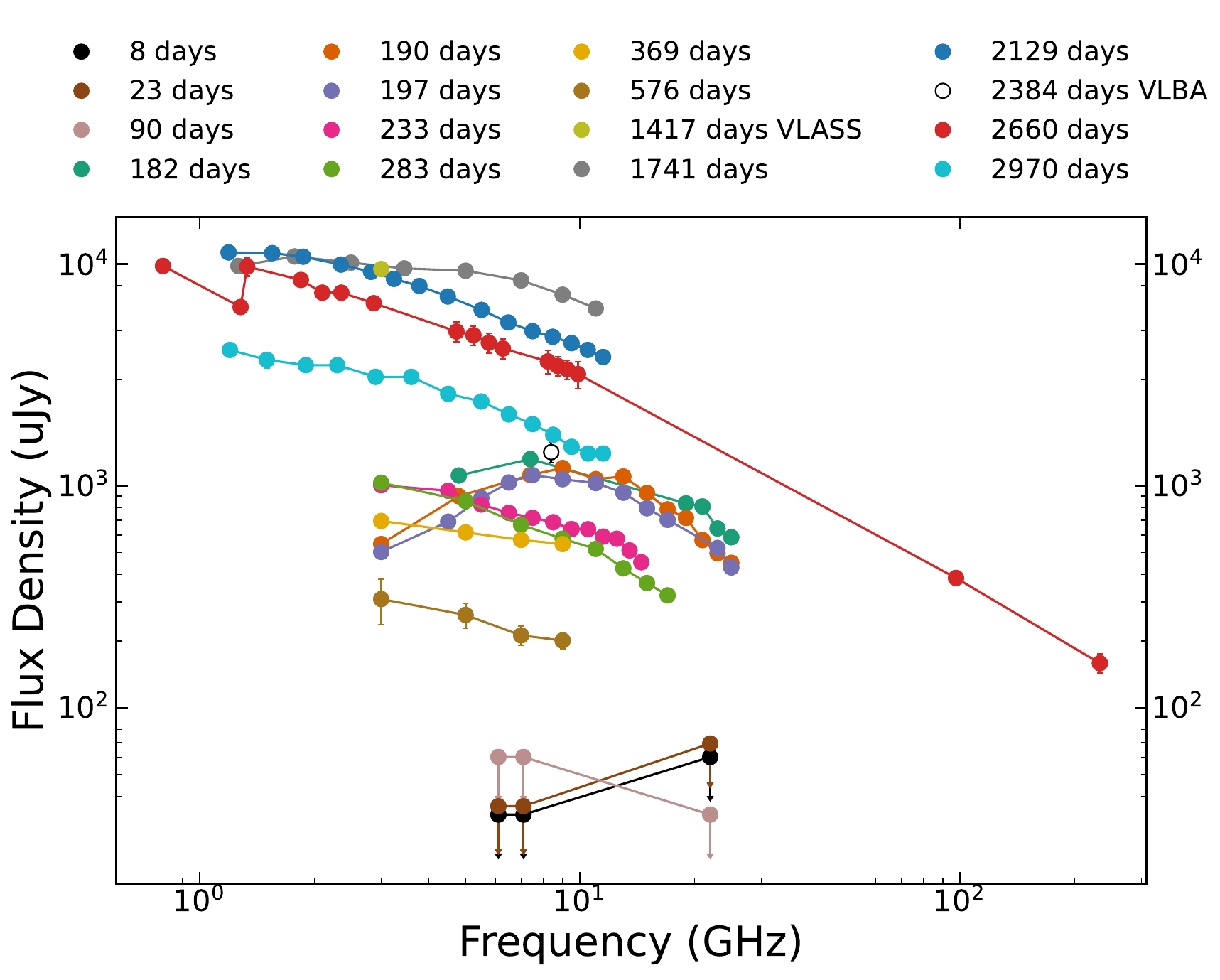}
    \caption{Evolution of the emission from ASASSN-15oi across the radio spectrum from the time of the first radio detection at $\delta t = 8\,$d extending to $\delta t = 2970\,$d.   The data at $\delta t\ge 1741\,$d are presented here for the first time. All the flux densities are reported in Table \ref{tab:radiodata}. The single frequency VLASS and VLBA observations are highlighted separately in the plot legend.}
    \label{fig:radioseds}
\end{figure}

\subsection{Radio: ATCA} \label{SubSec:DataATCA}
We acquired observations of ASASSN-15oi (Project code: C3325; PI: Alexander) on 2022 September 30 ($\delta t = 2596$\,d) with ATCA. The observations were conducted in the 6D configuration, consisting of all six 22 m diameter antennae \citep{wilson2011}.  The observing bands were centered at frequencies 2.1, 5.5, and 9\,GHz, with a total bandwidth of 2\,GHz in each of the bands. For calibration purposes, PKS\,1934$-$638 served as the primary calibrator for all observing bands, whereas PKS\,1921$-$293 (J2024$-$3253) was observed in $\sim$2.5 minute scans enclosing each scan of ASASSN-15oi, and was used as the secondary calibrator to calibrate the time-varying complex gains in the 2.1\,GHz (5.5 and 9\,GHz) observing band(s).

We carry out the data reduction following the standard calibration procedures in \textsc{Miriad} \citep{miriad}, with some additional manual 
flagging. 
We also split the data into sub-bands, followed by imaging using inverted visibilities, and deconvolving the images using the multi-frequency synthesis \texttt{CLEAN} algorithm \citep{Hogbom1974, Clark1980, Sault1994}, and calculated the flux density by fitting a point source model to the target in the image plane. 
We add 10\% systematic uncertainty in quadrature to our measurements \citep{atca_uncertainty}.

\subsection{Radio: MeerKAT} \label{SubSec:DataMeerKAT}

We obtained observations with the MeerKAT radio telescope in 580 -- 1015 MHz (UHF) and 0.8 -- 2 GHz (L) bands on 2022 December 9 (SCI-20220822-YC-01, PI: Cendes; $\delta t = 2674$\,d) and 2023 January 5 (DDD-20220414-YC-01, PI: Cendes; $\delta t = 2701$\,d).  We used ICRF J193925.0$-$634245
as the flux density calibrator and 
QSO PKS J2052$-$3640 as the phase and gain calibrator.
We use the calibrated images obtained from the SARAO Science Data Processor pipeline (SDP)\footnote{\url{https://skaafrica.atlassian.net/wiki/spaces/ESDKB/pages/338723406/}}.  We measured the flux density of ASASSN-15oi using the {\tt imtool} package within {\tt pwkit} \citep{Williams17}. Since these observations are close in time to the ATCA observations (\S\ref{SubSec:DataATCA}) with a mean observation epoch $\approx 2660$\,d, we model them together in \S\ref{Sec:RadioModeling}.

\subsection{Radio: ALMA} \label{SubSec:DataALMA}

Finally, we obtained mm-wavelength observations of ASASSN-15oi with ALMA under program 2019.1.01166.T (PI: Alexander). A single epoch of observations at Bands 3 and 6 (mean frequencies 97.5 GHz and 232 GHz respectively) was obtained on 2022 September 29 ($\delta t \approx 2128$\,d). The data were reduced and imaged using the standard NRAO ALMA pipeline in \textsc{CASA}. We add a $3\%$ and $6\%$ systematic uncertainty in quadrature to our Band 3 and Band 6 measurements, respectively. 

All the radio flux densities are recorded in Table \ref{tab:radiodata}. The radio light curve and SEDs are shown in Figures \ref{fig:radioLCcomparisonAH} and \ref{fig:radioseds}, respectively. In Figure \ref{fig:radioLCcomparisonAH}, we note that the VLBA measurement of $1.4 \pm 0.3$\,mJy, at $8.30$\,GHz is lower by $5 \sim
6 \sigma$ than the flux densities of $4.7 \pm 0.2$ and $4.2\pm0.4$\,mJy
measured with the VLA and ATCA at comparable times and frequencies (\S
\ref{SubSec:DataVLA} and \ref{SubSec:DataATCA}, respectively; see Table \ref{tab:radiodata}). This discrepancy could indicate true radio variability at timescales of $\sim 300$\,d at later stages of ASASSN-15oi's evolution, however a low signal-to-noise ratio of the VLBI image (as mentioned in \S\ref{SubSec:DataVLBI}) and a relatively low source declination ($\sim -30{\arcdeg}$) for VLBA\footnote{\url{https://science.nrao.edu/facilities/vlba/docs/manuals/oss/vlba-declination-limits}} may impact this flux density measurement. In addition, since the radio emission measured by other telescopes in the nearby epochs show a simple trend of $t^{-2}$, the case of variability less likely. We therefore, do not include the VLBA measurement in our analysis. To alleviate such issues in the future, we encourage simultaneous observations with the VLBA and other radio telescopes, e.g., VLA, to get full spectral information.   
We finally also note the discrepancy between the MeerKAT and ATCA data in the $2660$\,d SED at $\sim 2$GHz in Figure \ref{fig:radioseds}.  
However, these are marginally consistent within $3\sigma$, and the difference could be due to inter-calibration issues between the two telescopes. In the next section we model the radio SEDs and for this particular frequency and epoch, we proceed with assuming $3\sigma$ uncertainties on the MeerKAT measurement, instead of the usual $1\sigma$ for all others.

\section{Radio Data Modeling} \label{Sec:RadioModeling}
\subsection{Fitting of Individual SEDs}\label{subsec:sedfit}
We show the radio SEDs of ASASSN-15oi in Figure \ref{fig:radioseds}. Spectral evolution is evident, with some SEDs described by a single power-law ($F_{\nu} \propto \nu^{\alpha}$), and others requiring up to three power-law segments. To fit the SEDs containing \emph{one} break we use the following parametric model:

\begin{equation}
\label{eq:BPL}
    F_\nu = F_{b_1} \Big[ \Big( \dfrac{\nu}{\nu_{b_1}}\Big) ^{\alpha_i/s_1} + \Big( \dfrac{\nu}{\nu_{b_1}}\Big) ^{\alpha_j/s_1} \Big] ^{s_1}
\end{equation}

\noindent and to accommodate \emph{two} breaks, we modify the above model as follows:

\begin{equation}
    \label{eq:BPL_mod}
    \begin{split}
    \rm{Equation\,\ref{eq:BPL}} \times \left[ 1 + \Big( \frac{\nu }{\nu_{b_2}} \Big) ^{(\alpha_k -\alpha_j)/s_2}\right]^{s_2},
    \end{split}
\end{equation}

\noindent where $\nu_{b_{1,2}}$ are the two break frequencies, $s_{1,2}$ define the sharpness of the two breaks (larger $|s|$ represents a smoother break), and $\alpha_{i,j,k}$ are the power-law indices corresponding to the three spectral segments. 
The optically thick emission manifests as a spectrum with positive slope ($\alpha > 0$) that we label $\alpha_{\rm thick}$. The optically thin emission has a negative slope ($\alpha < 0$) and is labelled with $\alpha_{\rm thin}$. If both of these segments are present, their asymptotic intersection is the peak of the SED and is represented by $F_{\rm pk} = F_{\rm b_1}$ and $\nu_{\rm{pk}} = \nu_{\rm b_1}$. If there is an additional break frequency $\nu_q$ on the optically thin section of the SED, the power-law index of the steeper segment at higher frequencies is referred to as $\alpha_q$. 

We perform a Markov Chain Monte Carlo (MCMC) fit (using the \texttt{emcee} python package; \citealt{emcee}) to the individual radio SEDs. For SEDs featuring both $\alpha_{\rm thin}$ and $\alpha_q$ spectral segments, we introduce an additional constraint to our fitting procedure i.e. $\alpha_q - \alpha_{\rm thin} = -0.5$\footnote{
This assumption is valid for the physical models used in \S\ref{subsec:syncphysparams} and \S\ref{subsec:offaxisjetparams}.}. We provide the details of the fits and our assumptions in Appendix \ref{sec:AICBIC} in \S\ref{subsec:firstflaresedfit} and \S\ref{subsec:secondflaresedfit}. Due to the complex spectral shapes of the SEDs, we fit them with different models and calculate the Akaike Information Criterion (AIC) and the Bayesian Information Criterion (BIC) values for comparison. The details of our model selection are given in \S\ref{subsec:aicbic} and Table \ref{tab:modelselection}. 

We briefly summarize our results for each SED fit here. For the first radio flare SEDs $\#1,2,3$ (see Table \ref{tab:radiosedfitting}),
we perform a fit with both Equation \ref{eq:BPL} (as was also done by \citealt{Horesh21}) and Equation \ref{eq:BPL_mod}.  Based on the AIC and BIC values in Table \ref{tab:modelselection}, both fits to these SEDs are statistically equivalent. However, as the fits with Equation \ref{eq:BPL} have already been presented in \cite{Horesh21}, here we proceed with the best-fitting parameters from Equation \ref{eq:BPL_mod}.  We individually fit for $\alpha_{\rm thick,thin}$, $F_{\rm pk}$, $\nu_{\rm pk}$ and $\nu_{q}$ in SEDs $\#2$ and $3$, resulting in $\alpha_{\rm thick} \sim +1$. We follow a similar process for SED $\#1$ but fix $\alpha_{\rm thick} = 1$ as there are too few measurements below $\nu_{\rm pk}$ to allow an independent constraint. The optically thick segments of the SED is similarly not sampled in SEDs $\#4$ and $5$. These epochs show a statistically strong preference for Equation \ref{eq:BPL} fit with the spectral break $\nu_{\rm b_1} = \nu_{q}$. 
SEDs $\#6,7$ are best fitted with a simple power-law. 

\begin{figure*}[tp]
    \centering
    \includegraphics[scale=0.25]{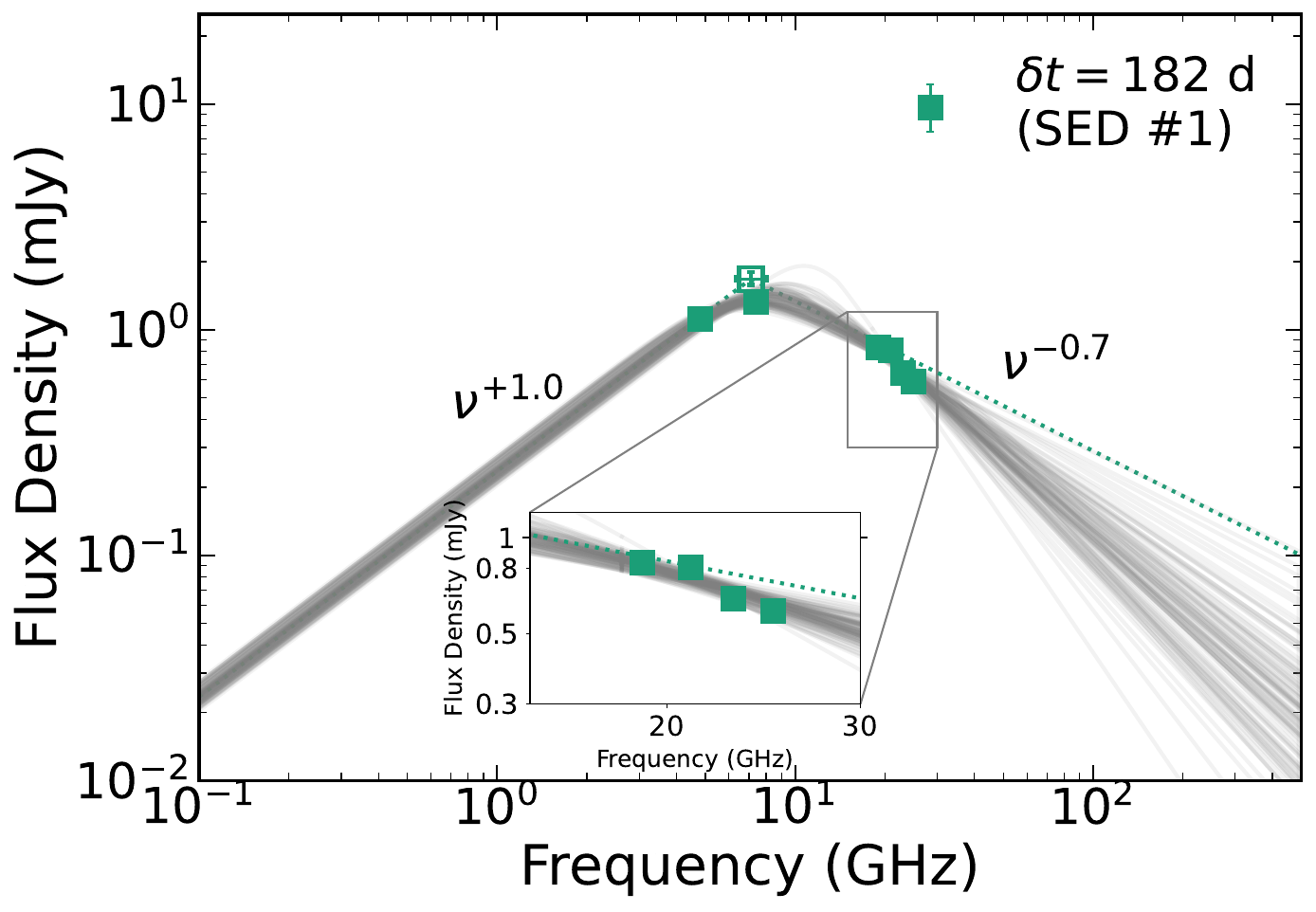}
    \includegraphics[scale=0.25]{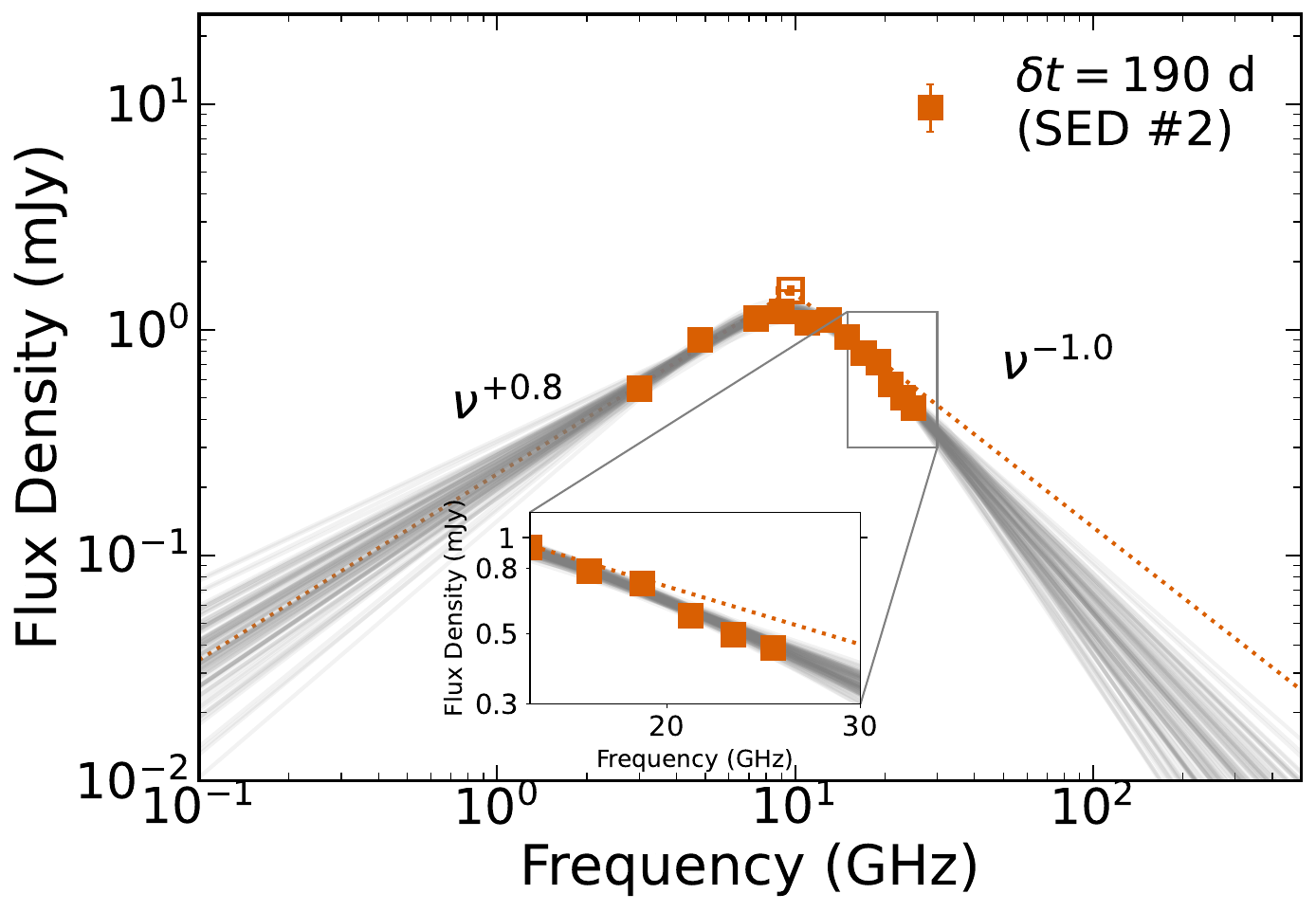}
    \includegraphics[scale=0.25]{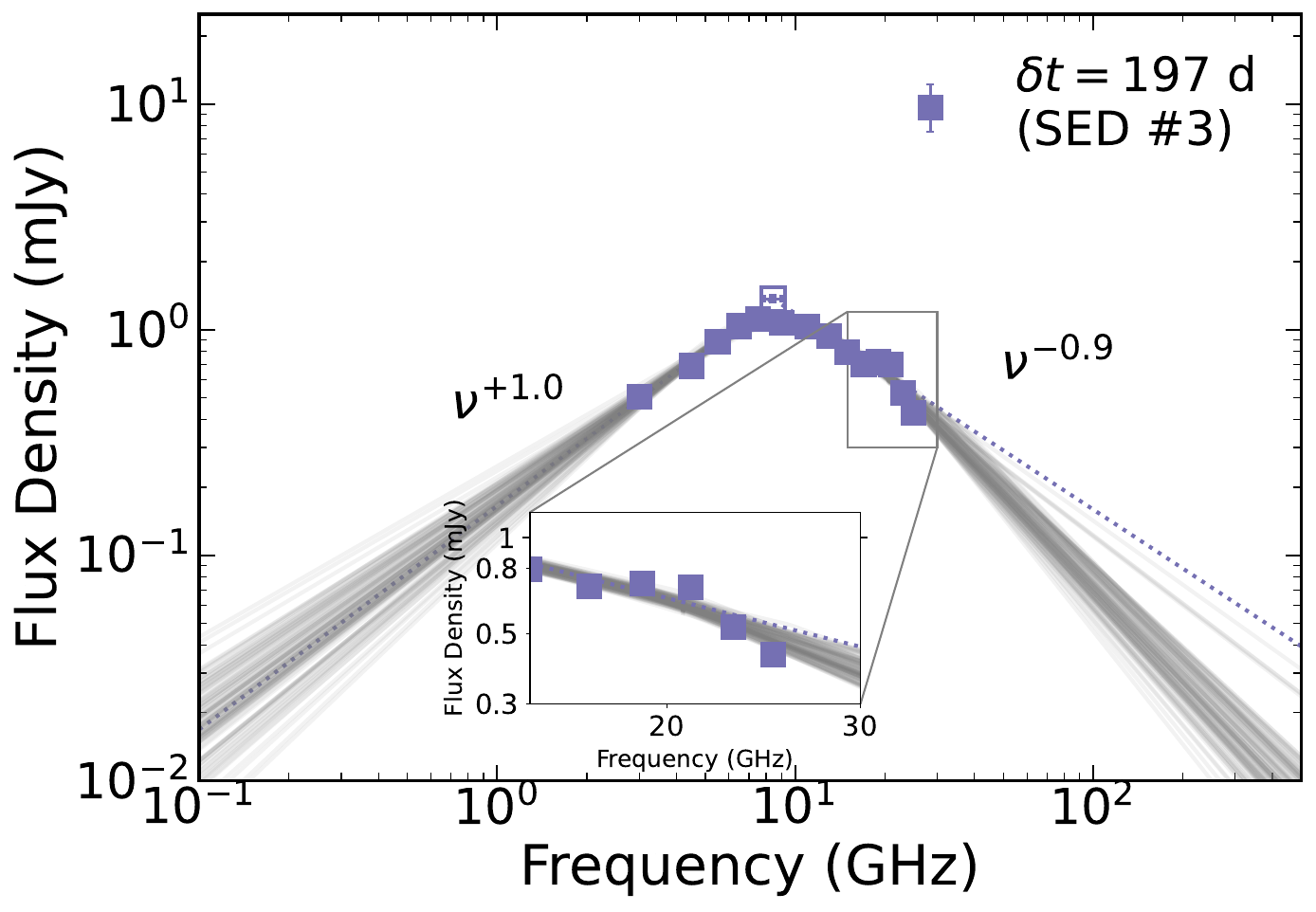}
    \includegraphics[scale=0.25]{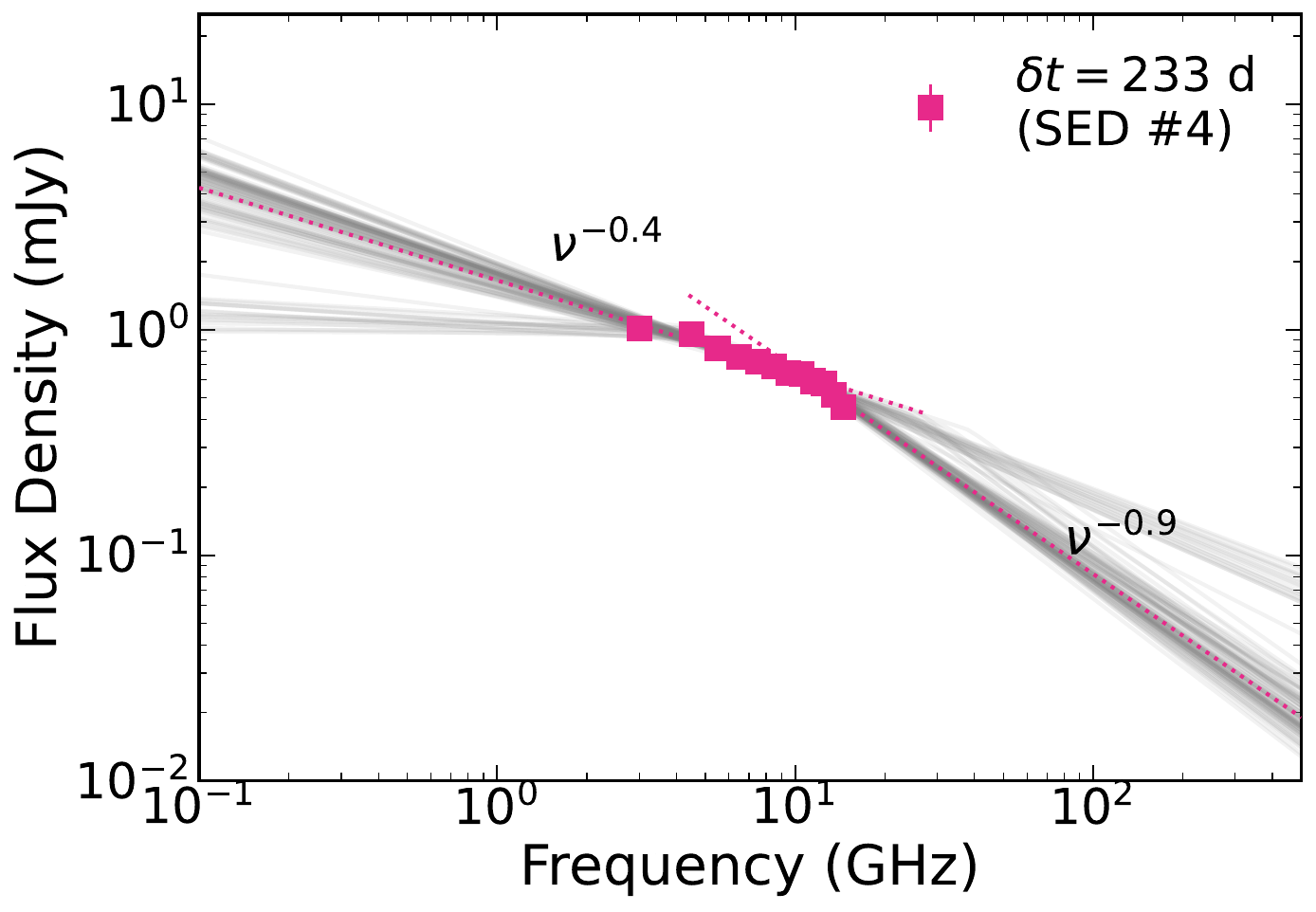}
    \includegraphics[scale=0.25]{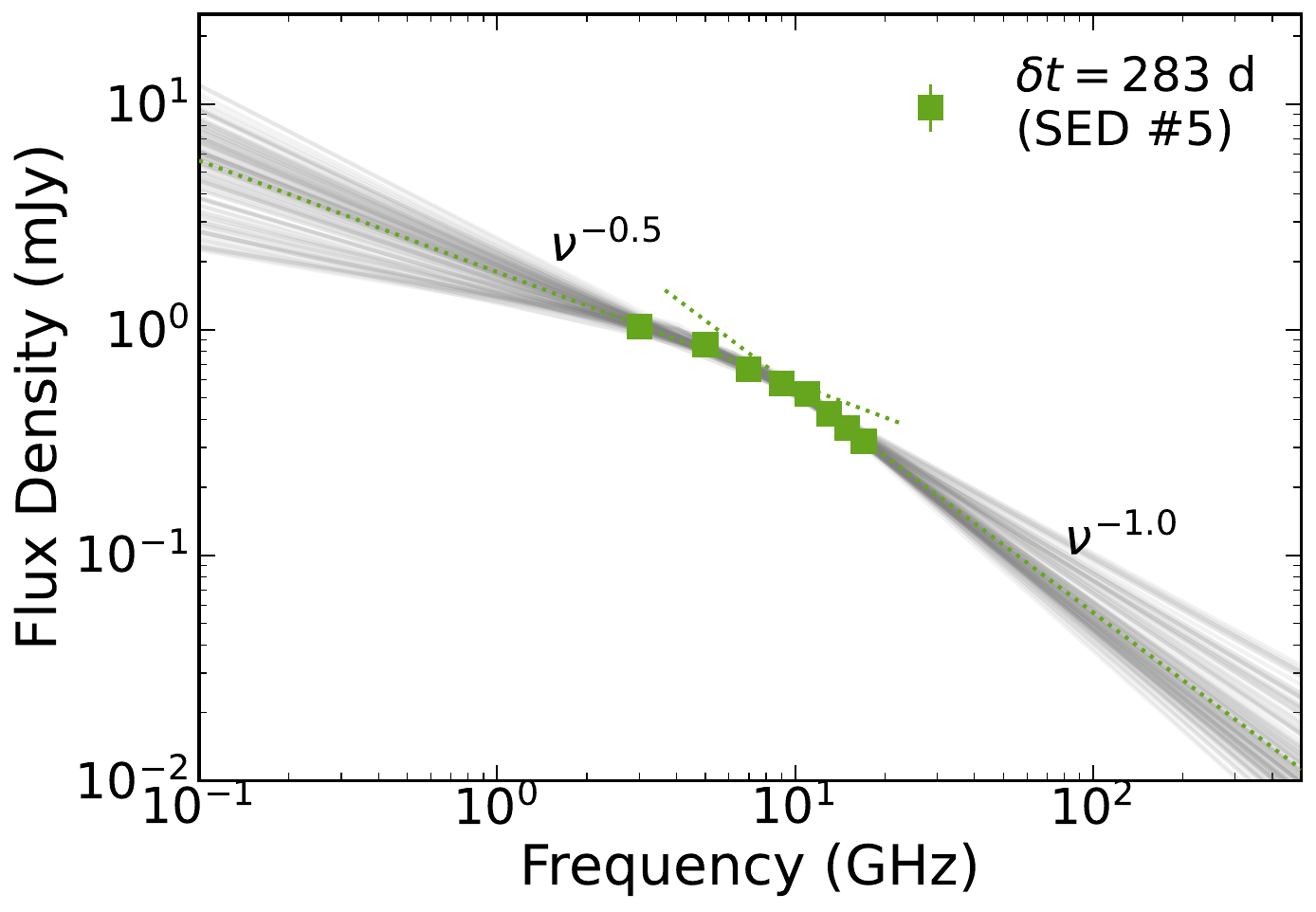}
    \includegraphics[scale=0.25]{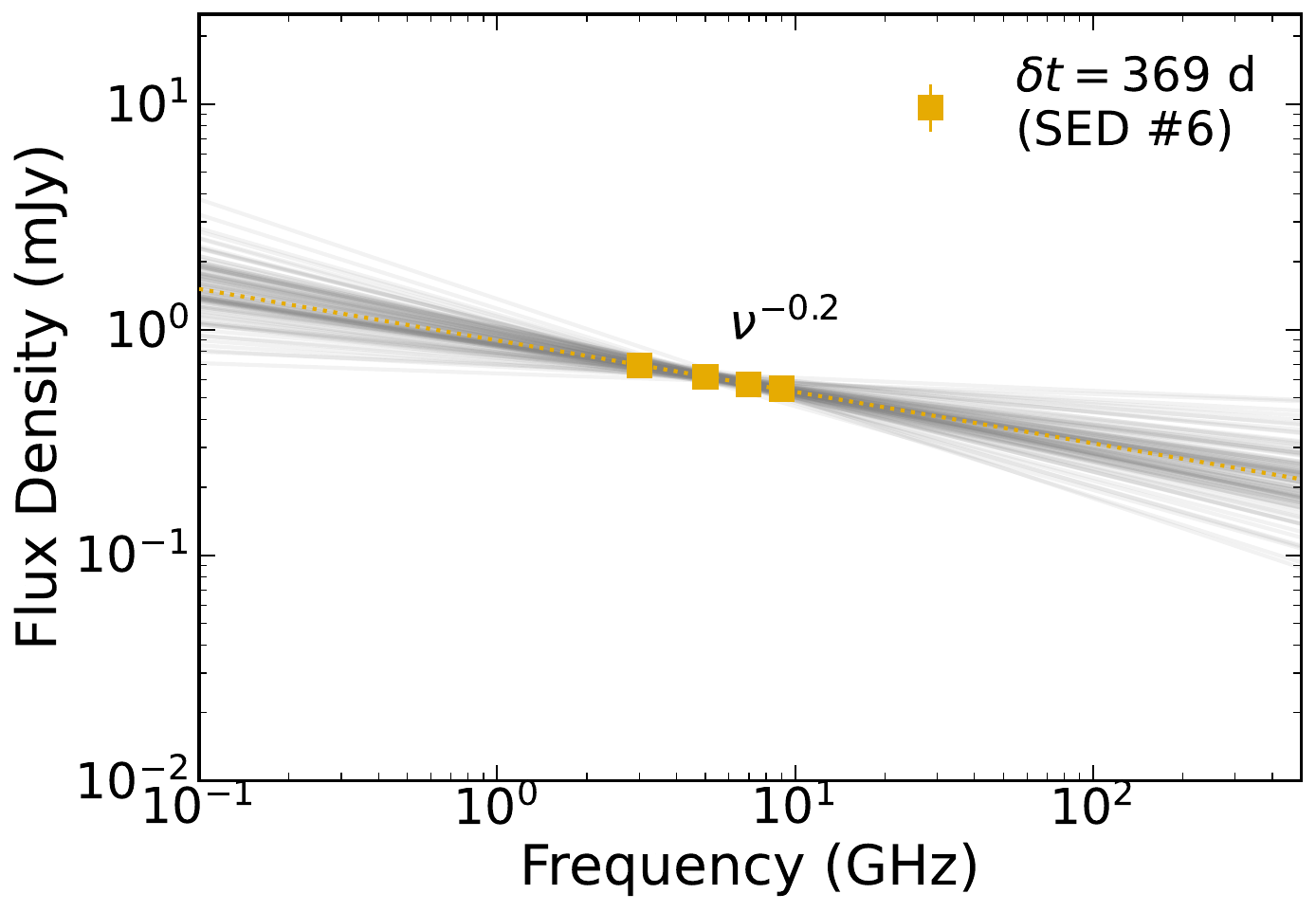}
    \includegraphics[scale=0.25]{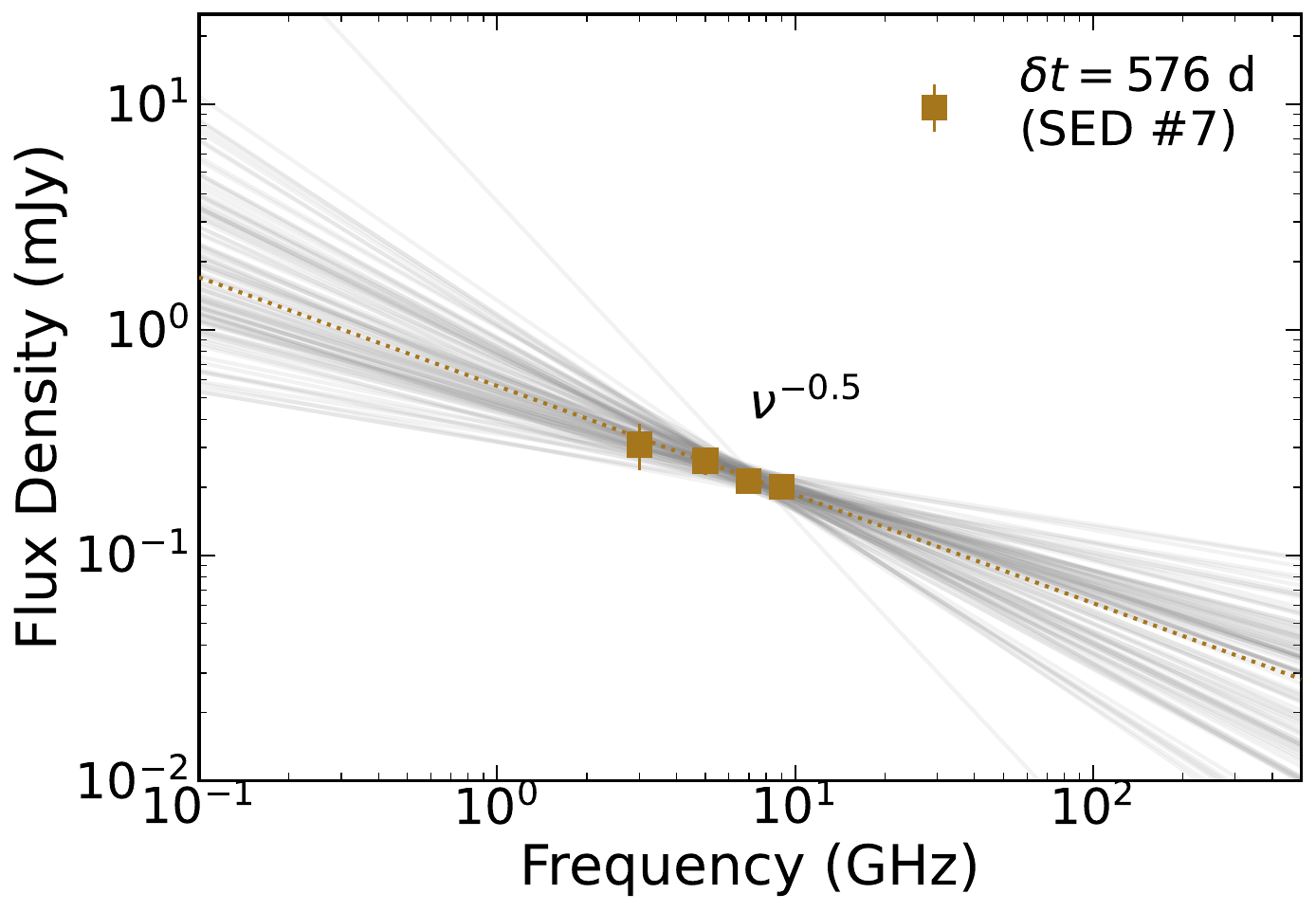}
    \includegraphics[scale=0.25]{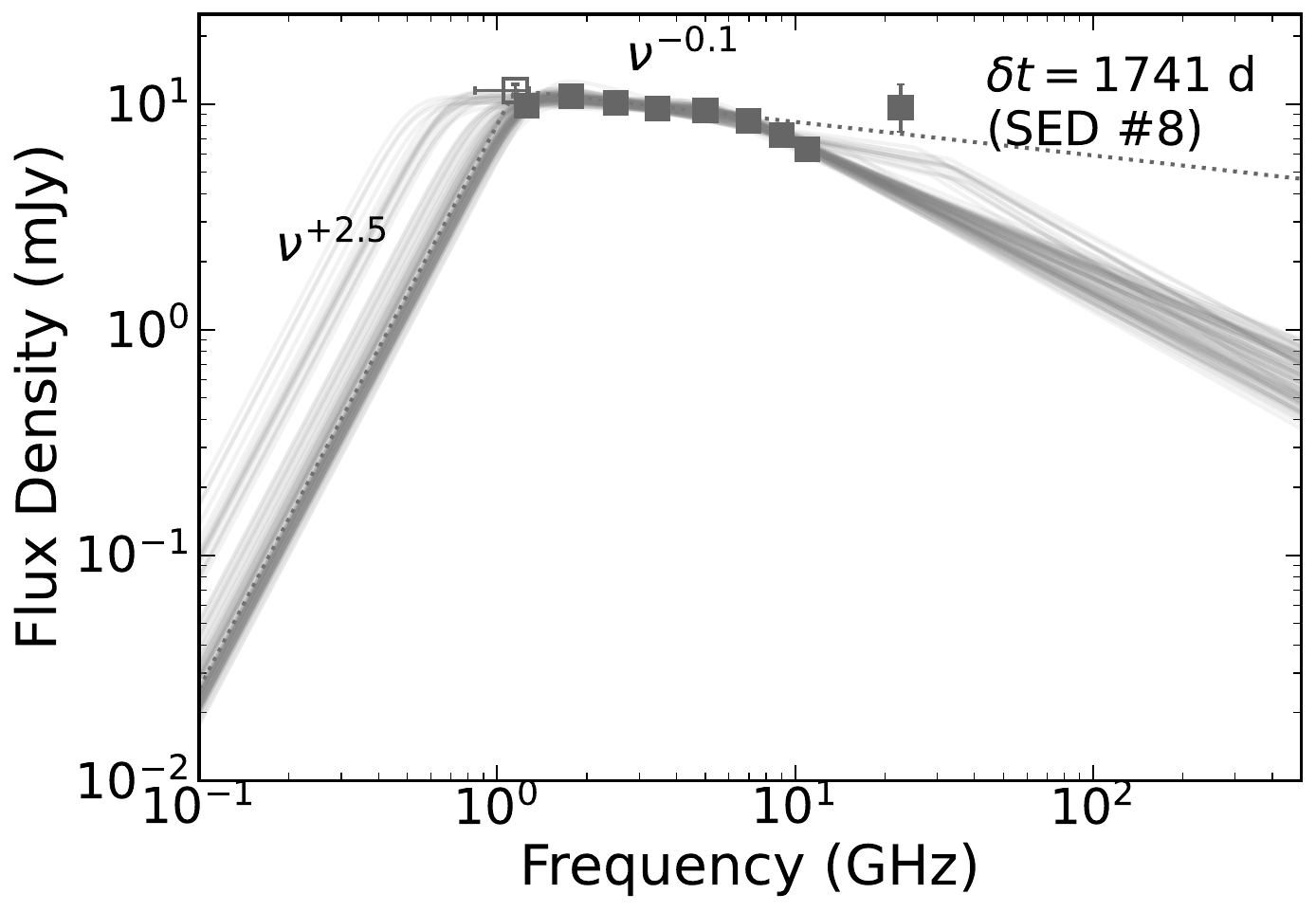}
    \includegraphics[scale=0.25]{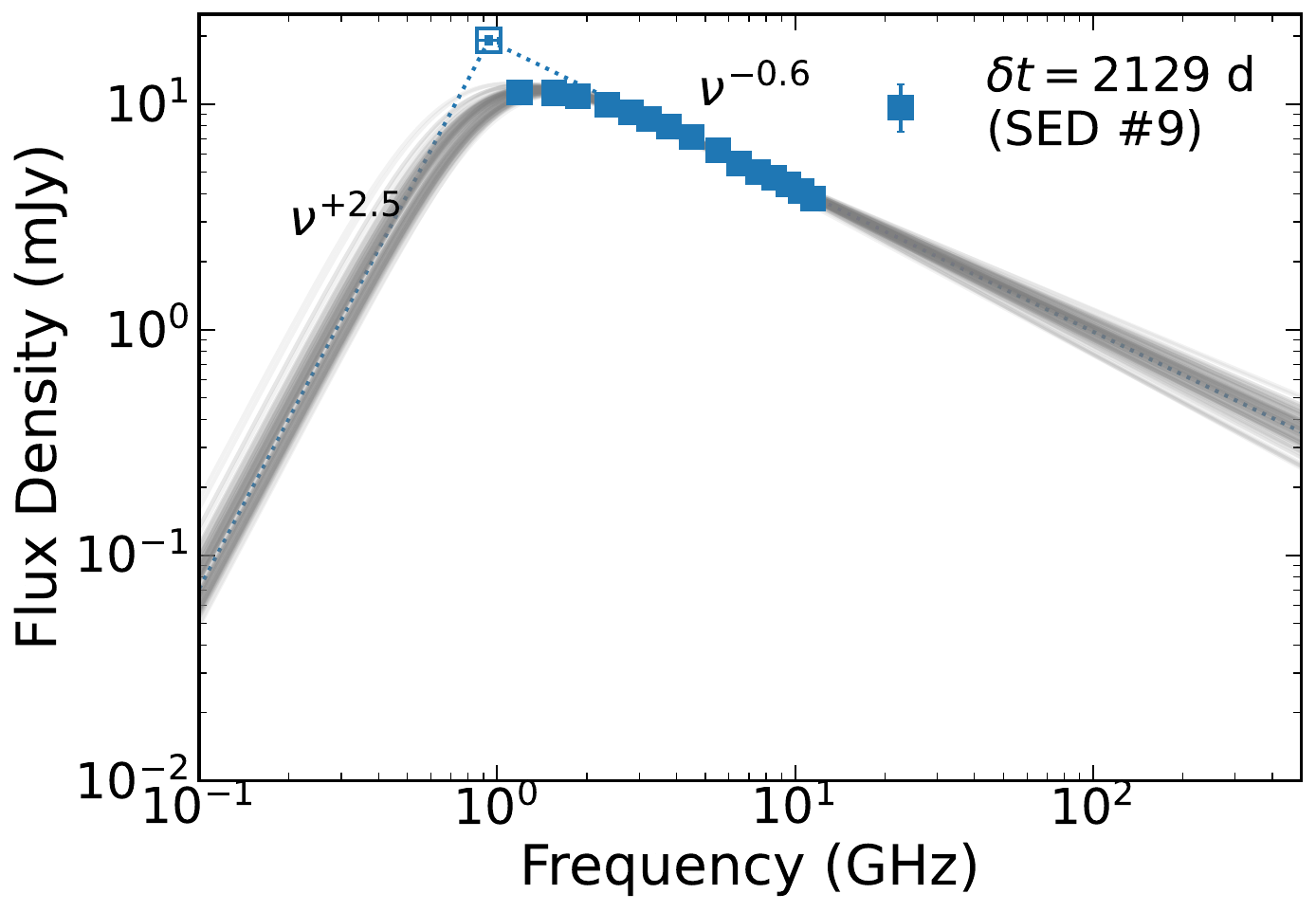}
    \includegraphics[scale=0.25]{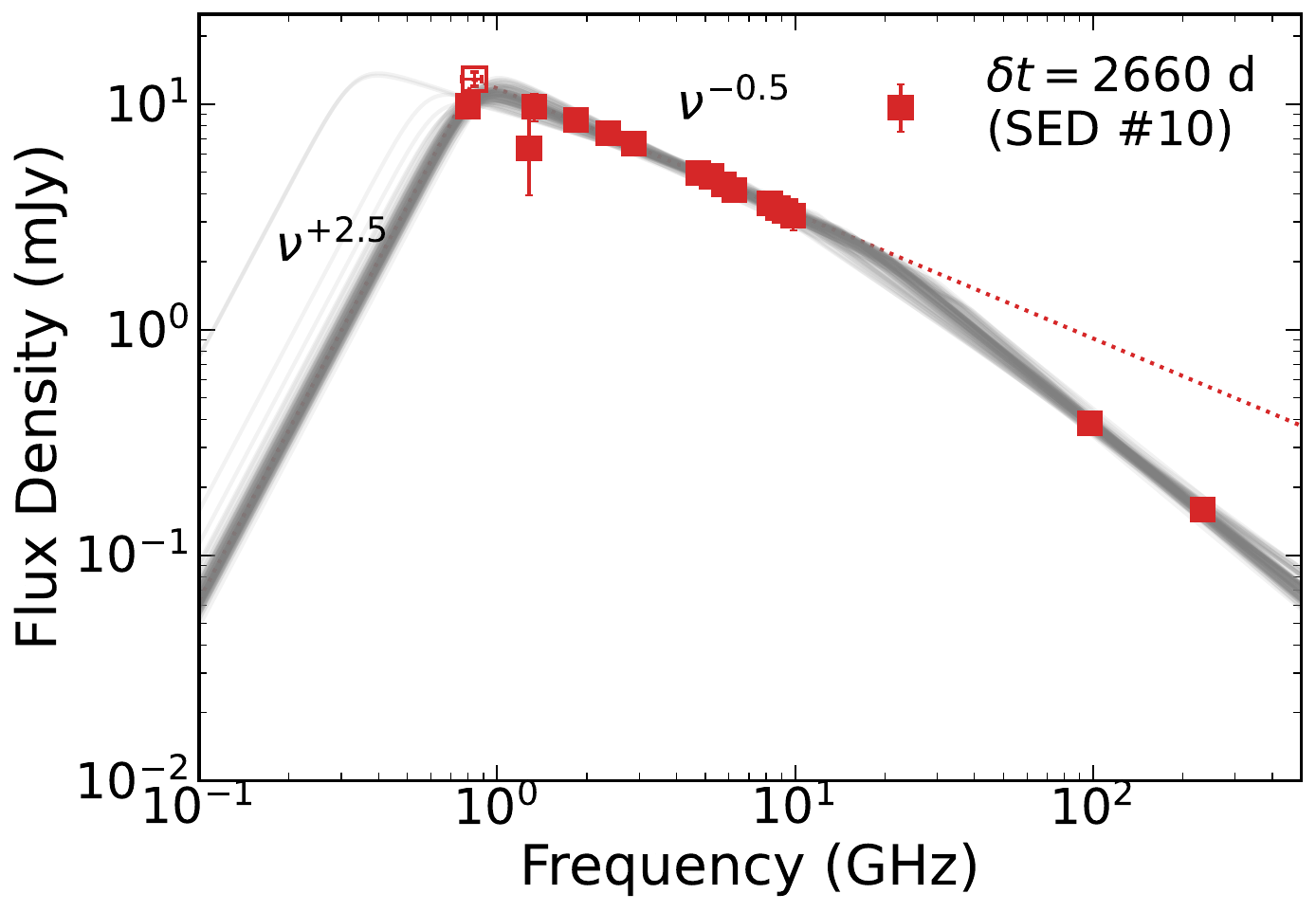}
    \includegraphics[scale=0.25]{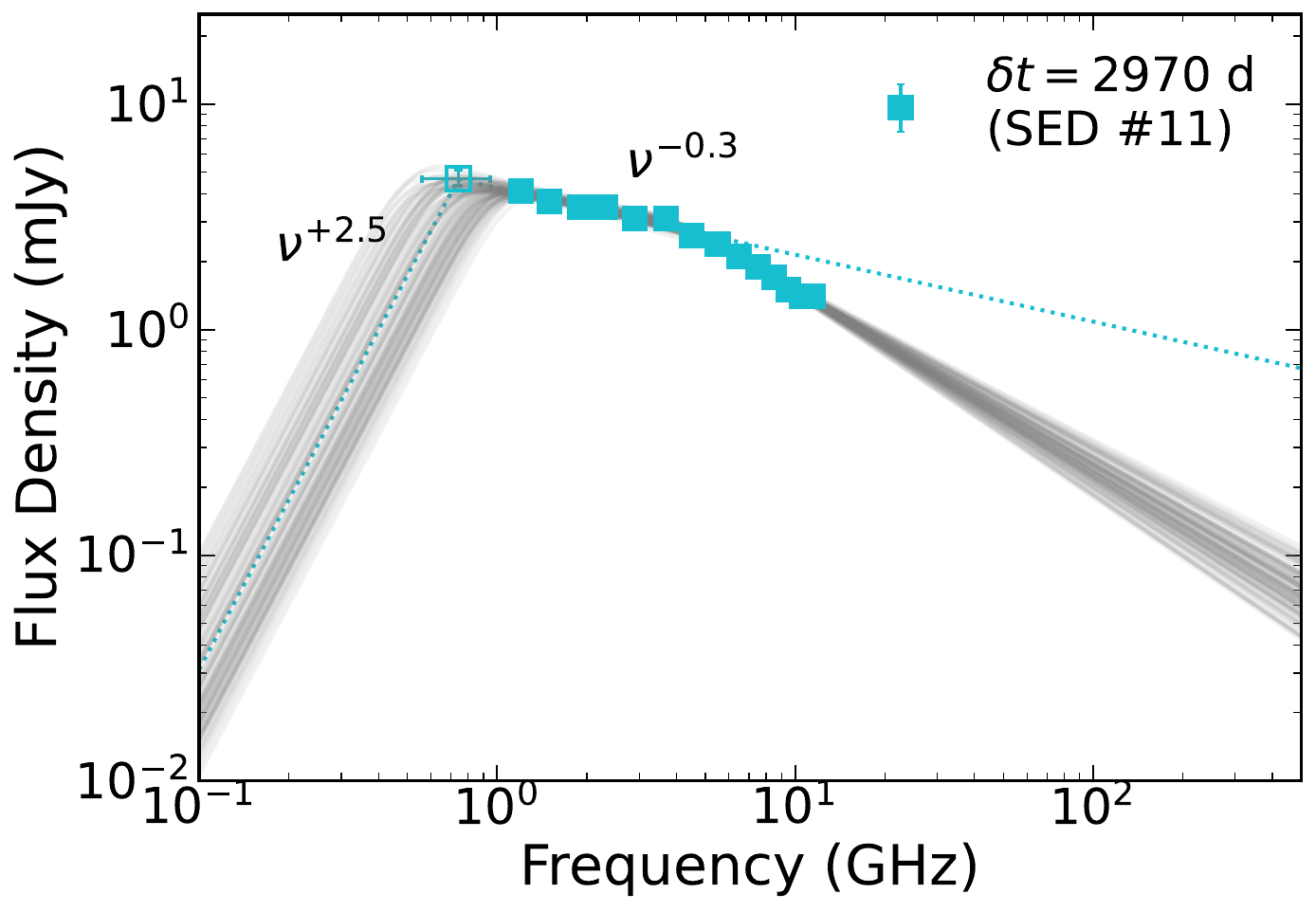}
    \caption{Radio SEDs of ASASSN-15oi and the best-fitting models: SED $\#1,2,3,8,10,11$ are fitted with Equation \ref{eq:BPL_mod}, while SED $\#4,5,9$ with Equation \ref{eq:BPL}. SED $\#6,7$ are fitted with a simple power-law. 
    The solid grey curves are a selection of 50 models randomly sampled from the posterior distribution. Where applicable, dotted lines show the asymptotic power-laws with the best-fitting spectral indices (the medians of the posterior distributions), and the empty squares are the intersection of the $\alpha_{\rm thick}$ and $\alpha_{\rm thin}$ segments in the respective SEDs. For clarity, we have also zoomed-in on the frequencies around $\nu_q$ in some SEDs.}
    \label{fig:radiosedfits6a}
\end{figure*}

\begin{figure}
    \centering
    \includegraphics[scale=0.43]{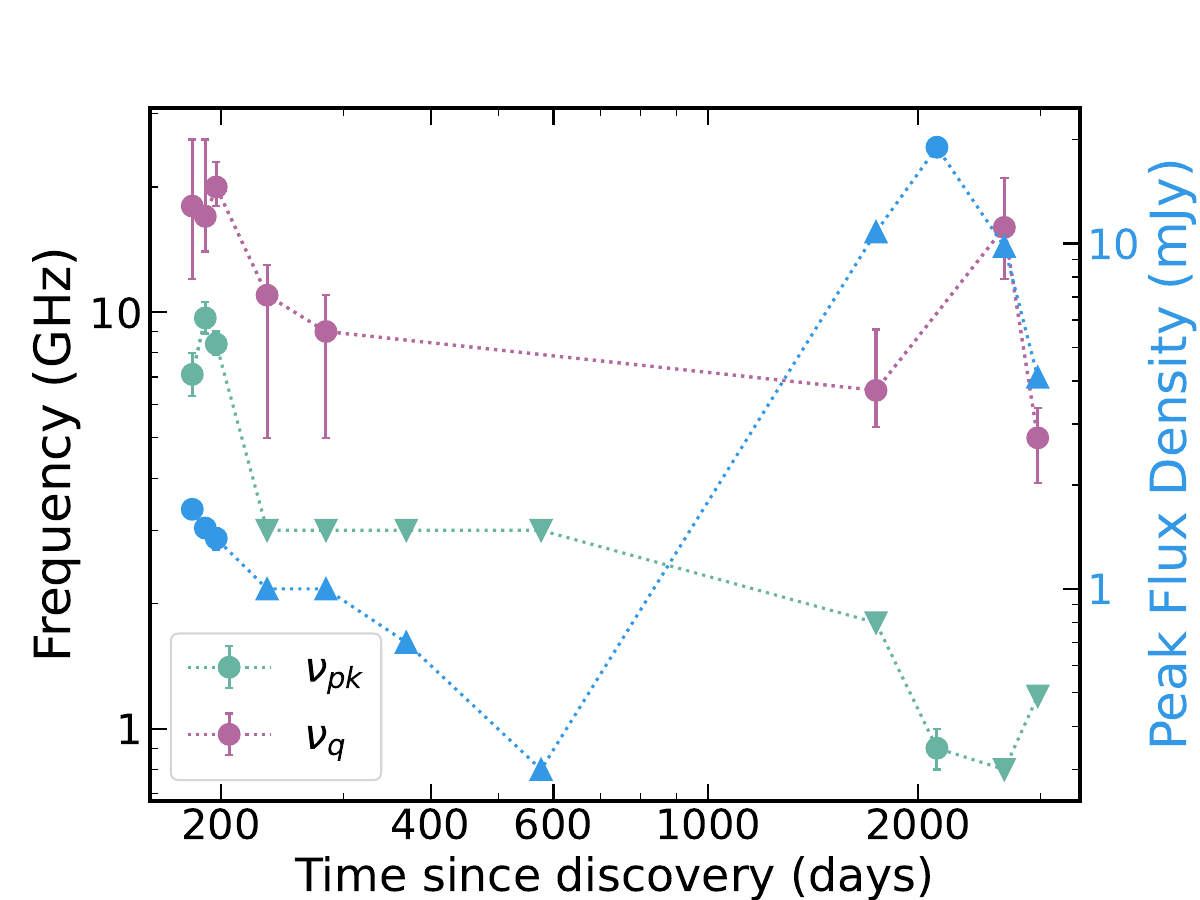}
    \caption{The evolution of the radio SED peak flux density $F_{\rm pk}$ and the break frequencies $\nu_{\rm pk}$ and $\nu_q$ with time.} 
        \label{fig:spectralpeakevol}
\end{figure}

For the second radio flare (\S\ref{subsec:secondflaresedfit}) SEDs, we fix $\alpha_{\rm thick} = 5/2$ as is typically done for TDEs. Equation \ref{eq:BPL_mod} is the statistically preferred model for SEDs $\#8,10,11$, while Equation \ref{eq:BPL} is preferred for SED $\#9$. A peak can only be constrained for SEDs $\#1,2,3$, and $9$. For every other SED ($\#4,5,6,7,8,10,11$) where a peak cannot be constrained, we assume a lower-limit on $F_{\rm pk}$ and an upper-limit on $\nu_{\rm pk}$ as the maximum observed flux density and its corresponding frequency, respectively. We report the best-fitting parameters for all SEDs in Table \ref{tab:radiosedfitting}, present the fits in Figure \ref{fig:radiosedfits6a}, and show the evolution of the SED peaks and $\nu_q$ in Figure \ref{fig:spectralpeakevol}.

\setcounter{table}{0}
\setlength{\tabcolsep}{5pt}
\renewcommand{\arraystretch}{1.2}
\begin{deluxetable}{c|cccccc}
\tablecolumns{7}
\tablecaption{Results of the MCMC fitting of the individual radio SEDs with $> 2$ data points. We fit different models to the SEDs and only report the parameters of the preferred model based on the AIC and BIC values reported in Table \ref{tab:modelselection} (see \S\ref{Sec:RadioModeling} for details).
 For all SEDs except $\delta t=2129$, we assume $s = -0.3$, and $s_2 = -0.003$; we instead use $s = -1$ to accommodate the broader peak of the $\delta t = 2129$ SED. Where $\nu_q$ is identified, we use $\alpha_q = \alpha_{\rm thin} - 0.5$. The central values reported here are the median of the posterior distribution, and the errors denote the 68\% credible interval.\label{tab:radiosedfitting}}
\tablehead{SED & Epoch & $\alpha_{\rm thick}$ & $\alpha_{\rm thin}$ & $F_{\rm pk}$ & $\nu_{\rm pk}$ & $\nu_q$ \\
$\#$ & (d) & & & (mJy) &(GHz) & (GHz)}
\startdata
\hline
\multicolumn{7}{c}{SEDs associated with the First Radio Flare }\\
\hline
1 & 182 & $1$ & $-0.7^{+0.2}_{-0.2}$ & $1.7^{+0.1}_{-0.1}$ & $7.1^{+0.9}_{-0.8}$ & $18^{+8}_{-6}$\\
2 &190 & $0.8^{+0.1}_{-0.1}$ & $-1.0^{+0.1}_{-0.2}$  & $1.5^{+0.1}_{-0.1}$ & $9.7^{+0.9}_{-0.8}$ & $17^{+9}_{-3}$\\
3 &197 & $1.0^{+0.1}_{-0.1}$ & $-0.9^{+0.1}_{-0.1}$  & $1.4^{+0.1}_{-0.1}$ & $8.4^{+0.6}_{-0.5}$ & $20^{+3}_{-2}$\\
4 &233$^{\rm{\displaystyle[{a,b}]}}$ & -- & $-0.4_{-0.1}^{+0.3}$  & $> 1.0$ & $ < 3$ & $11_{-6}^{+2}$ \\
5 &283$^{\rm{\displaystyle[{a,b}]}}$ & -- & $-0.5_{-0.1}^{+0.2}$  & $> 1.0$ & $ < 3$ & $9_{-4}^{+2}$\\
6 &369$^{\rm{\displaystyle[{b}]}}$ & -- & $-0.2^{+0.1}_{-0.1}$  & $>0.7$ & $< 3$ & --$^{\rm{\displaystyle[{c}]}}$\\
7 & 576$^{\rm{\displaystyle[{b}]}}$ & -- & $-0.5^{+0.2}_{-0.2}$  & $>0.3$ & $< 3$  & --$^{\rm{\displaystyle[{c}]}}$ \\
\hline
\multicolumn{7}{c}{SEDs associated with the Second Radio Flare$^{\rm{\displaystyle[{d}]}}$}\\
\hline
8 & 1741$^{\rm{\displaystyle[{b}]}}$ & $5/2$ & $-0.15_{-0.09}^{+0.07}$ & $< 10.8 $ & $< 1.8$ & $6.8_{-1.2}^{+2.6}$ \\
9 & 2129 & $5/2$ & $-0.64_{-0.03}^{+0.03}$ & $19_{-1}^{+1}$ &  $0.9_{-0.1}^{+0.1}$ & --$^{\rm{\displaystyle[{c}]}}$ \\
10 & 2660$^{\rm{\displaystyle[{b}]}}$ & $5/2$ & $-0.54^{+0.04}_{-0.04}$  & $< 9.8$ & $< 0.8$ & $16^{+5}_{-4}$\\
11 & 2970$^{\rm{\displaystyle[{b}]}}$ & $5/2$ & $-0.32_{-0.06}^{+0.08}$ & $< 4.1 $ & $ < 1.2$ &  $4.9^{+0.9}_{-1.1}$\\
\enddata
\tablecomments{{
${\rm{\displaystyle[{a}]}}$ These SEDs show no low frequency turnover, but show a steepening at high frequencies. We therefore fit them with Equation (\ref{eq:BPL}), where the two spectral indices are given by $\alpha_{\rm thin}$ and $\alpha_{\rm q} = \alpha_{\rm thin} - 0.5$.\\ 
${\rm{\displaystyle[{b}]}}$ In these SEDs, we use the maximum observed $F_{\nu}$, and the corresponding frequency as the limits on $F_{\rm{pk}}$ and $\nu_{\rm pk}$.\\
${\rm{\displaystyle[{c}]}}$ $\nu_q$ is not observed in these SEDs.\\
${\rm{\displaystyle[{d}]}}$ For these SEDs, the optically thick part of the spectrum cannot be constrained. We fix $\alpha_{\rm thick} = 5/2$.}}
\vspace{-0.5cm}
\end{deluxetable}

\setlength{\tabcolsep}{10pt}
\renewcommand{\arraystretch}{1.2}
\begin{deluxetable*}{c|cc|ccccc}
\label{tab:radiophysparams}
\tablecolumns{8}
\tablecaption{Physical parameters inferred from the \textit{equipartition} modeling  of the radio data, detailed in \S\ref{Sec:RadioModeling}, assuming $\epsilon_{\rm e} = \epsilon_{\rm B} = 0.1$. Here, $p$ is derived from $\alpha_{\rm{thin}}$ resulting from the MCMC fitting of the observed SEDs. $\alpha_{\rm{thin}}$, $F_{\rm pk}$, and $\nu_{\rm pk}$ values reported in Table \ref{tab:radiosedfitting} are used to derive the physical parameters according to the model described by \cite{Chevalier98}. The derivations of the physical parameters in Equations (\ref{eq:B}) -- (\ref{eq:ne}) is dependent on $p > 2$. 
For SEDs with $p < 2$, we use Equations (\ref{eq:B_pless2}) -- (\ref{eq:U_pless2}).
The values in the table are sensitive to $F_{\rm pk}$, and $\nu_{\rm pk}$, and therefore to how smoothly the power-laws are joined in the broken power-law model (i.e., to the value of $|s|$).}
\tablehead{SED & $\delta t$ & $p$ & $B_{\rm eq}$ & $U_{\rm eq}$ & $R_{eq}$ & $n_{\rm e, eq}$ & $v_{\rm eq}$$^{\rm{\displaystyle[{a}]}}$ \\
$\#$ & (d) &  & (G) & ($\times 10^{49}$\,erg) & ($\times 10^{16}$\,cm) & ($\times 10^3$\,cm$^{-3}$) & ($\times$speed of light)}
\startdata
\hline
\multicolumn{8}{c}{Modeling of SEDs associated with the First Radio Flare. }\\
\hline
1 & 182 & $2.3^{+0.3}_{-0.3}$ & $0.41^{+0.08}_{-0.06}$ & $0.9^{+0.5}_{-0.2}$ & $4.0^{+0.3}_{-0.3}$  & $9^{+4}_{-2}$ &  $0.084^{+0.007}_{-0.006}$\\
2 & 190 & $3.0^{+0.5}_{-0.3}$ & $0.74^{+0.34}_{-0.18}$ & $1.9^{+3.5}_{-0.9}$ & $3.5^{+0.4}_{-0.3}$ & $38^{+30}_{-16}$ & $0.070^{+0.006}_{-0.007}$ \\
3 & 197 & $2.7^{+0.2}_{-0.2}$ & $0.55^{+0.09}_{-0.08}$ & $1.0^{+0.5}_{-0.3}$ & $3.4^{+0.2}_{-0.2}$ & $24^{+9}_{-6}$ & $0.066^{+0.003}_{-0.003}$ \\
4 & 233 &  $\sim 1.8$ & {$<$ 0.16} & {$> 0.45$} & {$>$ 6.0} & {$<$ 0.9} & {$> 0.1$}\\
5 & 283  &  $\sim 1.9$ & $< 0.16 $ & $> 0.54$ & $> 6.3$ & $< 1.2$ & $> 0.1$\\
6 & 369     &   $\sim 1.4$
& {$<$ 0.18} & {$>$ 0.25} & {$>$ 4.4} & {$<$ 5.4} & {$>$0.04}\\
7 & 576    &   $\sim 1.9$
& {$<$ 0.18} & {$>$ 0.13} & {$>$ 3.6} & {$<$ 20} & {$>$0.02}\\
\hline
\multicolumn{8}{c}{Modeling of SEDs associated with the Second Radio Flare}\\
\hline
8 & 1741 &  $\sim1.3$ & $< 0.08$ & $> 9$ & $> 26$ & $< 0.72$ & $> 0.06$\\
9 & 2129 & $2.2^{+0.1}_{-0.1}$ 
& $0.04^{+0.01}_{-0.01}$ & $93^{+9}_{-6}$  & $90^{+8}_{-5}$ & $0.02^{+0.01}_{-0.01}$ & $0.16^{+0.01}_{-0.01}$\\
10 & 2660 & $2.1^{+0.1}_{-0.1}$ & $< 0.04$ & $> 52$ & $> 75$ & $< 0.04$ & $> 0.10$\\
11 & 2970 & $\sim 1.6$ & $< 0.06$ & $> 6$ & $>27$ & $ < 0.89$ & $> 0.03$
\enddata
\tablecomments{${\rm{\displaystyle[{a}]}}$ The velocity is measured with respect to days since the optical discovery. We note that velocities of the outflow at one epoch relative to the other epoch can be calculated as $\Delta R / \Delta t$.\\\vspace{-1cm}}
\end{deluxetable*}

\subsection{Equipartition Physical Parameters from a Quasi-spherical Outflow Model}\label{subsec:syncphysparams}
Using the inferred SED peak and $\alpha_{\rm thin}$, we can now estimate the physical parameters of the outflow responsible for the radio emission. To that end we adopt a synchrotron self-absorption (SSA) model given by \cite{Chevalier98}, as commonly used in the TDE literature \citep[e.g.,][and references therein]{Alexander20}. This model assumes that the synchrotron radiation is powered by the relativistic electrons accelerated by a shock interacting with the ambient medium, resulting in a broadband SED.
Below the SED peak ($\nu < \nu_{\rm pk}$), the synchrotron radiation is self-absorbed resulting in $F_{\nu} \propto \nu^{5/2}$. At $\nu > \nu_{\rm pk}$, the spectrum becomes optically-thin and follows $F_\nu \propto \nu^{-(p-1)/2}$, where $p$ is the power-law index of the population of accelerated electrons. In the context of our model parameters in \S\ref{subsec:sedfit}, the spectral indices become $\alpha_{\rm thick} = 5/2$ and $\alpha_{\rm thin} = -(p - 1)/2$. The indices $p$ derived from the best-fitting $\alpha_{\rm thin}$ for each SED are reported in Table \ref{tab:radiophysparams}. Clearly, the abrupt changes in $p$ between epochs are unphysical for a single population of radiating electrons, suggesting the presence of multiple synchrotron sources instead. This interpretation will be corroborated further as we proceed with our analysis. 

For SEDs $\#1,2,$ and $3$, we have $\alpha_{\rm thick} \sim 1 \neq 5/2$. A shallow optically thick slope can possibly be explained by a clumpy medium \citep[e.g.,][]{weiler2002,Bjornsson2017}, a possibility also explored by \cite{Horesh21}. These inhomogeneities result in a superposition of several SSA spectra with different peaks.  In this case, the physical parameters derived from the \cite{Chevalier98} analysis are representative of the population of electrons that dominate the peak of the SED. 

We adopt a volume filling factor of $f=0.5$ in the equations presented in \cite{demarchi2022} to constrain the post-shock magnetic field $B$, the radius of the emitting region $R$, the internal energy $U$, the electron number density of the ambient medium $n_e$, and the average velocity of the shock as $v = R/\delta t$.
The equations in \citet{demarchi2022} generalize those originally presented in \cite{Chevalier98} for an arbitrary value of $p$ (see also \citealt{Chevalier2017}). The expressions for $B, R, U$ and $n_e$ are presented in Appendix \ref{sec:demarchi_equations}. Equations (\ref{eq:B} -- \ref{eq:ne}) are derived assuming $\gamma_{\rm{max}} \to\infty$, where $\gamma_{\rm max}$ is the maximum Lorentz factor attained by the electrons accelerated by the shock. In this case, for the energy density in the electrons to be finite, $p > 2$ is required. Conversely, a $p < 2$ requires finite $\gamma_{\rm{max}}$. We derive equations for this regime in Appendix \ref{sec:pless2equations}. We note that multi-dimensional particle-in-cell simulations suggest $\gamma_{\rm max} \lesssim 10^8$ \citep{Sironi13_maxenergy}. $B$, $R$ and $U$ are all proportional to $\gamma_{\rm max}$. Here, we assume $\gamma_{\rm max} = 10^3$ (with these electrons emitting synchrotron radiation at $\sim 3$\,THz for a non-relativistic shock). All the parameters are in c.g.s.\ units.

As this model has many free parameters, a unique solution is generally only possible if multiple spectral breaks of the synchrotron spectra can be measured directly from the data. In the absence of other constraints, the system is assumed to be in equipartition, i.e., the fraction of internal energy in the relativistic electrons $(\epsilon_{\rm e}$) is equal to the fraction of internal energy of the post-shock magnetic field ($\epsilon_{\rm B}$). We note that there is no observational support for such a choice, but this is a standard assumption in the field. This enables us to estimate the physical parameters even if only $F_{\rm pk}$, $\nu_{\rm pk}$ and $p$ can be measured directly. Here we assume $\epsilon_{\rm e} = \epsilon_{\rm B} = 0.1$, as is typically assumed in the TDE literature, to estimate the physical parameters of the synchrotron-emitting region. Where the peak was not constrained, the limits on $F_{\rm pk}$ and $\nu_{\rm pk}$ translate into limits on the inferred physical quantities. We denote parameters ($X$) that were calculated assuming equipartition conditions as $X_{eq}$.

\begin{figure}
    \centering
    \includegraphics[scale=0.56]{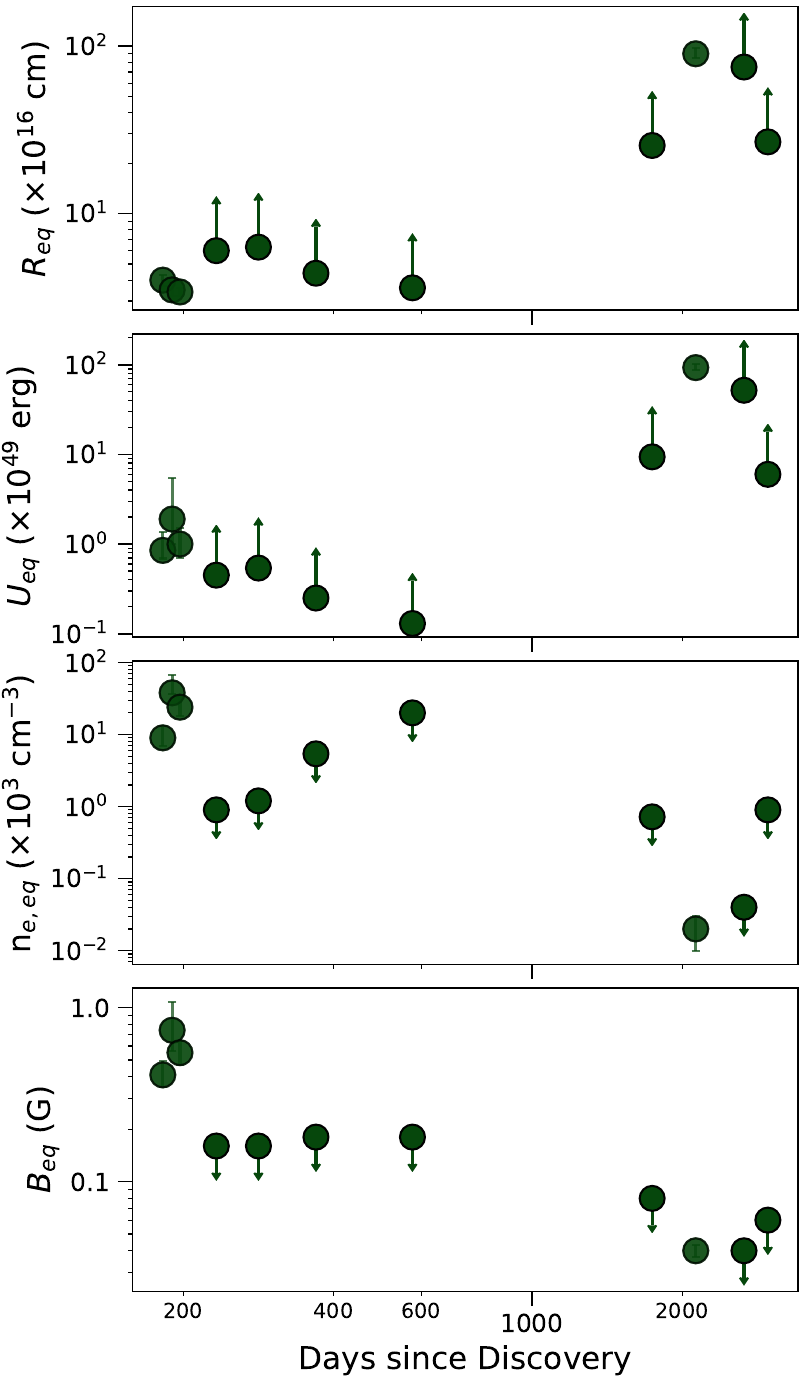}
    \caption{Physical parameters from the modeling of the radio data assuming equipartition. The dramatic jumps in the magnetic field and density estimate from the SEDs within the first radio flare suggest an inhomogeneous medium, or multiple synchrotron sources. A significant jump in the energy inferred for the first and second radio flare hints at two distinct sources of emission for the two flares. }
    \label{fig:radiomodelparams}
\end{figure}

The resulting values are reported in  Table \ref{tab:radiophysparams}, and the evolution of these parameters are visualized in Figure \ref{fig:radiomodelparams}. We note 
the dramatic increase in the inferred energy between the two flares from $U_{\rm eq} \sim 10^{49}$\,erg at $\delta t = 182$\,d to $U_{\rm eq} \sim 10^{51}$\,erg at $\delta t = 2129$\,d. This implies that a different, more powerful source is needed to explain the second radio flare. We explore a scenario of a relativistic jet as a source of emission for the second radio flare in \S\ref{subsec:offaxisjetparams}. In the next section we briefly discuss the inferences that can be drawn if we assume that $\nu_{q}$ is the synchrotron cooling break.

\subsubsection{Synchrotron Cooling Break}\label{subsubsec:synccool}

\setlength{\tabcolsep}{17pt}
\renewcommand{\arraystretch}{1}
\begin{deluxetable}{c|ccc}
\label{tab:newphysparams_fromnuc}
\tablecolumns{4}
\tablecaption{We measure $B_{\rm c}$ from Equation (\ref{eq:nucool}) using the constrained $\nu_q$, then solve for $\epsilon_{\rm B, c}$ from Equation \ref{eq:B} or \ref{eq:B_pless2}, assuming $\epsilon_{\rm e,c} \geq 10^{-4}$ as long as the resulting $\epsilon_{\rm B,c} < 1$.} 
\tablehead{SED & $\delta t$  & $B_{\rm c}$ & $\epsilon_{\rm B, c}$\\
$\#$ & (d) & (G) & } 
\startdata
\hline
1 & 182
& $0.72$ & $\approx \epsilon_{\rm{e},c}$ \\
2 & 190 & $0.73$ & $\approx \epsilon_{\rm{e},c}$ \\
3 & 197 & $0.66$ & $\approx \epsilon_{\rm{e},c}$ \\
4 & 233 & $0.72$ & $\geq 0.052$ \\
5 & 283  & $0.68$ & $\geq 0.045$\\
8 & 1741 & $0.23$   & $\geq 0.005$\\
10 & 2660  & $0.12$ & $\geq 0.018$\\
11 & 2970  & $0.17$ & $\geq 0.009$ \\
\enddata
\vspace{-1cm}
\end{deluxetable}

We can relax the assumption of equipartition, i.e. partially break the degeneracy between $\epsilon_e$ and $\epsilon_B$, if we can measure another characteristic of the synchrotron spectrum, such as the synchrotron cooling break $\nu_{\rm cool}$. This enables a more accurate estimate of the outflow parameters. For ASASSN-15oi, the break $\nu_{{q}}$ constrained in \S\ref{subsec:sedfit} for some SEDs may be interpreted as $\nu_{\rm cool}$. 

For a non-relativistic shock, $\nu_{\rm{cool}}$ is calculated as:
\begin{equation}
\nu_{\rm{cool}} = \frac{18 \pi m_{\rm{e}} c q_{\rm{e}}}{\sigma_{T}^2 {B_c}^3 (t_{\rm{dyn}})^2}
\label{eq:nucool}
\end{equation}
where $m_{\rm{e}}$ and $q_{\rm{e}}$ are the mass and the charge of the electron, $c$ is the speed of light, $\sigma_{T}$ is the Thomson cross-section, $B_c$ is the magnetic field, and $t_{\rm{dyn}}$ is the dynamical time. 
Note that we represent the physical parameters here as $X_c$ to differentiate these estimates from the equipartition ones above. 

SEDs $\#1,2$ and $3$ have two break frequencies (see \S\ref{subsec:sedfit}). To test whether the second break frequency could be the synchrotron cooling frequency, we assume $t_{\rm dyn} = \delta t$ and $B_{c} = B_{\rm eq}$ in Equation \ref{eq:nucool}, and find that $\nu_{\rm cool} \approx \nu_{q}$ within $1\sigma$ for SEDs $\#2$ and $3$ and within $3\sigma$ for SED $\#1$. This supports our findings that there are two break frequencies as well as the initial equipartition assumption. However, we note that the uncertainties on the best-fitting $\nu_q$ are large so we cannot conclusively claim that these are the cooling breaks. 

We fitted for $\nu_q$ in SEDs $\#4,5,8,10,11$ in \S\ref{subsec:sedfit}. For each of these SEDs, if we assume, $t_{\rm dyn} = \delta t$ and $\nu_{\rm cool} = \nu_q$, the measured $B_c$ from Equation \ref{eq:nucool} (in Table \ref{tab:newphysparams_fromnuc}) is different from $B_{\rm eq}$ in Table \ref{tab:radiophysparams}. Using $B_c$, we can solve for $\epsilon_{\rm{B},c}/\epsilon_{\rm{e},c}$ using either Equation \ref{eq:B} or \ref{eq:B_pless2}. With our initial assumption of $\epsilon_{\rm e,c} = 0.1$, we obtain $\epsilon_{\rm{B},c} > 1$. This is clearly unphysical. Smaller values of $\epsilon_{\rm{e},c}$ (specifically $10^{-4} \leq \epsilon_{\rm{e},c} < 0.1 $) would yield acceptable $\epsilon_{\rm{B},c}$ values. Here, we adopt $\epsilon_{\rm{e},c} \geq 10^{-4}$, where this lower-bound is set by the simulations in e.g., \cite{Park2015} for non-relativistic collisionless shocks. The upper-bound on $\epsilon_{\rm e,c}$ is set by the condition $\epsilon_{\rm B,c} < 1$. Both $B_c$ and the allowed range of $\epsilon_{\rm{B},c}$ are reported in Table \ref{tab:newphysparams_fromnuc}. As can be seen from Table \ref{tab:newphysparams_fromnuc}, our choices yield extreme ratios of the microphysical shock parameters $\epsilon_{\rm{B},c}/\epsilon_{\rm{e},c} \lesssim 500$.

There is a third possibility that the breaks constrained here are not synchrotron cooling breaks. Indeed, the shallow $\alpha_{\rm thin}$ found in most of these SEDs hints at a contribution from multiple synchrotron shocks. Several SSA spectra with different $p$ values can superimpose to create a shallow ($p < 2$) optically thin segment, e.g. as also seen in some AGNs \citep{Kellermann1969}. This may indeed be true as we see $\alpha_{q}$ in SEDs $\#4,5,8,10,11$ resembles the $\alpha_{\rm thin}$ expected from a standard SSA spectrum, i.e. only a single synchrotron component dominates at higher frequencies. Finally, the steep jumps, particularly in the inferred density $n_{\rm e,eq}$ between $\delta t \approx 197$\,d and $233$\,d also suggest an inhomogeneous medium. Thus, a likely unified scenario where the SED properties and the abrupt jumps in the physical parameters of the system can be explained is that of `multiple outflows launched in an inhomogeneous medium'. We discuss this scenario further in \S\ref{subsec:nonthermaldisc}.

\subsection{Physical Parameters From An Off-Axis Relativistic Jet}\label{subsec:offaxisjetparams}
Here, we examine if an off-axis relativistic jet could power the second radio flare at $\delta t \gtrsim 1400$\,d. This scenario is commonly invoked for explaining the delayed radio emission in TDEs (see e.g., \citealt{Perlman_igr_2022,Cendes_18hyz_2022,Matsumoto2023_equip, Sfaradi_18hyz_2024, Christy2024}).
\cite{Matsumoto2023_equip} provide a generalized treatment of jets viewed at arbitrary angles in the context of TDEs. Compared to the \cite{Chevalier98} model, in a jetted relativistic regime three additional variables come into play: the jet Lorentz factor, $\Gamma$, the jet opening angle $\theta_{\rm{jet}}$, and the viewing angle, $\theta_{\mathrm{obs}}$, i.e., the angle between the jet axis and the observer's line of sight. The jet energy per solid angle is assumed to be constant for $\theta < \theta_{\rm{jet}}$ and 0 at larger angles (a so-called ``top-hat" jet). The equations derived in \cite{Matsumoto2023_equip} are applicable for a system in equipartition
and we follow \cite{Christy2024} to incorporate the corrections outlined in \S3 and \S4.2.2 of \cite{BarniolDuran13} for a more general system. Proceeding as in \S\ref{subsubsec:synccool} and assuming our measured high-frequency break ($\nu_q$) as the relativistic synchrotron cooling break ($\nu_{\rm cool, jet}$), allows us to relax the assumption of equipartition. This results in $\epsilon_{\rm B, jet} < 1$ for $\epsilon_{\rm{e,jet}} = 0.1$, reported in Table \ref{tab:radiomatsumoto}, in contrast to the non-relativistic uncollimated blastwave scenario. 

Here, we consider a dynamically spreading jet whose opening angle $\theta_{\rm jet} = 1/\Gamma$ grows as the jet decelerates. 
This is equivalent to considering only the energy within an angle of $1/\Gamma$ from our line of sight. As material at larger angles has a negligible contribution to the observed emission due to relativistic beaming, this provides the minimum energy necessary to explain our observations at each epoch. In contrast, a narrow jet with constant opening angle $\theta_{\rm jet} \sim 10^\circ$ (typical of AGN jets), which is more physically likely, would require higher Lorentz factors and correspondingly larger emission radii to produce the same observed emission.

As \cite{Matsumoto2023_equip}'s analysis only applies for $p > 2$, we perform this analysis only on the $\delta t = 2129$ and $2660$\,d SEDs. For the $\theta_{\rm jet} \geq 1/\Gamma$ case, we calculate the radius that minimizes the total energy under a $p = 2.1$ power law distribution of electrons (resulting from the best-fitting $\alpha_{\rm thin}$ in \S\ref{subsec:sedfit}). We assume equipartition, with $\epsilon_{\rm e, jet} = \epsilon_{\rm B, jet} = 0.1$ for the $2129$\,d SED and break the degeneracy in the $2660$\,d SED using the constrained $\nu_{\rm cool, jet}$. We carry out MCMC analysis as before, and all the best-fitting values are reported in Table \ref{tab:radiomatsumoto} for two illustrative scenarios: a promptly-launched jet with $t_{\rm dyn}=\delta t$ (top rows) and a delayed jet launched with $t_{\rm dyn} = \delta t - 1471$\,d (bottom rows). Here the delayed jet's launch date is chosen to be the time of the first detection of the second flare; the true launch date is presumed to be between these two extremes. Times measured since these jet launch dates are represented as $\delta t_{\rm jet}$. For each launch date, we also consider two possible observer viewing angles: $\theta_{obs}=30^{\circ}$ and $\theta_{obs}=90^{\circ}$. 

Now we discuss the fitting results and the associated caveats. For the most off-axis case ($\theta_{\rm obs} \approx 90{\arcdeg}$) at the latest possible jet launch time, the jet models require $\Gamma \gtrsim 3$. This is an unusually large Lorentz factor for an outflow $\sim 4$ years after its launch, given the high-density nuclear environments in which TDEs occur. 
The required jet energy is also high in most cases ($U_{\rm jet}\gtrsim10^{52}$ erg, Table \ref{tab:radiomatsumoto}), but is similar to the energies previously inferred for radio-luminous TDEs with powerful on-axis relativistic jets \citep[e.g.,][]{Berger12,Zauderer13,Brown15,Cenko12}. 
Note that the lowest energy is inferred for the $\delta t_{\rm jet} = 658$\,d SED with $\theta_{\rm obs} = 90\arcdeg$ but this is the equipartition energy; if the system is not in equipartition the true energy will be even higher. With the equipartition $R_{\rm jet}$ and $\Gamma$ for this epoch, the emitting region has a transverse size of $R_{\rm jet}/\Gamma \sim 6\times10^{17}$\,cm. In comparison, our VLBA observation at $= 2384$\,d in \S\ref{SubSec:DataVLBI} gives a $3\sigma$ upper limit to the physical source size of $< 9 \times 10^{18}$\,cm, which is not constraining. The less off-axis cases all require even higher jet Lorentz factors, which we consider to be physically unlikely. Furthermore, if the emission was powered by an off-axis jet, $\Gamma_{\rm jet}$ is expected to decrease with time. However, in Table \ref{tab:radiomatsumoto}, we do not see such a trend. A single-component top-hat jet is thus unlikely to explain the second radio flare of ASASSN-15oi.

\setlength{\tabcolsep}{2.5pt}
\renewcommand{\arraystretch}{1.1}
\begin{deluxetable*}{c|cccccccccc}
\label{tab:radiomatsumoto}
\tablecolumns{11}
\tablecaption{Physical Parameters derived from the off-axis jet model of \S\ref{subsec:offaxisjetparams}. The constraint on the SED peak translates into limits on physical parameters. We break the degeneracy using the constrained $\nu_q = \nu_{\rm cool,jet}$ to solve for $\epsilon_{\rm B,jet}$, and assuming $\epsilon_{\rm e,jet} = 0.1$. The last column gives the transverse size of the emitting region. As can be seen, the VLBA $3\sigma$ upper-limit on the source size of $<9 \times 10^{18}$\,cm is non-constraining in all the cases.}
\tablehead{SED & $\delta t_{\rm jet}$ & $\theta_{\rm obs}$ & $\epsilon_{\rm B,jet}$ & & $B_{\rm jet}$ & $U_{\rm jet}$ & $R_{\rm jet}$ & $n_{\rm e, jet}$ & $\Gamma_{\rm jet}$ & $R_{\rm{jet}}/\Gamma$ \\
$\#$ & (d) & (\degr) & & & (G) & ($\times 10^{52}$\,erg) & ($\times 10^{18}$\,cm) & (cm$^{-3}$) & & ($\times 10^{17}$\,cm)}
\startdata
\hline
\multicolumn{10}{c}{Times measured as days since discovery}\\
\hline
9 & $2129^*$ & $30$ & $0.1$ & & $0.22^{+0.02}_{-0.02}$ & $33.7^{+3.7}_{-3.4}$ & $49.5^{+0.1}_{-0.1}$ & $0.15^{+0.06}_{-0.04}$ & $59^{+4}_{-5}$ & $\sim 8.4$\\
10 & $2660$ & $30$ & $0.003$ & & $< 0.04$ & $> 51$ & $> 37.1$ & $> 0.3$ & $> 41$ & $\sim 9.0$\\
\hline
\hline
9 & $2129^*$ & $90$ & $0.1$ & & $0.25^{+0.03}_{-0.03}$ & $6.0^{+0.5}_{-0.6}$ & $6.58^{+0.02}_{-0.02}$ & $0.9^{+0.2}_{-0.2}$ & $8^{+1}_{-1}$ & $\sim 8.6$\\
10 & $2660$ & $90$ & $0.015$ & & $< 0.07$ & $> 5.2$ & $> 6.24$ & $> 0.8$ & $> 6$ & $\sim 1.0$\\
\hline
\multicolumn{10}{c}{Times measured as days elapsed since T$_0 = 1471$\,d$^{[a]}$}\\
\hline
9 & $658^*$ & $30$ & $0.1$ & & $0.14^{+0.01}_{-0.01}$ & $2.0^{+0.2}_{-0.2}$ & $15.23^{+0.03}_{-0.03}$ & $0.13^{+0.04}_{-0.03}$ & $25^{+2}_{-2}$ & $\sim 6.0$\\
10 & $1189$ & $30$ & $0.051$ & & $< 0.07$ & $> 4.7$ & $> 24.9$ & $> 0.1$ & $> 26$ & $\sim 9.6$\\
\hline
\hline
9 & $658^*$ & $90$ & $0.1$ & & $0.15^{+0.02}_{-0.02}$ & $0.32^{+0.03}_{-0.04}$ & $1.94^{+0.02}_{-0.02}$ & $0.7^{+0.2}_{-0.2}$ & $3.2^{+0.2}_{-0.3}$ & $\sim 6.1$\\
10 & $1189$ & $90$ & $0.293$ & & $<0.14$ & $> 1$ & $> 4.1$ & $> 0.1$ & $> 4$ & $\sim 1.0$\\
\enddata
\tablecomments{\small{$^*$ These values are equipartition values, and any deviation from equipartition will result in an increase in the energy and density estimate.\\}}
\vspace{-1cm}
\end{deluxetable*}

\section{Discussion} \label{Sec:Discussion}
\begin{figure*}
    \centering
    \includegraphics[scale=0.45]{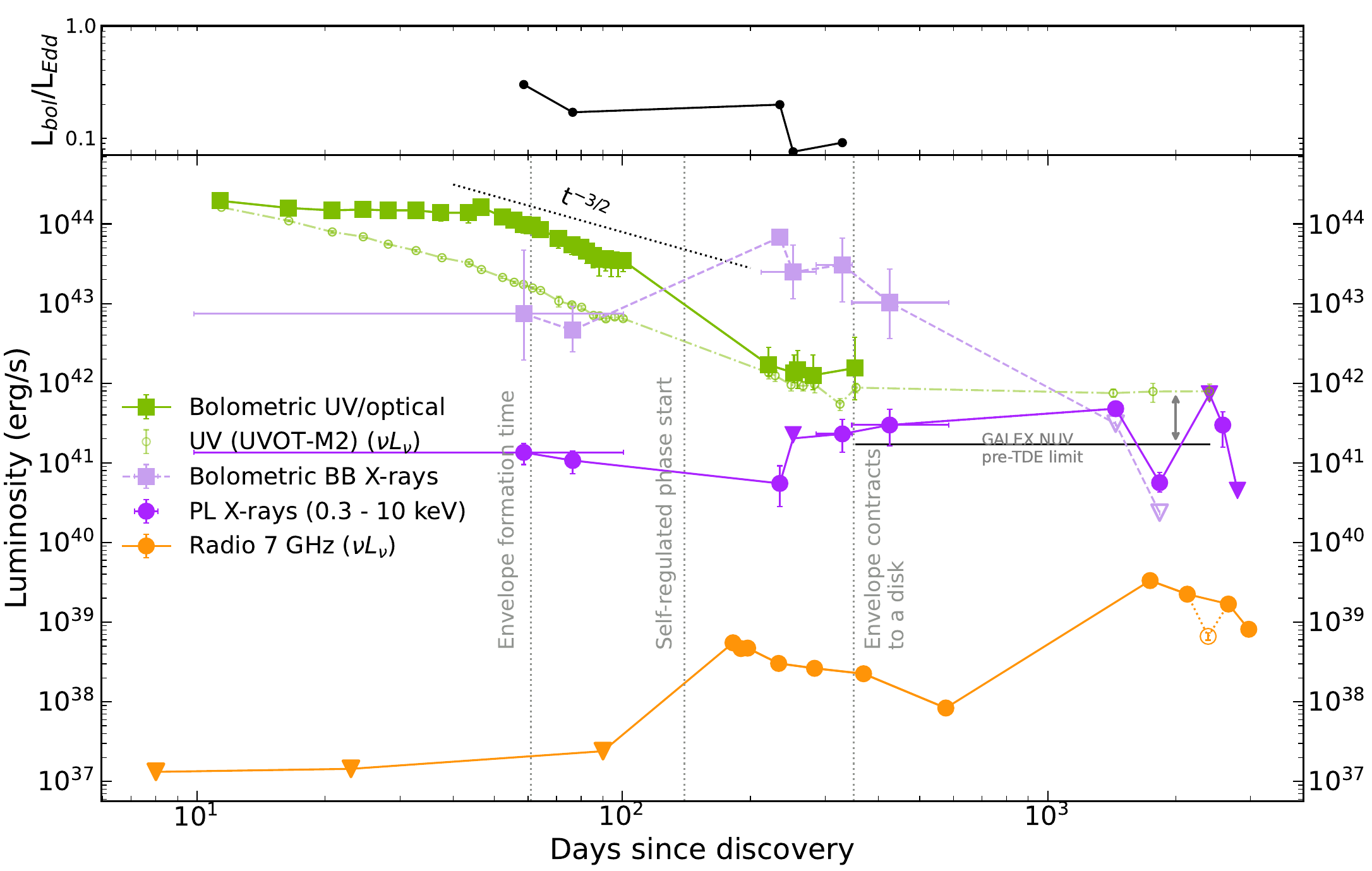}
    \caption{\textit{Bottom Panel: }Multi-band light-curve evolution of ASASSN-15oi from the time of discovery until $\sim 3000$\,d. The thermal emission is represented by squares, and the non-thermal emission in circles. We plot the bolometric UVO emission in green. The UVO bolometric luminosity was calculated only for the epochs where enough spectral coverage was present across the UV and optical bands. At $\sim 3000$ d post-disruption, the UVOT-m2 band luminosity ($\nu L_{\nu}$ at $\approx 1.5\times10^{15}\,$Hz; small empty green circles) converges to a constant value, in contrast with continuing variability in the non-thermal X-rays. We show a dotted line with the trend $\propto t^{-3/2}$ predicted by the cooling envelope model for the optical luminosity. The thermal bolometric X-ray luminosity is presented with light purple squares. The non-thermal power-law (PL) component of the X-rays in the energy range 0.3--10 keV is plotted with dark purple circles. The non-thermal radio ($\nu L_{\nu}$ at $\nu \sim 7.5$\,GHz) is plotted in orange, where the VLBA observation is presented with an empty circle. \emph{Top Panel:} We plot the ratio of the total observed thermal bolometric luminosities $L_{X,BB} + L_{UVO}$ to $L_{\rm{Edd}}$ at $\lesssim 300$\,d. We note that it remains constant between $\delta t \approx 80 - 230$\,d, with the rise in thermal X-ray emission compensating for the decline in the UVO emission.} 
     \label{fig:multibandlc}
\end{figure*}

Figure \ref{fig:multibandlc} shows the contemporaneous evolution of the emission associated with ASASSN-15oi at all wavelengths. Here is a brief overview of how the luminosity changes over time at each wavelength:
\begin{enumerate}\label{list:multiwavoverview}
    \item\label{lt:uvo_disclist} The UV/optical (UVO) bolometric luminosity ($L_{\rm UVO,bol}$) remains stable for the first $\sim 50$\,d (with the luminosity consistent within $1\sigma$) and subsequently shows a steep decline. It plateaus starting at $\delta t \approx 200$\,d for $\sim 100$\,d before the optical luminosity fades to the host galaxy level. In contrast, the UV luminosity continues to plateau out to $\sim 3000$\,d as an excess of emission compared to the host galaxy.
    \item\label{lt:thermal_x_disclist} The X-ray luminosity has two components characterized by a blackbody and a power-law spectrum, respectively. The X-ray blackbody flux shows little to no variation for the first $\sim 80$\,d. Over the following $\sim 150$\,d, it brightens significantly  by a factor of $\sim 6$, accompanied by an increase in $R_{\rm BB}$ but not $T_{\rm BB}$. Similar to the UVO emission, the X-ray blackbody flux shows little variation from $\delta t \sim 200 - 300$\,d. The blackbody X-rays then fade to non-detectable levels at $\delta t > 1400$\,d.
    \item\label{lt:bolandeddlum} The top panel of Figure \ref{fig:multibandlc} shows the total observed thermal bolometric luminosity, $L_{\rm bol} = L_{\rm{UVO,bol}} + L_{\rm{X, BB, bol}}$, where $L_{\rm{UVO,bol}}$ is adopted from \cite{Hinkle21} and $L_{\rm{X,BB,bol}}$ is calculated using the best-fitting $R_{\rm BB}$ and $T_{\rm BB}$ in Table \ref{tab:xraydata}. As can be seen, the thermal bolometric luminosity remains constant during the period $\delta t \sim 80 - 230$\,d before abruptly declining. As noted in the above points, the UVO and thermal X-ray luminosities show simultaneous opposite trends between $\sim 80 - 230$\,d, and add up to a constant $L_{\rm bol}$, suggesting that the UVO and thermal X-ray emission are related.
    \item\label{lt:nonthermalxray} The non-thermal X-ray flux remains stable (with luminosity consistent within $1\sigma$) up to $\delta t \approx 230$\,d (this timescale coincides with the blackbody X-ray peak), thereafter increasing as $\propto t^{4}$. The luminosity shows a further plateau starting from $\delta t \approx 330$\,d to $\delta t \approx 1440$\,d. At $\delta t > 1440$\,d,  there is significant variability with fluctuations as $\sim t^{\pm9}$, observed at timescales of $\sim 1$ year.  
    In our last observation, the luminosity fades below detectable levels, exhibiting a $\propto t^{-24}$ decline following the $\delta t \approx 2580$\,d detection.
    \item\label{lt:radio1st} In the radio, ASASSN-15oi exhibits \textit{two} distinct flares, with the first radio detection at $\delta t = 182$\,d. It rises from the earlier non-detections as $F_{\nu} \propto t^{4.5}$ and subsequently declines as $\sim t^{-1}$.
    \item \label{lt:radio2nd}The second radio flare was first detected at $\delta t \approx 1471$\,d (with a rise of $\sim t^{4}$ from the last observation of the first flare) and subsequently fades as $\propto t^{-1.5}$ from $\delta t \gtrsim 1741 - 2660$\,d.  
    Between the $2660$\,d and $2970$\,d observations, the light curve declines steeply as $t^{-7}$. 
\end{enumerate}

Observations at UVO wavelengths and of the blackbody component of the X-ray radiation can be best described as thermal emission with maximum temperatures of $\approx 5 \times 10^4$\,K, and $\approx 6 \times 10^5$\,K, respectively. The associated radius of the thermal X-ray emission is $\sim (2 - 13)\times 10^{11}\,\rm{cm}$, which is comparable to the inner regions ($\lesssim 4 r_{\rm{g}}$) of an accretion disc around a $M_{\rm BH} = 2.5 \times 10^{6}\,M_{\odot}$ non-rotating SMBH\footnote{We provide other host galaxy mass estimates of ASASSN-15oi that have been reported in the literature in Appendix \ref{sec:smbhmass}, and note that using these other estimates has no major impact on our conclusions.}. This $M_{\rm BH}$ was reported by (\citealt{Gezari17}); and we adopt this value for our subsequent analysis. The UVO emission originates from a region extending out to $\sim 10^{15}\,\rm{cm} >> r_{\rm g}$. 
Conversely, both the emission detected at radio wavelengths, and the contribution of the power-law component to the observed X-rays, can be characterized as non-thermal. Below we discuss the possible physical origin of each emission component.

\subsection{Thermal Emission from ASASSN-15oi as explained by the ``cooling envelope'' model}\label{subsec:thermaldisc}
In a previous investigation of ASASSN-15oi \citep{Gezari17} that extended to $\delta t \sim 600$\,d, it was suggested that the prompt UVO emission originated in shocks resulting from stream-stream collisions during the process of circularization, but owing to inefficiencies in this process, the formation of the accretion disc was postponed causing a delayed rise in thermal X-rays (at $\delta t \sim 100$\,days). Motivated by points $\#$\ref{lt:uvo_disclist}, \ref{lt:thermal_x_disclist}, and \ref{lt:bolandeddlum} in the above list, we explore an alternate scenario of a cooling envelope model proposed by \cite{Metzger2022}, where the UVO and thermal X-rays can be explained in a unified scenario. It is similar to the assumptions made by \cite{Gezari17} in that the accretion is delayed. However, the difference is that in the cooling envelope model, the debris circularizes promptly and efficiently (e.g., as seen in recent simulations, \citealt{Steinberg2024}), and forms a quasi-spherical pressure-supported envelope. The envelope undergoes Kelvin-Helmholtz contraction and cools radiatively resulting in a fading UV and optical luminosity, as observed. As the envelope is contracting, material is gradually accreted onto the SMBH causing the delayed rise in the X-rays. This scenario is more appealing for explaining the decline in the UV/optical bolometric luminosity together with the rise in thermal X-ray emission and the subsequent UV/optical/X-ray luminosity plateaus. 

In the cooling envelope model, the debris stream promptly circularizes on the fallback timescale of the most tightly bound debris according to
\begin{equation}
    t_{\rm{fb}} \simeq 41\,\rm{d} \left(\frac{\emph{k}}{0.8}\right)^{-3/2}\,M_{\rm BH,6}^{1/2}\,m_{*}^{1/5}
\end{equation}
where $k$ is the constant related to stellar structure and rotation prior to disruption, and is taken as $k = 0.8$ for a $\gamma = 5/3$ polytropic star (as assumed in \citealt{Metzger2022}); $M_{\rm BH,6}$ is the mass of the SMBH in units of $10^{6} M_{\odot}$ and $M_{*} = m_{*}M_{\odot}$ is the mass of the disrupted star. 
Assuming that the disruption occurs at the time of the optical discovery, when a star with mass $m_{*} = 1\,M_{\odot}$ is disrupted on a path with a penetration factor $\beta = 1$ (with $k \simeq 0.8$), a pressure-supported quasi-spherical envelope is formed at $t_{\rm fb} \approx 60$\,d \footnote{If the timescale of envelope formation, and hence $t_{\rm{fb}}$, is the same as the time at which the UVO luminosity starts decaying at $t = 45$\,d, the disrupted star will have m$_{*} \approx 0.2\,\rm{M_{\odot}}$.
On the other hand, if we assume $m_{*} = 1\,\rm{M_{\odot}}$, we get M$_{\rm{BH}} \approx 1.2\times10^6 \,\rm{M_{\odot}}$, lower than most of the mass estimates in the literature, but consistent with the studies by \cite{Wevers19, Wevers20Erratum, Mummery2024_galscaling}. None of these mass estimates will significantly change our results.} (plotted in grey dotted line in Figure \ref{fig:multibandlc}).

The cooling envelope model predicts an initial photospheric radius of $R_{\rm{ph}} \approx 6 \times 10^{14}\,\rm{cm}$ (eq. 17 in \citealt{Metzger2022}), similar to the observed $R_{\rm ph} \sim 2 \times 10^{14}$\,cm of the optical emission from \cite{Hinkle21} at $t = t_{\rm{fb}}$.
The initial optical luminosity of $2.5 \times 10^{43}\,\rm{erg/s}$ at $\nu = 6.8 \times 10^{14}\,$Hz predicted by this model (given by their eq. 32 for mass estimates used here) is in good agreement with the observed optical luminosity of $\sim 1.5 \times 10^{43}\,\rm{erg/s}$ at $t = t_{\rm{fb}}$.
As the envelope contracts, the optical luminosity decreases while the accretion rate increases, leading to a delayed rise in the X-rays. We see both the decrease in optical luminosity and rise in X-ray luminosity over a period of $\sim 150$\,days.\footnote{We note that the declining opacity of the envelope and/or a widening of the accretion funnel may further contribute to the increased X-ray luminosity.}

If there is sufficient energy feedback from accretion, the envelope remains ``puffed up'', i.e., depending on how efficiently accretion energy goes into the envelope vs. being radiated, it can result in a flat light curve at all frequencies. We calculate the start time of this phase using the envelope-radius evolution eq.\ 27 in \citealt{Metzger2022} (also plotted in grey dotted line in Figure \ref{fig:multibandlc}).  The observed R$_{\rm{ph}} \approx 8 \times 10^{13}$\,cm during this phase matches the model's prediction (given by their eq. 37).  
Another robust prediction of the {cooling envelope} model is for the total radiated luminosity to remain constant at the Eddington luminosity ($L_{\rm Edd}$) while the envelope is contracting. In ASASSN-15oi the total thermal bolometric luminosity ($L_{\rm bol}$) indeed remains constant (from $t_{\rm fb}$ until the accretion peak at $230$\,d; top panel in Figure \ref{fig:multibandlc}), although at sub-Eddington levels with $L_{\rm bol} \sim (0.2 - 0.3)L_{\rm{Edd}}$ which are consistent with other TDEs \citep[e.g.,][]{Wevers2017}. This can be reconciled with the expectations if either most of the radiation is emitted at unobservable EUV wavelengths, and/or if the uncertainty in $M_{\rm BH}$ leads to an error in the calculation of $L_{\rm Edd}$.

With this set-up, the {cooling envelope} model predicts that the envelope would be completely accreted by $\delta t \sim 350$\,d (their eq. 38; plotted in grey dotted line in Figure \ref{fig:multibandlc}).  However, more realistically, the envelope might settle into an unobscured disc post X-ray peak, which can be modeled as a viscously spreading disc \citep[e.g.,][]{Cannizzo_1990tde,Shen14,Mummery2021_arXiv}. The resulting decrease in the inner disc temperature would account for the exponential decline in thermal X-rays observed after $\sim 350$\,d. As the disc spreads outwards to larger radii, the emitting area increases while it cools, resulting in an almost constant emission that dominates the UV wavelengths (as these wavelengths lie near the Rayleigh-Jeans part of the disc spectrum). This is indeed the case in ASASSN-15oi, even at very late-times ($\delta t \approx 2900$\,d). Previous studies by \cite{vanVelzen2019}, \cite{Mummery2020_uv, Mummery2024_galscaling}, and \cite{Guolo2023}, have also demonstrated that the late-time UV flattening observed in many other TDEs (e.g., ASASSN-14li, iPTF16fnl) is disc dominated and can be explained by longer ($\sim \mathcal{O}(10^3)$\,d) viscous timescales at these larger radii.

Finally, while the predictions of the {cooling envelope} model and the subsequent evolution to a bare accretion disc matches well with most of the thermal emission, certain other features observed in ASASSN-15oi remain unexplained. Particularly, the constant $L_{\rm UVO, bol}$ at $\delta t \lesssim t_{\rm fb} = 60$\,d is not explained by the current analytical model. One can speculate that this flattening results from the complex processes taking place prior to/during the envelope formation, perhaps even due to stream-stream collisions as proposed by \cite{Gezari17}. Regarding the X-ray emission, the low S/N in the individual \emph{Swift}-XRT observations during this period limits our understanding of the X-ray emission behavior at early-times. 
We also currently lack an explanation for the observed temporal evolution of the temperature and radius of the thermal X-ray emitting region. A gradual rise in the accretion rate is expected to result in an increase in the temperature and a decrease in the radius. Here instead, the X-ray blackbody temperature exhibits little variation (marginally constant or else gradually decreasing with time as $T_{\rm BB} \propto t^{-0.2}$) and the best-fitting blackbody radius ($R_{\rm BB}$) seems to increase ($\sim 1.3 \times 10^{12}$\,cm) near the epochs of the peak accretion before returning to the pre-peak levels ($\sim 0.3 \times 10^{12}$\,cm). This would be a problem if we were looking at a bare accretion disc at early times, however, as the geometry of the cooling envelope model is complex, we can expect an unconventional evolution of the thermal X-ray parameters depending on our observing angle.

\subsection{Non-thermal Emission from ASASSN-15oi}\label{subsec:nonthermaldisc}

\begin{figure}
    \centering
    \includegraphics[scale=0.25]{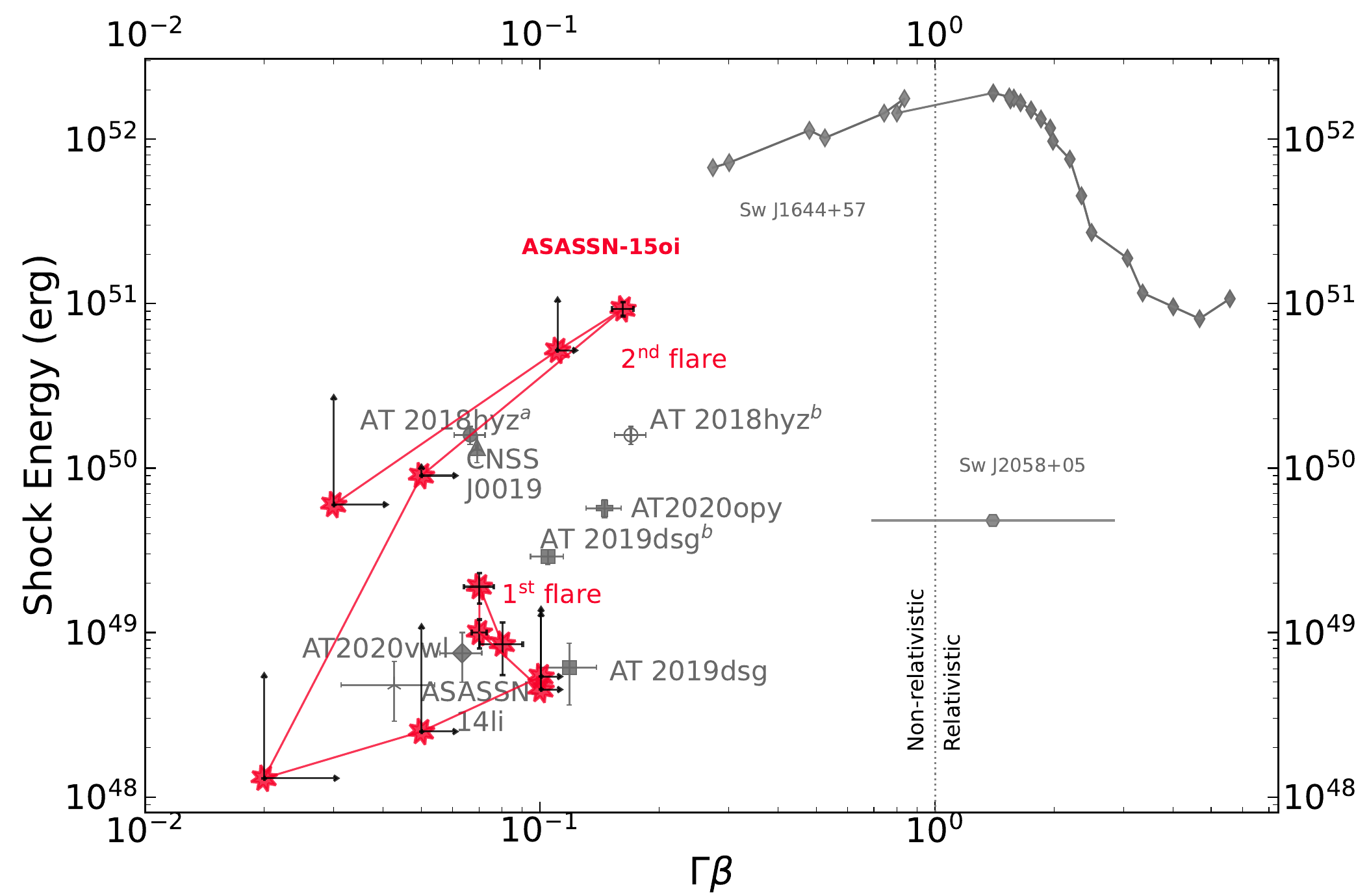}
    \caption{ASASSN-15oi in the context of other TDEs in the literature. The first radio flare of ASASSN-15oi shows energetics ($U_{\rm eq}$ from \S\ref{subsec:syncphysparams}) similar to other non-relativistic TDEs, while the second flare is the most energetic of the subset. For the purpose of comparison, we re-calculate the energetics of other TDEs using the same \cite{Chevalier98} formulation that we used for ASASSN-15oi along with assuming equipartition. The two estimates for AT\,2018hyz and AT\,2019dsg are due to two different inferences of the outflow launch time.
    We also include some jetted TDEs that make-up a very small fraction of all TDEs. Other references: ASASSN-14li (\citealt{Alexander16}); AT\,2018hyz \cite{Cendes_18hyz_2022}; CNSS\,J0019+00 \citep{Anderson2020_cnss}; AT\,2019dsg \citep{cendes21_dsg}; AT\,2020opy \citep{Goodwin_2020opy_2022}; AT\,2020vwl \citep{Goodwin_2020vwl_2023}.}
    \label{fig:radioenergy}
\end{figure}

Unlike the thermal emission, the spectral and temporal evolution of the non-thermal emission (see points \#\ref{lt:nonthermalxray}, \#\ref{lt:radio1st} and \#\ref{lt:radio2nd} in the overview list of \S\ref{Sec:Discussion}) is extremely complex. 
In this section, we briefly speculate on different physical processes/origins and additionally investigate whether the radio and non-thermal X-rays are physically related.

\subsubsection{Possible Origin of the First Radio Flare}\label{subsubsec:disc_firstradio}

\begin{figure}
    \includegraphics[scale=0.35]{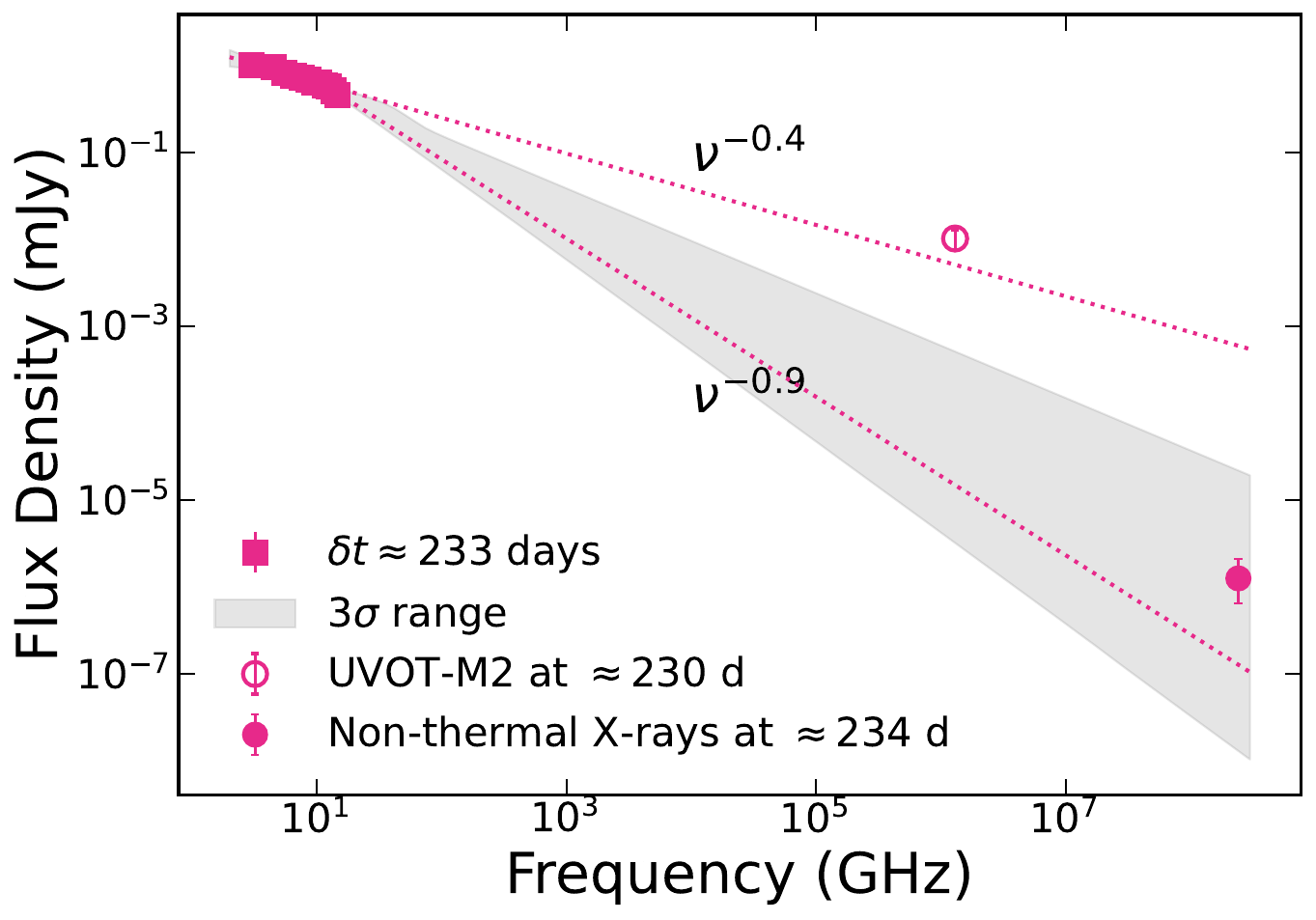}
    \includegraphics[scale=0.35]{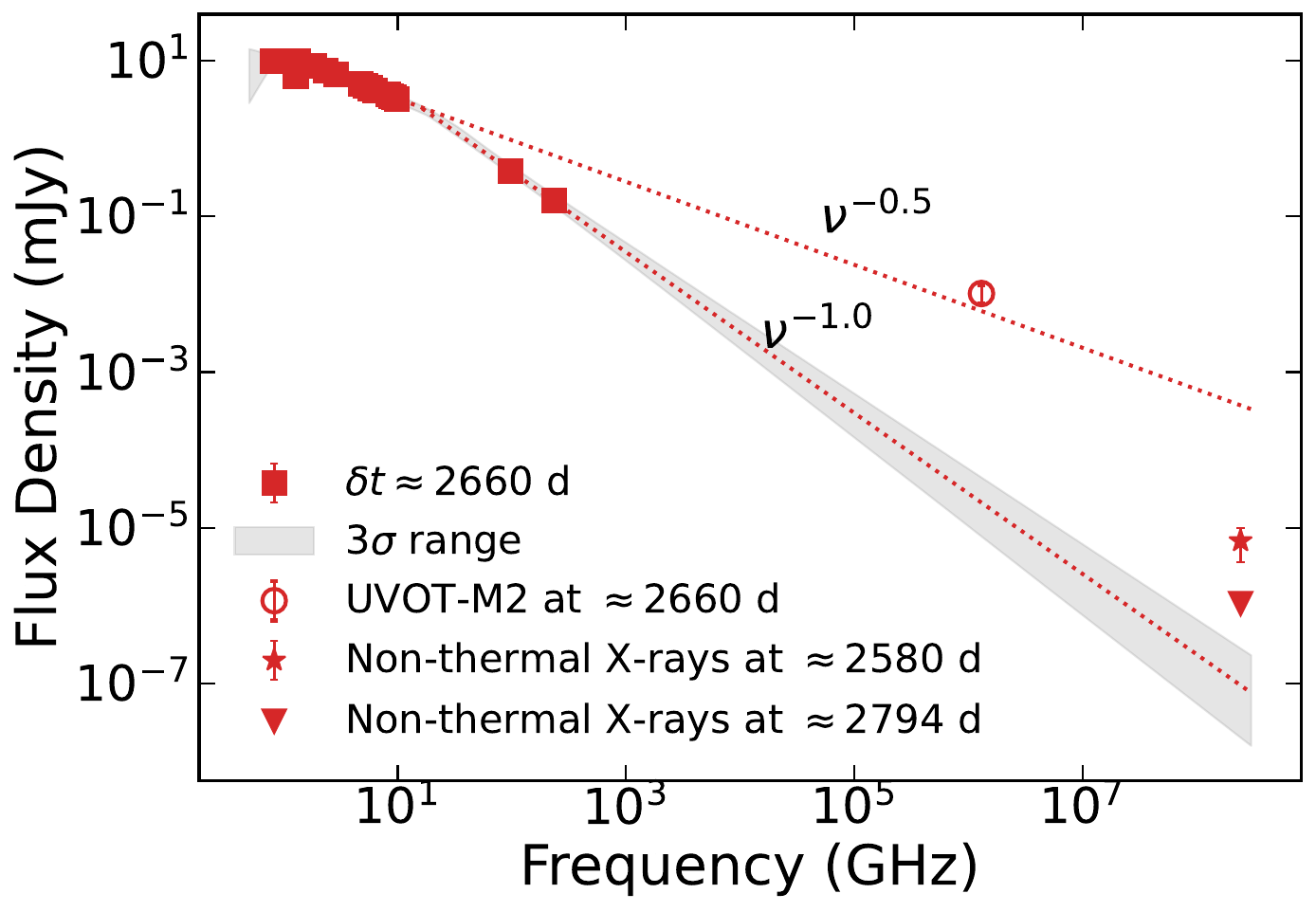}
    \caption{Broad-band radio-to-X-ray SEDs of ASASSN-15oi associated with the radio flares at $\delta t \approx 233$\,d (top, also coinciding with the peak of the X-rays) and $\delta t \approx 2660$\,d, (bottom). Grey shaded areas are the $3\sigma$ ranges of the best-fitting spectra extrapolated from the radio observations to the X-rays. We also plot the median $\alpha_{\rm thin}$ and $\alpha_q$ segments as dotted lines.
    The open circles are UVOT-M2 observations. The plotted X-ray observations are the power-law component. At $233$\,d, the X-rays are consistent within $3\sigma$ of the extrapolation. However, the X-rays at $2580$\,d (closer in epoch to the radio observations) are clearly in excess. The X-ray upper-limit at $\approx 2794$\,d, the next closest epoch to the radio, is also shown.
    }
    \label{fig:radiotoxraysed}
\end{figure}

The first radio flare was analysed in detail by \cite{Horesh21}. Our results are broadly consistent with their study. A particular feature of the system parameters determined here is the significant jump in the density estimates between the SEDs at $\delta t \approx 197$\,d and $233$\,d, similar to the inferences by \cite{Horesh21}. \cite{Horesh21} suggested that the outflow responsible for the first radio flare was not likely to be launched at the time of disruption but was delayed as a consequence of an accretion state transition. \cite{Zhuang2024_15oi} proposed that the first radio flare is powered by the interaction of an outflow with \textit{three} different clouds present around the SMBH at $\sim$ parsec scales. Their model explains the observed temporal evolution, although a close examination of their fig. 13 reveals discrepancies with the observed SEDs. Here, we discuss our interpretation of the power source of the first radio flare. 

The energy and the velocity of the outflow that powers the initial radio flare are similar to other non-relativistic TDEs (as seen in Figure \ref{fig:radioenergy}). However, the timescale at which the first radio emission appears in ASASSN-15oi and the steep rise ($F \propto t^4$) from the prior non-detections are unusual compared to non-relativistic TDEs with ``prompt'' emission (e.g., AT\,2019dsg in Figure \ref{fig:radioLCcomparisonAH}). Among commonly invoked scenarios that can also explain the somewhat higher kinetic energy in this case (e.g., compared to the energy in ASASSN-14li as inferred by \citealt{Krolik16}), we disfavor the scenario whereby the radio emission is powered by the interaction of a super-Eddington accretion-driven wind outflow with the ambient medium \citep{Alexander16, cendes21_dsg}. This would require large accretion rates that are not inferred in ASASSN-15oi until \emph{after} the appearance of the first radio flare. 

Instead, we find the case of a stream-stream Collision Induced Outflow (CIO) proposed by \cite{LuBonnerot2020_CIO} more likely. 
Episodic mass ejections during the envelope formation (at timescales $\lesssim t_{\rm fb} \sim 60$\,d) are expected due to the repeated orbits of the self-colliding streams around the SMBH. This process creates a clumpy medium with fast moving material driving multiple shocks of varying energies (and possibly different microphysical paramaters). For a significant amount ($> 20\%$) of initially unbound tidal debris moving at $0.01 - 0.1\,c$, CIO is expected to dissipate $10^{49} - 10^{52}$\,erg kinetic energy over a diverse set of timescales. It is subsequently possible in this scenario that an outflow could decelerate over $\sim 100$\,d and power the first radio flare. 
The radio SEDs and derived velocity and energy from the first flare ( see \S\ref{subsec:syncphysparams}) are consistent with this scenario involving multiple outflows in an inhomogeneous medium. 
Detailed modeling for this scenario is left for future work.

\subsubsection{Possible Origin of the Second Radio Flare}\label{subsubsec:disc_secondradio}

As seen in Figure \ref{fig:radioenergy}, the outflow that powers the second radio flare carries substantially more energy than the first flare, suggesting distinct physical origins. In the recent study by \cite{Sato2024_15oi}, these authors modeled the two flares (with observations at $\delta t < 1500$\,d) with a structured jet model consisting of a wide jet (responsible for the first flare) and a narrow jet (responsible for the second flare) with different shock microphysics, essentially the same scenario as the two distinct outflows that we propose here. Their conclusions are based solely on the temporal evolution. Their narrow jet (with $\theta_{\rm jet} \sim 6\arcdeg$ and $\theta_{\rm obs} \sim 70\arcdeg$) component that is responsible for the second radio flare carries kinetic energy ($E_{\rm{k}} \sim 10^{52}$\,erg) similar to other TDEs in the literature. It is unclear, however, if their model can explain the observed spectral evolution.  

Here, we propose that the second radio flare is powered by an outflow different from that of the first radio flare, and was likely launched when the accretion rate peaked (at $\gtrsim 200$\,d) in the context of the {cooling envelope} model discussed in \S\ref{subsec:thermaldisc}. The outflow(s) may manifest as a relativistic jet or a quasi-spherical mildly-relativistic wind interacting with the ambient medium. However, the details of the outflow powering the second radio flare are not well captured by simplified models like those presented in \S\ref{Sec:RadioModeling}, invoking the need for a more intricate system, perhaps similar to the one proposed for the first radio flare in the previous section, but with a more powerful source that is launched significantly later after the disruption. 

\subsubsection{Possible Origin of the Non-thermal X-rays}\label{subsubsec:disc_xrays}

If the radio emission is indeed powered by synchrotron emission, it may extend to X-ray energies as well. We verify here if the non-thermal X-rays and first radio flare share the same synchrotron spectrum. As can be seen in the top panel of Figure \ref{fig:radiotoxraysed}, the non-thermal X-rays fall within the $3\sigma$ region of the extrapolated models\footnote{As discussed in \S\ref{subsubsec:synccool}, the break in the radio SED is likely not $\nu_{\rm cool}$. The real synchrotron cooling break can possibly lie between radio and X-rays and cause an even steeper spectral segment which when extrapolated to X-ray wavelengths might make the observed X-rays appear in excess.} at $\sim 233$\,d (the only epoch when both X-rays and radio were observed simultaneously). However, X-ray emission existed prior to the first radio flare  making them causally unrelated, unless the radio emission was absorbed at frequencies at least as high as $22$\,GHz for the first $180$\,d. An extremely large $n_{\rm e} \sim 10^{7}\,\rm{cm^{-3}}$ is required if the source was synchrotron self-absorbed and its SSA spectrum were to peak at $\sim 22$\,GHz with $F_{\nu} \sim 0.06$\,mJy at $\lesssim 180$\,d, which is unlikely. The higher densities found at $\sim 180$\,d in \S\ref{subsec:syncphysparams} could also potentially result in free-free absorption of the radio emission \citep{weiler2002}. The free-free absorption optical depth is given by $\tau_{\rm FFA} \propto T^{-1.35}\nu^{-2.1} n_e^2 l$ (\citealt{Mezger1967_freefree}), where $T$ is the temperature of the absorbing gas and $l$ is the optical path along the line of sight.  With the estimated $n_{e}$ for $\delta t = 182$\,d SED in Table \ref{tab:radiophysparams}, we find $\tau_{\rm FFA} < 1$ at $\sim 22$\,GHz. Therefore, if there was emission present at this frequency, it is unlikely to be affected by either synchrotron self-absorption or free-free absorption. Finally, the evolution of power-law X-rays and the radio emission is distinct, as can be seen in Figure \ref{fig:multibandlc}, challenging this interpretation further.

Furthermore, there is no strong evidence that the second radio flare and the late-time non-thermal X-ray emission share the same synchrotron spectrum because: (1) as seen in the lower panel of Figure \ref{fig:radiotoxraysed}, the $\delta t= 2580$\,d X-rays (closest to the epoch of the $\delta t \approx 2660$\,d radio SED) are in excess of the extrapolation of the best-fitting radio model beyond $3\sigma$. Therefore, the X-rays likely require a different mechanism. (2) In the latest observations, the steep decline in the X-ray emission at $= 2790$\,d precedes the steep decline in the radio emission at $= 2970$\,d, also making them temporally inconsistent. 

A common source for non-thermal X-rays in TDEs, independent of synchrotron emission that powers radiation at radio wavelengths, is the inverse Compton scattering of the lower-energy disc photons by an electron corona. Such coronae were identified in AT\,2019ehb \citep{Yao_21ehb}, in AT\,2018fyk \citep{Wevers_2018fyk_corona} and a few others \citep{Guolo2023}, and can form/become strong during soft-to-hard state transitions \citep[e.g.,][]{Done2007_XRBs}, similar to X-ray binary systems (XRBs). The phenomenology of ASASSN-15oi does hint at state transitions of hard $\rightarrow$ soft $\rightarrow$ hard (top panel in Figure \ref{Fig:XrayLC}), and a direct comparison to the above studies suggest the presence of a corona at least at $\delta t \gtrsim 300$\,d, after the soft-to-hard transition. A hard-to-soft transition was observed in AT\,2019azh in \cite{Sfaradi_19azh}, followed by a late-time radio flare at $\sim 200$\,d, similar to timescales of the first radio flare of ASASSN-15oi. Here, the first radio flare occurs during the hard-to-soft transition, more alike XRBs, as also invoked by \cite{Horesh21}.

Assuming that the early- and late-time non-thermal X-rays share the same physical origin, we find several temporal features that can also support the corona (or a near-disc) origin. First, the power-law component declining when the blackbody component (originating near the inner disc regions) showed an increase at $\sim 230$\,d hints that the non-thermal X-rays are affected by the accretion process in the inner disc region. Second, the non-thermal X-rays showing significant variability at $> 1400$\,d may hint at changes in the disc (and/or corona). We note that the outer disc is likely stable, as the late-time UV emission originating in the outer regions (as discussed in \S\ref{subsec:thermaldisc}) continues to plateau. And finally, the lag between the steep decline in the non-thermal X-rays and the steep decline in the radio at $\sim 3000$\,d can possibly be explained as a delay caused by the propagation of information from the inner regions of the disc to the radio-emitting components at larger distances from the SMBH \citep{Fabio2012}. 

We briefly speculate on one possible scenario where all these characteristics can be accounted for. We base our arguments on the premise that physical processes governing accretion are universal around BHs of different masses, and therefore, we draw inspiration from the phenomenon observed in AGN and XRB systems. The scenario is that of a hybrid disc, with the inner disc transitioning from a disc/corona system (possibly representative of the system at $\lesssim 1400$\,d) to an Advection-Dominated Accretion Flow (ADAF; e.g., \citealt{Narayan94, Czerny2000}). The outer disc is likely a geometrically thin, optically thick stable standard disc. This transition in the inner regions can be triggered by a decrease in the mass-accretion rate beyond a critical value, as often seen in XRBs in their low-hard state \citep[e.g.,][]{esin1997,xrayvariability_xrbs_2004}. In XRBs/AGNs, such a transition typically occurs at $\lesssim 10^{-2}\dot{M}_{\rm Edd}$, where ${\dot{M}}_{\rm Edd}$ is the Eddington mass accretion rate, which for our SMBH mass would translate to $\sim 10^{42}$\,erg/s, as observed. Furthermore, at the interface between the inner ADAF and the outer regions, there may exist a narrow unstable zone where radiation pressure instabilities can induce the observed X-ray variability. The instabilities are driven on viscous timescales which can be shortened depending on the size of this unstable zone, and can range over timescales of $\sim$ months - years \citep[e.g.,][]{Sniegowska2020},  aligning with our observed non-thermal X-ray variability.
As an example, a recent study by \cite{Veronese2024} attributed the long-term X-ray variability (at timescales of $\sim 10$ years) in Mrk\,1018, a changing-look AGN, to a similar accretion flow transition in the inner disc.

There are other physical processes such as wind-torus interaction proposed by \citet{Mou2021_xray_windtorusint} and \cite{ Mou2021_radio_windtorusint} that can produce variability in the non-thermal emission at late-times on $\sim$ months - years timescales. (Such torii are usually present at $\sim 0.1$ pc around AGNs.) In this case, the model-predicted luminosity $\gtrsim 10^{41}$\,erg/s matches the observed power-law X-ray luminosity. However, first, there are no infrared observations of ASASSN-15oi that could have confirmed a presence of a torus. Furthermore, the archival mid-infrared WISE colors of the host galaxy also do not support the presence of a strong AGN around which such torii are usually present. And second, if this were the case, that would imply that the early- and late-time non-thermal X-rays are powered by different sources. We also want to emphasize that other TDEs have also shown late-time X-ray variability, albeit on shorter timescales ($\lesssim$ days in e.g., in AT\,2018fyk,\citealt{Wevers_2018fyk_corona}). Due to our sparse temporal coverage, our observations are not sensitive to variability on $\sim$day timescales, if it existed. Finally,  we stress that our discussion here is only speculative, and we leave a more in-depth quantitative analysis, including testing for variability of the late-time non-thermal X-rays at shorter timescales to future work.

\section{Summary and Conclusions} \label{Sec:Conclusions}
In this study we have presented one of the most extensive datasets for a TDE, namely ASASSN-15oi, and our analysis thereof. ASASSN-15oi was observed across the electromagnetic spectrum using an array of telescopes including the VLA, VLBA, MeerKAT, ATCA, ALMA, \textit{Swift}-UVOT, \textit{Swift}-XRT, and \textit{XMM}-Newton. We re-reduced and re-analysed the previously reported UV and X-ray observations here to provide a homogeneous dataset. 
We also present new observations obtained through our coordinated multi-wavelength campaigns, particularly following the occurrence of the second radio flare,
at $\delta t > 1400$\,d until $\sim 8$ years. We note that ASASSN-15oi was the first TDE observed to have multiple radio flares and highlights the importance of long-term monitoring of
TDEs. We study the emission at different wavelengths individually, but 
also investigate any physical connection between them within the framework of existing models. We summarize our findings below.

We divide the observed emission of ASASSN-15oi into thermal and non-thermal emission.
We treat the emission dominating the optical and UV wavelengths and the blackbody component of the X-rays as thermal.
We establish (in \S\ref{subsec:thermaldisc}) that the \emph{cooling envelope} model proposed by \cite{Metzger2022} explains the evolution of the thermal emission in ASASSN-15oi extremely well. In this model, the radiative cooling of a quasi-spherical envelope formed after the stellar disruption drives the optical/UV emission, different from the interpretations of typical TDEs, where the optical/UV emission is either produced in stream-stream collisions, or via reprocessing of the X-rays by the surrounding material. This model naturally explains the delayed rise of the thermal X-rays as more and more material gets accreted onto the SMBH. The envelope then transforms into a bare accretion disc that spreads outwards on long viscous timescales, explaining the eventual decline in the thermal X-rays in the inner regions and the late-time persisting UV plateau emitting from the outer disc. 

Unlike the thermal emission, qualitatively explaining the non-thermal emission, i.e. the power-law component of the X-rays and the radio emission, is more challenging. In \S\ref{subsec:nonthermaldisc}, we posit the first ($\delta t \lesssim 600$ days) and second ($\delta t \gtrsim 1400$ days) radio flares to have distinct origins. A simplified approach of either a spherical blastwave, or an ultrarelativistic off-axis jet interacting with the ambient medium alone cannot explain the observed spectral and temporal radio evolution without requiring unphysical conditions. Therefore, to account for the spectral variations and the sharp jumps in the physical parameters of the system (as derived in \S\ref{subsec:syncphysparams}) within the first radio flare, we invoke a more complex system with `multiple shocks driven in a clumpy medium' such as that produced by stream-stream collision induced outflows (discussed in \S\ref{subsec:nonthermaldisc}) during the \textit{envelope} formation. The second radio flare is possibly produced by a similarly complex, but more powerful outflow launched as a result of the accretion rate reaching its peak in the context of the {cooling envelope} model.
Finally, the non-thermal X-rays and the radio emission do not arise from same synchrotron spectra. 
In \S\ref{subsubsec:disc_xrays}, we briefly discuss the near-disc origin of the non-thermal X-rays. 
We emphasize that this is purely speculative, and future monitoring of ASASSN-15oi and developing more sophisticated models will be crucial to confirm the origin of its late-time emission. 

In conclusion, the multiwavelength observations of ASASSN-15oi reveal a complex picture of TDEs and their long-term evolution. The detailed dataset allows us to probe the different emission mechanisms at various frequencies and phases in the ``life'' of a TDE. This study emphasizes the importance of late-time observations across the electromagnetic spectrum to better understand TDEs. As a final example, the late-time monitoring may help address the "missing energy" problem in TDEs (as detailed in \citealt{Lu18}) where the observed radiated energy in most TDEs is $\approx 10^{51}$\,erg \citep[e.g.,][]{Mockler2021} falling short of the expected $10^{52} - 10^{53}$\,erg.  Since its discovery, ASASSN-15oi has radiated {$2 \times 10^{51}$\,erg in X-rays, UV and optical, with the first radio flare contributing an additional a few $\times 10^{49}$\,erg of kinetic energy. We further find that the second radio flare carries an energy $\gtrsim 10^{51}$\,erg (in \S\ref{subsec:syncphysparams}), potentially more if a jet is involved. This suggests that kinetic energy of the outflows is a substantial fraction of the overall energy budget and that multiple delayed radio flares could be more common in TDEs (as indeed found in \citealt{Cendes2023_latetimeTDEs}) harboring the energy deficit.
Thus, long-term monitoring of TDEs is crucial, not only to understand their complexities, but also to uncover hidden or delayed emission sources that may resolve the missing energy problem. Exploring the late-time phase in TDEs has only been a recent occurrence, and ASASSN-15oi has clearly demonstrated its importance, but also a need for more complex and nuanced models to incorporate the diverse and evolving nature of TDEs.

\section*{Acknowledgments}
The authors thank Clement Bonnerot, Damiano Caprioli, Brian Metzger, and Juergen Ott from the NRAO helpdesk for helpful discussions. We acknowledge the use of public data from the \textit{Swift} data archive. \textit{XMM-Newton} is an ESA science mission
with instruments and contributions directly funded by
ESA Member States and NASA. The National Radio Astronomy Observatory is a facility of the National Science Foundation operated under cooperative agreement by Associated Universities, Inc. The MeerKAT telescope is operated by the South African Radio Astronomy Observatory, which is a facility of the National Research Foundation, an agency of the Department of Science and Innovation. The Australia Telescope Compact Array is part of the \href{https://ror.org/05qajvd42}{Australia Telescope National Facility} which is funded by the Australian Government for operation as a National Facility managed by CSIRO. This paper makes use of the following ALMA data: ADS/JAO.ALMA\#2019.1.01166.T. ALMA is a partnership of ESO (representing its member states), NSF (USA) and NINS (Japan), together with NRC (Canada), NSTC and ASIAA (Taiwan), and KASI (Republic of Korea), in cooperation with the Republic of Chile. The Joint ALMA Observatory is operated by ESO, AUI/NRAO and NAOJ.

A.~Hajela was supported by a Future Investigators in NASA Earth and Space Science and Technology (FINESST) award \#80NSSC19K1422 during the initial stages of this project. A.~Hajela also acknowledges support from the Carlsberg Foundation Fellowship Programme by Carlsbergfondet. K.~D.~Alexander and C.~T.~Christy acknowledge support provided by the National Science Foundation through award SOSPA9-007 from the NRAO and award AST-2307668. R.~Margutti acknowledges support by the National Science Foundation under award No. AST-2221789 and AST-2224255.
The TReX team at UC Berkeley is partially funded by the
Heising-Simons Foundation under grant 2021-3248 (PI: Margutti). M.~Bietenholz is grateful for computing and office support from York University, Canada. A.~Horesh is grateful for the support by the the United States-Israel Binational Science Foundation (BSF grant 2020203) and by the Sir Zelman Cowen Universities Fund. This research was supported by the ISRAEL SCIENCE FOUNDATION (grant No. 1679/23). 

\clearpage
\appendix
\counterwithin*{equation}{section}
\renewcommand\theequation{\thesection\arabic{equation}}

\vspace{-0.5cm}

\section{Broadband Observations of ASASSN-15\lowercase{oi}}

\setcounter{table}{0}
\renewcommand\thetable{\thesection.\arabic{table}}
\renewcommand*{\theHtable}{\thetable}

\setlength{\LTcapwidth}{\textwidth}
\begin{longtable*}{ccccc|>{\hspace*{.01\linewidth}}c<{\hspace*{.01\linewidth}}|ccccc}
\caption{\textit{Swift-}UVOT Photometry. All the magnitudes and the corresponding errors are reported in the AB magnitude system, and are corrected for galactic extinction. 
\label{tab:uvotphot}}
\\
\toprule
Epoch (MJD) & $\delta t$ (d) & Filter & Magnitude & Error & & Epoch (MJD) & $\delta t$ (d) & Filter & Magnitude & Error\\\midrule\endhead
\hline\multicolumn{3}{r}{{Continued on next page}} & & & & & & \multicolumn{3}{r}{{Continued on next page}} \\ \hline
\endfoot

\hline \hline
\endlastfoot
57259.41 & 11.41 & v & 15.974 & 0.051 & & 57336.40 & 88.4& uvw1 & 19.039 & 0.099 \\
57264.42 & 16.42 & v & 16.298 & 0.056 & & 57339.43 &91.43& uvw1 & 18.991 & 0.101 \\
57270.71 & 22.71 & v & 16.664 & 0.057 & & 57342.10  &94.10& uvw1 & 19.059 & 0.124 \\
57276.13 & 28.13 & v & 16.781 & 0.070 & & 57345.79 &97.79& uvw1 & 19.261 & 0.150 \\
57280.72 & 32.72 & v & 16.831 & 0.071 & & 57348.41 &100.41& uvw1 & 19.319 & 0.113 \\
57288.50 & 40.5  & v & 16.793 & 0.073 & & 57468.80 &220.80& uvw1 & 20.265 & 0.197 \\
57300.57 & 52.57 & v & 16.953 & 0.054 & & 57484.95 &236.95& uvw1 & 20.532 & 0.153 \\
57318.89 & 70.89 & v & 17.061 & 0.059 & & 57508.58 &260.58& uvw1 & 20.416 & 0.103 \\
57340.96 & 92.96 & v & 17.066 & 0.061 & & 57527.29 &279.29& uvw1 & 20.450 & 0.113 \\
57501.11 &253.11 & v & 17.016 & 0.064 & & 57572.84 & 324.84 & uvw1 & 20.596 & 0.127 \\
57572.85 &324.85 & v & 17.135 & 0.064 & & 57602.41 &354.41& uvw1 & 20.714 & 0.148 \\
57602.42 &354.42& v & 17.195 & 0.074 & & 58674.29 &1426.29& uvw1 & 20.633 & 0.090  \\
58674.29 &1426.29& v & 17.139 & 0.041 & &  59020.30 &1772.30& uvw1 & 20.684 & 0.203\\
59649.93 &2401.93& v & 17.234 & 0.088 & & 59649.93 &2401.93& uvw1 & 20.698 & 0.200 \\
59784.74 & 2536.74 & v & 17.109 & 0.105 & & 59828.39 & 2580.39 & uvw1 & 21.003 & 0.235\\
57259.40 & 11.40 &  b & 15.995 & 0.039 & & 57259.41 & 11.41 &  uvm2 & 15.590 & 0.041 \\
57259.42 & 16.42 &  b & 16.374 & 0.042 & & 57259.43 & 16.43 &  uvm2 & 16.017 & 0.042 \\
57268.75 &20.75&b & 16.542 & 0.043 & & 57268.79 &20.79& uvm2 & 16.366 & 0.052 \\
57272.57 & 24.57 & b & 16.808 & 0.047 & & 57272.58 &24.58& uvm2 & 16.516 & 0.044 \\
57276.12 & 28.12 & b & 16.919 & 0.049 & & 57276.13 & 28.13 & uvm2 & 16.750 & 0.046 \\
57280.71 & 32.71 & b & 17.046 & 0.051 & & 57280.72 & 32.72 & uvm2 & 16.951 & 0.047 \\
57285.64 &37.64& b & 17.147 & 0.065 & & 57285.64 &37.64& uvm2 & 17.174 & 0.045 \\
57291.48 &43.48& b & 17.241 & 0.067 & & 57291.59 &43.59& uvm2 & 17.340 & 0.054 \\
57294.61 &46.61& b & 17.358 & 0.074 & & 57294.62 &46.62&uvm2 & 17.545 & 0.047 \\
57300.26 & 52.26 & b & 17.355 & 0.071 & & 57300.27 & 52.27 & uvm2 & 17.793 & 0.048 \\
57303.66 &55.66& b & 17.406 & 0.081 & & 57303.66 &55.66& uvm2 & 17.945 & 0.053 \\
57306.52 & 58.52&  b & 17.345 & 0.075 & & 57306.52 & 58.52&  uvm2 & 18.014 & 0.052 \\
57309.48 &61.48& b & 17.462 & 0.080 & & 57309.48 &61.48& uvm2 & 18.122 & 0.054 \\
57312.20 &64.20& b & 17.349 & 0.086 & & 57312.20 &64.20& uvm2 & 18.203 & 0.056 \\
57318.77 & 70.77 & b & 17.573 & 0.085 & & 57318.87 & 70.87 & uvm2 & 18.537 & 0.176 \\
57324.08 &76.08& b & 17.630 & 0.123 & & 57324.09 &76.09& uvm2 & 18.653 & 0.071 \\
57328.14  &80.14& b & 17.644 & 0.093 & & 57328.17 &80.17& uvm2 & 18.730 & 0.066 \\
57333.48  &85.48& b & 17.634 & 0.100 & & 57333.52 &85.52& uvm2 & 18.984 & 0.077 \\
57336.40  &88.40& b & 17.649 & 0.116 & & 57336.41 &88.41& uvm2 & 19.000 & 0.072 \\
57340.73 & 92.73 & b & 17.687 & 0.089 & & 57339.44 &91.44& uvm2 & 19.091 & 0.077 \\
57347.12 &99.12& b & 17.817 & 0.098 & & 57343.93 &95.93& uvm2 & 19.031 & 0.085 \\
57501.10 &253.10 & b & 17.914 & 0.065 & & 57348.42 &100.42& uvm2 & 19.084 & 0.077 \\
57572.84 & 324.84 & b & 17.907 & 0.061 & & 57468.80 &220.80& uvm2 & 20.788 & 0.175 \\
57602.41 &354.41& b & 17.861 & 0.064 & & 57476.65 &228.65& uvm2 & 20.874 & 0.178 \\
58674.29 &1426.29& b & 17.931 & 0.041 & & 57497.32 &249.32& uvm2 & 21.166 & 0.180 \\
59649.93 &2401.93& b & 17.890 & 0.077 & & 57514.00 &266& uvm2 & 21.197 & 0.153 \\
59784.74 & 2536.74 & b & 17.803 & 0.095 & & 57531.25 &283.25& uvm2 & 21.153 & 0.234\\
57259.40 & 11.40 &  u & 15.972 & 0.037 & &  57572.85 & 324.85 & uvm2 & 21.767 & 0.185\\
57259.41 & 16.41 & u & 16.434 & 0.040 & &  57602.42 &354.42& uvm2 & 21.251 & 0.154\\
57268.75 &20.75&u & 16.728 & 0.041 & & 58674.29 &1426.29& uvm2 & 21.423 & 0.126 \\
57272.57 &24.57& u & 16.964 & 0.044 & &  59016.54 &1768.54& uvm2 & 21.373 & 0.290\\
57276.12 & 28.12 & u & 17.199 & 0.047 & & 59649.93 &2401.93& uvm2 & 21.368 & 0.247 \\
57280.71 & 32.71 & u & 17.460 & 0.050 & & 59828.39 & 2580.39 & uvm2 & 21.488 & 0.313\\
57285.64 &37.64& u & 17.572 & 0.062 & & 57257.83 & 9.83 & uvw2 & 15.957 & 0.039 \\
57291.48 &43.48& u & 17.697 & 0.065 & & 57259.40 & 11.40 &  uvw2 & 15.986 & 0.039 \\
57294.61 &46.61& u & 17.933 & 0.075 & & 57259.42 & 16.42 &  uvw2 & 16.300 & 0.040 \\
57300.26 & 52.26 & u & 18.080 & 0.078 & & 57268.75 &20.75&uvw2 & 16.455 & 0.043 \\
57303.65 &55.65&  u & 18.048 & 0.084 & & 57272.57 &24.57& uvw2 & 16.656 & 0.041 \\
57306.52 & 58.52&  u & 18.165 & 0.085 & & 57276.12 & 28.12 & uvw2 & 16.842 & 0.041 \\
57309.48 &61.48& u & 18.283 & 0.090 & & 57280.72 & 32.72 & uvw2 & 17.044 & 0.042 \\
57312.20 &64.20& u & 18.331 & 0.109 & & 57285.64 &37.64& uvw2 & 17.157 & 0.044 \\
57318.77 & 70.77 & u & 18.338 & 0.093 & & 57291.49 &43.49& uvw2 & 17.312 & 0.046 \\
57326.21 &78.21& u & 18.413 & 0.084 & & 57294.62 &46.62&uvw2 & 17.331 & 0.045 \\
57335.04 &87.04& u & 18.452 & 0.086 & & 57300.27 & 52.27 & uvw2 & 17.550 & 0.046 \\
57340.73 & 92.73 & u & 18.731 & 0.113 & & 57303.66 &55.66& uvw2 & 17.577 & 0.048 \\
57347.11 &99.11& u & 18.787 & 0.119 & & 57306.52 & 58.52&  uvw2 & 17.756 & 0.049 \\
57460.27 &212.27& u & 19.321 & 0.063 & & 57309.48 &61.48& uvw2 & 17.815 & 0.049 \\
57501.10 &253.10 & u & 19.327 & 0.104 & & 57312.20 &64.20& uvw2 & 17.933 & 0.052 \\
57572.84 & 324.84 & u & 19.264 & 0.093 & & 57318.77 & 70.77 & uvw2 & 18.168 & 0.053 \\
57602.41 &354.41& u & 19.247 & 0.100 & & 57324.08 &76.08& uvw2 & 18.254 & 0.064 \\
58674.29 &1426.29& u & 19.264 & 0.057 & & 57328.14  &80.14& uvw2 & 18.342 & 0.058 \\
59649.93 &2401.93& u & 19.246 & 0.119 & & 57333.48  &85.48& uvw2 & 18.548 & 0.066 \\
59784.73 & 2536.73 & u & 19.074 & 0.142 & & 57336.41 &88.41& uvw2 & 18.545 & 0.062\\
57259.40 & 11.40 &   uvw1 & 15.770 & 0.040 & &  57339.43 &91.43& uvw2 & 18.606 & 0.066\\
57259.41 & 16.41 &  uvw1 & 16.153 & 0.041 & &  57342.10  &94.10& uvw2 & 18.603 & 0.077\\
57268.75 &20.75& uvw1 & 16.459 & 0.042 & &  57345.79 &97.79& uvw2 & 18.652 & 0.083\\
57272.57 &24.57&  uvw1 & 16.678 & 0.044 & &  57348.41 &100.41& uvw2 & 18.782 & 0.066\\
57276.12 & 28.12 &  uvw1 & 16.985 & 0.046 & &  57468.80 &220.80& uvw2 & 20.830 & 0.191 \\
57280.71 & 32.71 &  uvw1 & 17.207 & 0.047 & & 57476.65 &228.65& uvw2 & 20.702 & 0.165\\
57285.64 &37.64&  uvw1 & 17.419 & 0.050 & & 57493.43 &245.43& uvw2 & 20.916 & 0.207 \\
57291.48 &43.48&  uvw1 & 17.667 & 0.053 & & 57501.27 &253.27 & uvw2 & 21.062 & 0.175 \\
57294.61 &46.61&  uvw1 & 17.782 & 0.056 & & 57506.29 &258.29& uvw2 & 21.073 & 0.199\\
57297.05 &49.05&  uvw1 & 18.047 & 0.078 & &  57515.12 &267.12& uvw2 & 21.112 & 0.243\\
57300.26 & 52.26 &  uvw1 & 18.059 & 0.059 & & 57521.44 &273.44& uvw2 & 21.178 & 0.215 \\
57303.65 &55.65&   uvw1 & 18.239 & 0.069 & & 57530.88 &282.88& uvw2 & 21.300 & 0.218 \\
57306.52 & 58.52&   uvw1 & 18.313 & 0.067 & & 57585.00 &337& uvw2 & 21.259 & 0.105 \\
57309.47 &61.47&  uvw1 & 18.337 & 0.066 & & 57839.52 &591.52& uvw2 & 21.406 & 0.137 \\
57312.20 &64.20&  uvw1 & 18.496 & 0.076 & & 58674.29 &1426.29& uvw2 & 21.521 & 0.102 \\
57318.77 & 70.77 &  uvw1 & 18.590 & 0.072 & &59016.54 &1768.54& uvw2 & 21.468 & 0.246  \\
57324.08 &76.08 &  uvw1 & 18.861 & 0.104 & & 59649.93 &2401.93& uvw2 & 21.464 & 0.197 \\
57328.14  &80.14&  uvw1 & 18.967 & 0.088 & & 59828.39 & 2580.39 & uvw2 & 21.289 & 0.215 \\
57333.48  &85.48& uvw1 & 18.992 & 0.093 & & 
\end{longtable*}

\renewcommand{\arraystretch}{1.3}
\begin{deluxetable*}{|c|c|cc|ccc|c|c|c|}[h!]
\tabletypesize{\footnotesize}
\tablecolumns{8}
\tablecaption{Best-fitting results from the time-resolved X-ray spectral analysis of ASASSN-15oi using \emph{Swift}-XRT (upper part) and \textit{XMM-Newton} (lower part). The data require two components of emission using a spectral  model of the form \texttt{tbabs*(cflux*pow+cflux*bbody)} within \texttt{XSPEC}. We adopted $N_{\rm H,gal}=5.59\times10^{20}\,\rm{cm^{-2}}$, $N_{\rm H,int}=0\,\rm{cm^{-2}}$, redshift, $z=0.0484$. \textit{XMM-Newton} (\emph{Swift}-XRT) observations are fitted in 0.2--12\,keV (0.3--10\,keV) energy range.  Fluxes are reported in the 0.3--10\,keV energy range, uncertainties are reported at 68\% confidence ($1\,\sigma$) level, and  upper limits are reported at $3\,\sigma$ confidence level.  $\Gamma_X$ is the power-law photon index, and $kT_{\rm BB}$ is the observed blackbody temperature in keV units. For \textit{XMM-Newton}  we report the EPIC-pn exposure times after removal of the time intervals affected by high background.\label{tab:xraydata}}
\tablehead{
	\colhead{$\delta t^{a}$} &
	\colhead{Exposure} &
	\multicolumn{2}{c}{PL} &
	\multicolumn{3}{c}{BB} &
	\colhead{Total} &
	\colhead{Source}
	\\
 \cmidrule(lr){3-4}\cmidrule(lr){5-7}
	\colhead{} &
    \colhead{} &
    \colhead{$\Gamma_X$} &
    \colhead{Unabsorbed Flux} &
    \colhead{kT$_{\rm BB}$ } &
    \colhead{Unabsorbed Flux} &
    \colhead{R$_{\rm BB}$} &
    \colhead{Absorbed Flux} &
    \colhead{}
	\\
    \colhead{(d)} &
    \colhead{(ks)} &
    \colhead{} &
    \colhead{\scriptsize($\times 10^{-14}$\,erg/cm$^{2}$/s)} &
    \colhead{\scriptsize($\times 10^{-2}$\,keV)} &
    \colhead{\scriptsize($\times 10^{-13}$\,erg/cm$^{2}$/s)} &
    \colhead{\scriptsize($\times 10^{12}$\,cm)} &
    \colhead{\scriptsize($\times 10^{-14}$\,erg/cm$^{2}$/s)} &
    \colhead{}
}
\startdata
\multicolumn{9}{c}{\textit{Swift}-XRT}\\
\hline
$58.6$ & $61.7$ & $1.7^{+0.5}_{-0.5}$ & $2.4^{+0.7}_{-0.7}$ & $4.5^{+0.5}_{-0.6}$ & $1.3^{+0.2}_{-0.2}$ & $0.4^{+0.4}_{-0.1}$ & $7^{+1}_{-1}$ &  This Work\\
$252.1$ & $31$ & $2.0$ & 
$< 3.6 $ & $4.9^{+0.3}_{-0.3}$  & $6.5^{+0.7}_{-0.6}$ & $0.6^{+0.2}_{-0.1}$ & $29^{+1}_{-6}$ & This Work\\
$329.0$ & $11.7$ & \rdelim\}{2}{*}[$2.0$] & $4.1^{+2.1}_{-1.7}$ & \rdelim\}{2}{*}[$5.2^{+0.4}_{-0.4}$] & $8.9^{+1.2}_{-1.0}$ & $0.6^{+0.2}_{-0.2}$ & $44^{+5}_{-5}$ &This Work\\
$425.1$ & $6.1$ &  & $5.3^{+3.1}_{-2.4}$ &   & $2.9^{+0.9}_{-0.7}$ & $0.3^{+0.1}_{-0.1}$ & $18^{+5}_{-3}$&This Work\\
$1444.9$ & $25.3$ & $2.3^{+0.3}_{-0.3}$ & $8.5^{+1.4}_{-1.5}$ & $5.2$  & $< 0.6$ & -- & -- & This Work\\
$2401$ & 7.2 & $2$ & $< 17$ & -- & -- &  -- & -- &  This Work \\
$2580$ & 7.0 & $2$ & $5.3^{+2.5}_{-2.5}$ & -- & -- &  -- &  -- &  This Work \\
\hline
\multicolumn{9}{c}{\textit{XMM}-Newton$^{b}$}\\
\hline
$76.4$  
& $10.3$  & $2.5^{+0.8}_{-0.8}$ & --$^{c}$ & $4.7^{+0.2}_{-0.2}$ & $1.6$  & $0.4$ & $7$ &  \cite{Gezari17}\\
$76.4$ & 12.4 & $1.7^{+1.0}_{-0.8}$ & $2.3^{+0.8}_{-0.8}$ & $6.2^{+6}_{-6}$ & $1.2^{+0.5}_{-0.5}$ & $0.5^{+0.6}_{-0.6}$ & -- &  \cite{Holoien18} \\
$76.4$ 
& $9.7$  & $1.5^{+0.5}_{-0.4}$ & $1.9^{+0.6}_{-0.6}$ & $5.1^{+0.3}_{-0.3}$ & $1.3^{+0.1}_{-0.1}$ & $0.2^{+0.1}_{-0.1}$& $8^{+1}_{-1}$ & 
This Work\\
\hline
$234.5$
& $12.0$  & $3.3^{+1.3}_{-1.3}$ & -- & $4.2^{+0.1}_{-0.1}$ &  $18$ & $2.1$  & $37$ &  \cite{Gezari17}\\
$234.5$
& $14.0$  & $4.8^{+2.6}_{-1.2}$ & $1.5^{+0.8}_{-0.8}$ & $5.3^{+2}_{-2}$ &  $3.2^{+0.9}_{-1.0}$ &  $1.1^{+0.8}_{-0.8}$& -- &\cite{Holoien18}\\
$234.5$  & $10.9$  & $3.1^{+1.2}_{-0.9}$ & $0.9^{+
0.7}_{-0.5}$ & $4.2^{+0.1}_{-0.1}$ & $8.9^{+0.2}_{-0.2}$ & $1.3^{+0.1}_{-0.1}$ & $38^{+1}_{-1}$ &
This Work\\  
\hline
$1833$ & $12.5$ & $2.0^{+0.3}_{-0.3}$ & $1.0^{+0.4}_{-0.2}$ & $5.2$ & $< 4.2\times10^{-2}$ & -- & -- & This Work\\
$2794$ & $19.7$ & $2$ & $< 0.8$ & -- & -- & -- & -- & This Work \\
\enddata
\tablecomments{ $^a$ For \emph{Swift}-XRT we list the mean time of arrival of photons. From top to bottom, the interval of times of extraction of the spectra are $9.8-100.5$\,d,  	$212.2-285.5$\,d, 	$285.5-346.4$\,d, $346.4 - 585.7$\,d, $1394.9-1774.2$\,d, $2401$\,d and $2540 - 2620$\,d.\\ $^{b}$ For XMM, we provide the numbers reported in the earlier works of \cite{Gezari17}, and \cite{Holoien18} for the purpose of comparison. \\ $^{c}$ The columns with no values (--) were not reported in the previously published works. The empty columns in the last XMM and XRT observations where we report the values from this work are because of the unconstrained nature of the BB component at these epochs. \\}
\end{deluxetable*}

\clearpage
\setlength{\tabcolsep}{2pt}
\renewcommand{\arraystretch}{0.1}
\startlongtable
\begin{deluxetable*}{cccccc}
\tablecolumns{6}
\tablecaption{Radio Observations of the TDE ASASSN-15oi. The VLA observations at $\delta t \lesssim 600$\,d were reported in \cite{Horesh21}, however we include the contribution for systematic uncertainties here that accounts for the inaccuracies in the flux density calibration. A 5\%, 10\%, 3\%, and 6\% systematic uncertainty is added in quadrature to the uncertainties of all the flux densities measured by the VLA, ATCA, and ALMA Band 3 and ALMA Band 6, respectively. For MeerKAT, the reported uncertainties are a combination of 5\% systematic, image background rms, and the standard deviation over three methods ((i) peak brightness (interpolated
between pixels); (ii) fit beam-shaped Gaussian + zero-level; and (iii) total over
a 10 pixel region, corrected for the fitted zero level from previous fit) of extracting the flux density --- all added in quadrature. The upper limits are reported as 3 $\times$ the image background rms values. \label{tab:radiodata}}
\tablehead{
	\colhead{Date} &
	\colhead{Observatory} &
	\colhead{Project} &
	\colhead{$\nu$} &
	\colhead{F$_\nu$} &
	\colhead{Source}
	\\
    \colhead{} &
    \colhead{} &
    \colhead{} &
    \colhead{(GHz)} &
    \colhead{(mJy)} &
    \colhead{}
}
\startdata
\multicolumn{6}{c}{$\delta t = 8$\,d}\\
\cmidrule(lr){1-6}
2015-Aug-22 & VLA & 16A-422 & 6.1 & $<$0.03 & \cite{Horesh21} \\
& &(PI: Horesh)  & 7.1  &  $<$0.03 &	\\
& & & 22 & $<$0.06 &	\\
\hline
\multicolumn{6}{c}{$\delta t = 23$\,d} \\
\cmidrule(lr){1-6}
2015-Sep-06 & VLA &16A-422 & 6.1 & $<$0.04 &	\cite{Horesh21} \\
& &(PI: Horesh)  & 7.1 & $<$0.04 &	\\
& &	& 22 	& $<$0.07 &	\\
\hline
\multicolumn{6}{c}{$\delta t = 90$\,d} \\
\cmidrule(lr){1-6}
2015-Nov-12 & VLA &	16A-422& 6.1 &	$<$0.06 	&	\\
& &(PI: Horesh) 	& 7.1 &	$<$0.06 	&	\\
&  &	& 22  &	$<$0.03 	&	\\
\hline
\multicolumn{6}{c}{$\delta t = 182$\,d}\\
\cmidrule(lr){1-6}
2016-Feb-12 & VLA &16A-422 &4.8 & 1.11 $\pm$ 0.06	&	\cite{Horesh21} \\
& &(PI: Horesh)  	&7.4 & 1.32 $\pm$ 0.07	&\\
& & 	&19     & 0.83 $\pm$ 0.05	&\\
& & 	&21 	& 0.81 $\pm$ 0.04 &\\
& & 	&23 	& 0.64 $\pm$ 0.04	&\\
& & 	&25 	& 0.59 $\pm$ 0.04	&\\	
\hline
\multicolumn{6}{c}{$\delta t = 190$\,d}\\
\cmidrule(lr){1-6}
2016-Feb-20 & VLA &16A-422	&3 	&   0.55 $\pm$ 0.03 & \cite{Horesh21}\\
&  &(PI: Horesh)    &4.8 &	0.89 $\pm$ 0.05 &	\\
&     & &7.4 &   1.12 $\pm$ 0.06 &	\\
&     & &9   &   1.20 $\pm$ 0.06 &	\\
&     & &11  &   1.07 $\pm$ 0.06 &	\\	
&     & &13  &   1.10 $\pm$ 0.06 &	\\	
&     & &15  &   0.93 $\pm$ 0.05 &	\\		
&     & &17  &   0.78 $\pm$ 0.04 &	\\		
&     & &19  &   0.72 $\pm$ 0.04 &	\\	
&     & &21  &   0.57 $\pm$ 0.03 &	\\	
&     & &23  &   0.49 $\pm$ 0.03 &	\\	
&     & &25  &   0.45 $\pm$ 0.03 &	\\
\hline
\multicolumn{6}{c}{$\delta t = 197$\,d} \\
\cmidrule(lr){1-6}
2016-Feb-27 & VLA  &16A-422	&3 	&   0.50 $\pm$ 0.03	&	\cite{Horesh21}	\\
& &(PI: Horesh) &4.5 &	0.69 $\pm$	0.04	&		\\
&   &		&5.5 &	0.88 $\pm$	0.04	&		\\
&   &		&6.5 &	1.04 $\pm$	0.05	&		\\
&   &		&7.5 &	1.12 $\pm$	0.06	&		\\
&   &		&9 	 &  1.07 $\pm$	0.06	&		\\
&   &		&11  &  1.03 $\pm$ 0.05	&		\\
&   &		&13  &  0.93 $\pm$	0.05	&		\\
&   &		&15  &  0.79 $\pm$	0.04	&		\\
&   &		&17  &  0.70 $\pm$	0.04	&		\\
&   &		&19  &  0.72 $\pm$	0.04	&		\\
&   &		&21  &  0.69 $\pm$	0.04	&		\\
&   &		&23  &  0.53 $\pm$	0.03	&		\\
&   &		&25  &  0.43 $\pm$	0.03	&	     \\
\hline
\multicolumn{6}{c}{$\delta t = 233$\,d}\\
\cmidrule(lr){1-6}
2016-Apr-03 & VLA  &16A-422	&3 	 &   1.01 $\pm$ 0.06	&	\cite{Horesh21}\\	
& &(PI: Horesh)  		&4.5 &	0.95 $\pm$	0.05	&	\\	
&   &		&5.5  &	0.82 $\pm$	0.04	&	\\	
&   &		&6.5  &	0.76 $\pm$	0.04	&	\\	
&   &		&7.5  &	0.72 $\pm$	0.04	&	\\	
&   &		&8.5  &	0.69 $\pm$	0.04	&	\\	
&   &		&9.5  &	0.64 $\pm$	0.03	&	\\	
&   &		&10.5 &	0.64 $\pm$	0.03	&	\\	
&   &		&11.5 &	0.59 $\pm$	0.03	&	\\	
&   &		&12.5 &	0.58 $\pm$	0.03	&	\\
&   &		&13.5 &	0.51 $\pm$	0.03	&	\\
&   &		&14.5 &	0.45 $\pm$	0.03	&	\\
\hline
\multicolumn{6}{c}{$\delta t = 283$\,d}\\
\cmidrule(lr){1-6}
2016-May-23 & VLA  &16A-422	&3 &	1.03 $\pm$	0.06	&	\cite{Horesh21}\\	
& & (PI: Horesh)  		&5 &	0.85 $\pm$	0.04	&	\\	
&   &		&7  &	0.67 $\pm$	0.04	&	\\	
&   &		&9  &	0.58 $\pm$	0.03	&	\\	
&   &		&11 &   0.52 $\pm$	0.03	&	\\	
&   &		&13 &   0.43 $\pm$	0.02	&	\\
&   &		&15 &	0.36 $\pm$	0.02	&	\\	
&   &		&17 &	0.32 $\pm$	0.02	&	\\
\hline
\multicolumn{6}{c}{$\delta t = 369$\,d}\\
\cmidrule(lr){1-6}
2016-Aug-17 & VLA  &16A-422 &	3 &	0.69 $\pm$	0.04	&	\cite{Horesh21}	\\
&  &(PI: Horesh) 		&5 &	0.62 $\pm$	0.04	&		\\
&   &		&7 &	0.57 $\pm$	0.03	&		\\
&   &		&9 &	0.55 $\pm$	0.03	&	\\
\hline
\multicolumn{6}{c}{$\delta t = 576$\,d}\\
\cmidrule(lr){1-6}
2017-March-12 & VLA  &16A-422	&3 &	0.31 $\pm$	0.07	& \cite{Horesh21}\\	
&  &(PI: Horesh) 		&5 &	0.23 $\pm$	0.03	&		\\
&   &		&7 &	0.21 $\pm$	0.02	&		\\
&   &	    &9 &	0.20 $\pm$	0.02	&	\\
\hline
\multicolumn{6}{c}{$\delta t = 1417$\,d}\\
\cmidrule(lr){1-6}
2019-Jul-01 & VLA & VLASS & 3 & 9.5 $\pm$ 0.5 & \\
&   &	    &  &	&	\\
\hline
\multicolumn{6}{c}{$\delta t = 1741$\,d}\\
\cmidrule(lr){1-6}
2020-May-20& VLA & 20A-492 &  1.3 & 9.8 $\pm$	0.5	& This Work	\\
& &(PI: Alexander) &	1.8 & 10.8	$\pm$ 0.6 & \\
&&&	2.5  & 10.1  $\pm$ 0.5 & \\
&&&	3.5  & 9.6 $\pm$ 0.5  & \\
&&&	5	 & 9.3 $\pm$ 0.5 & \\
&&&	7	 & 8.4 $\pm$ 0.4 & \\
&&&	9	 & 7.3 $\pm$ 0.3 & \\
&&&	11	 & 6.3 $\pm$ 0.4 & \\
\hline
\multicolumn{6}{c}{$\delta t = 2129$\,d}\\
\cmidrule(lr){1-6}
2021-Jun-08 & VLA & 21A-303 & 1.2 & 11.3 $\pm$ 0.6 & This Work\\
& & (PI: Hajela) & 1.6 & 11.2 $\pm$ 0.6 & \\
& & & 1.9 & 10.8 $\pm$ 0.6 & \\
& & & 2.3 & 9.9 $\pm$ 0.6 & \\
& & & 2.8 & 9.2 $\pm$ 0.5 & \\
& & & 3.2 & 8.6 $\pm$ 0.4 & \\
& & & 3.8 & 7.9 $\pm$ 0.4 & \\
& & & 4.5 & 7.1 $\pm$ 0.4 & \\
& & & 5.5 & 6.2 $\pm$ 0.3 & \\
& & & 6.5 & 5.4 $\pm$ 0.3 & \\
& & & 8.5 & 4.7 $\pm$ 0.2 & \\
& & & 9.5  & 4.4 $\pm$ 0.2 & \\
& & & 10.5 & 4.1 $\pm$ 0.2 & \\
& & & 11.5 & 3.8 $\pm$ 0.2 & \\
\hline
\multicolumn{6}{c}{$\delta t = 2384$\,d}\\
\cmidrule(lr){1-6}
2022-Feb-22 & VLBI & BH238 & 8.3 & 1.4 $\pm$ 0.3 & This Work \\
& & (PI: Hajela) &&& \\
\hline
\multicolumn{6}{c}{$\delta t = 2660$\,d}\\
\cmidrule(lr){1-6}
2023-Jan-05 & MeerKAT & \footnotesize{DDD-20220414-YC-01} & 0.8 & 9.8 $\pm$ 0.5 & This Work \\
2022-Dec-09 &  MeerKAT & \footnotesize{SCI-20220822-YC-01}  & 1.3 & 6.4 $\pm$ 0.3 &  \\
& & (PI: Cendes) & \\
2022-Sep-30 & ATCA & C3325 & 1.3 & 9.7 $\pm$ 1.3 &  \\
 &  & (PI: Alexander) & 1.8 & 8.5 $\pm$ 0.9 &  \\
 &  &  & 2.4 & 7.4 $\pm$ 0.8 &   \\
 &  &  & 2.9 & 6.7 $\pm$ 0.7 &  \\
 &  &  & 4.7 & 4.9 $\pm$ 0.5 &  \\
 &  &  & 5.2 & 4.8 $\pm$ 0.5 &  \\
 &  &  & 5.8 & 4.4 $\pm$ 0.4 &  \\
 &  &  & 6.3 & 4.2 $\pm$ 0.4 &  \\
 &  &  & 8.2 & 3.6 $\pm$ 0.4 &  \\
 &  &  & 8.7 & 3.5 $\pm$ 0.3 &  \\
 &  &  & 9.3 & 3.4 $\pm$ 0.3 &  \\
 &  &  & 9.9 & 3.2 $\pm$ 0.4 &  \\
2022-Sep-29 & ALMA & 2019.1.01166.T & 97.5 & 0.38 $\pm$ 0.02 &  \\
 &  & (PI: Alexander) & 233 & 0.16 $\pm$ 0.02 & \\
 \hline
\multicolumn{6}{c}{$\delta t = 2970$\,d}\\
\cmidrule(lr){1-6}
2023-Oct-01 & VLA & 23A-241 & 1.2 & 4.1 $\pm$ 0.2& This Work \\
& & (PI: Cendes) & 1.5 & 3.7 $\pm$ 0.2& \\ 
& &  & 1.9 & 3.5 $\pm$ 0.2 & \\ 
& &  & 2.3 & 3.5 $\pm$ 0.2 & \\ 
& &  & 2.9 & 3.1 $\pm$ 0.2 & \\
& &  & 3.6 & 3.1 $\pm$ 0.2 & \\
& &  & 4.5 & 2.6 $\pm$ 0.1 & \\
& &  & 5.5 & 2.4 $\pm$ 0.1 & \\
& &  & 6.5 & 2.1 $\pm$ 0.1 & \\
& &  & 7.5 & 1.9 $\pm$ 0.1 & \\
& &  & 8.5 &  1.7 $\pm$ 0.1 &  \\ 
& &  & 9.5 & 1.5 $\pm$ 0.1 &  \\ 
& &  & 10.5 & 1.4 $\pm$ 0.1 &  \\
& &  & 11.5 & 1.4 $\pm$ 0.1 &  \\
\enddata
\end{deluxetable*}

\section{Radio Spectral Fitting and the Model Selection Criteria}\label{sec:AICBIC}

In sections \S\ref{subsec:firstflaresedfit} and \S\ref{subsec:secondflaresedfit}, we explain the fitting process in details, along with our underlying assumptions. In \S\ref{subsec:aicbic}, we give details on our model selection criteria.

\subsection{Fit of SED\lowercase{s associated with the first radio flare}}\label{subsec:firstflaresedfit}

We first fit the $\delta t =182$\,d, $ 190$\,d, and $197$\,d SEDs with Equation (\ref{eq:BPL_mod}), assuming 
values of $s_1 = -0.3$ and a sharper $s_2 = -0.003$. 
We set $\alpha_{\rm{thick,\,thin}}$ as free parameters for the $190$ and $197$\,d SEDs, resulting in $\alpha_{\rm thick} \sim +1$.  We fix 
$\alpha_{\rm thick}$ at +1 for the 182\,d SED,  at which
there are too few measurements below $\nu_{\rm pk}$ to allow an independent constraint.
A second break is weakly constrained in all three SEDs as $\nu_q \sim 18 - 20$\,GHz ({an example of the corner plot with posterior distributions of the fit parameters for $\delta t = 190$\,d is shown in Figure \ref{fig:posterior_190d}}). Due to large uncertainties on $\nu_q$ (in particular up to $\sim 40$\% for the $182$\,d SED), we also compute a fit without a $\nu_q$ break (i.e., Equation \ref{eq:BPL}) for each SED,
which was the model used by \citet{Horesh21}. Comparing the two models, we find that $\Delta$AIC$<2$ and $\Delta$BIC$<2$, indicating no strong preference for either model (Table \ref{tab:modelselection}). 

The SEDs at $\delta t = 233$\,d and $ 283$\,d do not show a strong evidence of a low-frequency turnover (i.e., no $\alpha > 0$ segment is present), but do exhibit a slight steepening at higher frequencies. We fit these SEDs with Equation \ref{eq:BPL} (with $\nu_{b_1} = \nu_{q}$, $\alpha_{i} = \alpha_{\rm thin}$ and $\alpha_{j} = \alpha_q$), and $\nu_q$ is constrained to $\sim 10$\,GHz in both SEDs. We also attempted a simple power-law fit of the form $F \propto \nu^{\alpha_{\rm thin}}$. Based on the AIC and BIC values (given in Table \ref{tab:modelselection}) for the two models, the model with \emph{one} break is marginally preferred over a simple power-law for $\delta t = 233$\,d SED (with $\Delta \rm{AIC} = 2.5$ and $\Delta \rm{BIC} = 2.1$) and is strongly preferred for $\delta t = 283$\,d (with $\Delta \rm{AIC} = 6.2$ and $\Delta \rm{BIC} = 6.1$). At $\delta t = 369$\,d and  $\delta t = 576$\,d, the SEDs show no low frequency turnovers or high frequency steepening, and are therefore fitted with a simple power-law of the form $F \propto \nu^{\alpha_{\rm thin}}$. 

\subsection{Fit of SED\lowercase{s associated with the second radio flare}}\label{subsec:secondflaresedfit}
We treat the SEDs taken during the second flare (i.e., at $\delta t \gtrsim 1400$\,d) independently of the earlier ones where we let $\alpha_{\rm thick}$ vary freely. Here, instead we try fitting the SEDs with both Equation \ref{eq:BPL_mod} (with $\alpha_{\rm thick, thin}$ spectral segments) and \ref{eq:BPL} with $\alpha_{\rm thick} = 5/2$ (as expected for standard self-absorbed synchrotron spectra in the context of TDEs; e.g., \citealt{Alexander20}). Similar to the previous SEDs, we 
adopt $s_1 = -0.03$ and $s_2 = -0.003$ (unless otherwise mentioned); and the AIC and BIC values are reported in Table \ref{tab:modelselection}.

For the SEDs at $\delta t = 1741$\,d, $ 2660$\,d and $2970$\,d, we find that the model with \textit{two} breaks is a better fit. Here $\nu_q$ is constrained to $\sim 7$\,GHz, $\sim 19$\,GHz and $\sim 5$\,GHz, respectively. In these SEDs, the peak's location largely mirrors the priors, and in particular, the lower bound on the peak frequency remains unconstrained, resulting in only limits on the SED peak. In contrast, the SED at $\delta t =2129$\,d prefers the model with only \textit{one} break and the peak is robustly constrained. However this SED exhibits a smoother turnover compared to others, requiring $|s_1| > 0.03$. We thus 
assume $s_1 = -1$ here and report the results accordingly.

\subsection{Model Selection Criteria}\label{subsec:aicbic}

We use the Akaike Information Criterion (AIC) and the Bayesian Information Criterion (BIC) in \S\ref{subsec:sedfit} as a means for model selection. These statistical measures compare how well the models fit the observed data, with both criteria penalizing the models that have more parameters. BIC imposes a stricter penalty, making it more conservative. The aim is to achieve an optimal balance between goodness of fit and the complexity of the model. 

The AIC is given by:
\begin{equation}
\text{AIC} = 2k - 2\ln(L),
\end{equation}

and the BIC is given by:
\begin{equation}
\text{BIC} = k\ln(n) - 2\ln(L)
\end{equation}
where $k$ is the number of free parameters used in our MCMC fitting, $L$ is the maximum likelihood of the model and $n$ is the number of observations. 

As mentioned in \S\ref{subsec:sedfit}, we use one or more models to fit for some SEDs. The models are as follows: (i) models with 0 breaks are represented by $F \propto \nu^{\alpha_{\rm thin}}$, (ii) models with 1 break are given by Equation \ref{eq:BPL}, and (iii) models with 2 breaks are given by Equation \ref{eq:BPL_mod}. We give the corresponding AIC and BIC in Table \ref{tab:modelselection}. Values of $\Delta AIC > 2$ and $\Delta BIC > 2$ show a statistically stronger preference for the model with the \emph{lower} AIC and BIC values.

\renewcommand\thetable{\thesection.\arabic{table}} 
\setcounter{table}{0}

\setlength{\tabcolsep}{7pt}
\renewcommand{\arraystretch}{1.2}
\begin{deluxetable}{ccccccc}[h!]
\label{tab:modelselection}
\tablecolumns{9}
\tablecaption{The values of AIC and BIC for the different models at various epochs are reported here. The three models compared are: (i) a model with 0 breaks of the form $F \propto \nu^{-\alpha_{\rm thin}}$, (ii) a model with 1 break of the form of Equation \ref{eq:BPL}, and (iii) a model with 2 breaks represented by Equation \ref{eq:BPL_mod}.}
\tablehead{
	\colhead{$\delta t^{a}$} &
	\multicolumn{2}{c}{0 Break} &
	\multicolumn{2}{c}{1 Break} &
        \multicolumn{2}{c}{2 Breaks}
	\\
 \cmidrule(lr){2-3}\cmidrule(lr){4-5}\cmidrule(lr){6-7}
	\colhead{(d)} &
    \colhead{AIC} &
    \colhead{BIC} &
    \colhead{AIC} &
    \colhead{BIC} &
    \colhead{AIC} &
    \colhead{BIC}
}
\startdata
182 & -- & -- & 17.0 & 16.4 & 15.9 & 15.1\\
190 & -- & -- & 23.3 & 25.2 & 22.5 & 24.9 \\
197 & -- & -- & 32.8 & 35.4 & 30.8 & 33.9 \\
233$^a$ & 17.4 & 18.4 & 14.9 & 16.3 & 16.8 & 18.7 \\
283$^a$ & 18.2 & 18.3 &  12.0 & 12.2 & 14.6 & 14.9 \\
1741 & 40 & 40.2 &  34.7 & 35.0 & 13.2 & 13.5\\
2129 & -- & -- & 10.3 & 12.4 & 13.4 & 15.5\\
2660 & -- & -- & 53.2 & 55.5 & 16.3 & 11.0\\
2970 & -- & -- & 25.6 & 27.5 & 16.1 & 18.7\\
\enddata
\tablecomments{$[a]$ In these SEDs, the `1 Break' model given by Equation \ref{eq:BPL} has $\nu_{b_1} = \nu_{q}$, $\alpha_{i} = \alpha_{\rm thin}$ and $\alpha_{j} = \alpha_{q}$.\\}
\end{deluxetable}

\section{Posterior distribution of fit parameters of the radio SED at \lowercase{$\delta t \approx$190 d}}
\renewcommand\thefigure{\thesection.\arabic{figure}} 
\setcounter{figure}{0}
\begin{figure}
    \centering
    \includegraphics[scale=0.3]{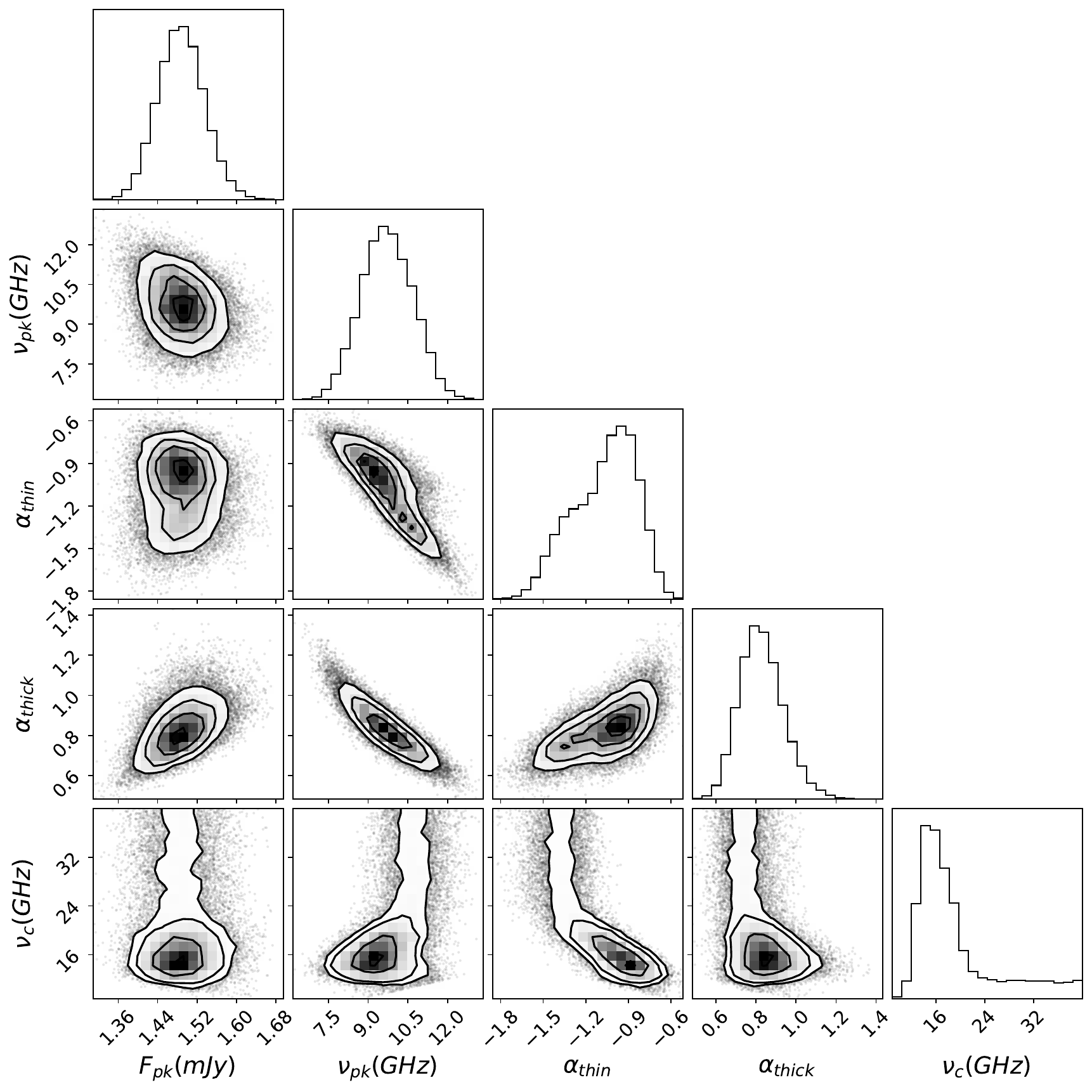}
    \caption{One- and two-dimensional projections of the posterior distributions of the free parameters resulting from the fitting of the radio SED at $\delta t = 190$\,d with the model: Equation (\ref{eq:BPL}) $\times$ Equation (\ref{eq:BPL_mod}). The contours are drawn at the 68\%, 95\%, and 99\% credible levels. As can be seen all the parameters are well constrained.}
    \label{fig:posterior_190d}
\end{figure}
We use the model: Equation (\ref{eq:BPL}) $\times$ Equation (\ref{eq:BPL_mod}) to fit the SEDs at $\delta t = 182 - 197$\,d. The following priors were used: $ 0 < F_{\rm pk} < 40\,\rm{mJy}$; $0 < \nu_{\rm pk} < 40\,\rm{GHz}$; $-3< \alpha_{\rm thin} < 0$; $0 < \alpha_{\rm thick} < 3$; $\nu_{\rm pk} < \nu_{q} < 40\,\rm{GHz}$. All priors have a uniform distribution. As an example, we present the posterior distributions of $F_{\rm pk}$, $\nu_{\rm pk}$, $\alpha_{\rm thin}$, $\alpha_{\rm thick}$, and $\nu_{q}$ resulting from the MCMC fitting of the SED at $\delta t = 190$\,d in Figure \ref{fig:posterior_190d}. 

\section{Expressions for Synchrotron Physical Parameters in case of \lowercase{$p > 2$}}\label{sec:demarchi_equations}
\cite{demarchi2022} derived the following expressions for the physical parameters in the synchrotron model, assuming $\gamma_{\rm max} \to\infty$. A key aspect of these equations is, therefore, $p > 2$. In \S\ref{Sec:RadioModeling} and Table \ref{tab:radiophysparams}, we use these equations to measure $B$, $R$, $U$, $n_{\rm{e}}$ for the SEDs where $p > 2$.
\begin{minipage}{.5\textwidth}
\begin{equation}
\label{eq:B}
\begin{split}
    B & = (2.50 \times10^9 \,\rm{G})
    \Big(\frac{\nu_{\rm{pk}}}{\rm{5\,GHz}} \Big) 
    \Big(\frac{1}{c_1}\Big)  \\
    & \times \Bigg[
    4.69\times10^{-23}
    \bigg(\frac{E_l}{\rm{erg}}\bigg)^{2(2-p)} \bigg(\epsilon_B/\epsilon_e\bigg)^2 c_5 \sin(\theta)^{\frac{1}{2}(-5-2p)} \\
    & \times (p-2)^{-2} \bigg(\frac{D}{\rm{Mpc}}\bigg)^{-2} \bigg(\frac{f}{0.5}\bigg)^{-2} \bigg(\frac{F_{\rm pk}}{\rm{Jy}}\bigg)^{-1}c_6^{-3}\Bigg]^{\frac{2}{13+2p}}
\end{split}
\end{equation}
\end{minipage}

\begin{equation}
\label{eq:R}
\begin{split}
R & = 
(2.50 \times10^9)^{-1}\,\rm{cm} \times 
c_1
\bigg(\frac{\nu_{\rm pk}}{\rm 5\,GHz}\bigg)^{-1} \\
& \times \bigg[
12  
\epsilon_B
c_5^{-(6 + p)}
c_6^{(5 + p)} 
(9.52\times10^{25})^{(6+p)}
\sin^2{\theta}  \\
& \times \pi^{-(5 + p)} \bigg(\frac{D}{\rm Mpc}\bigg)^{2 (6 + p)} \bigg(\frac{E_l}{\rm erg}\bigg)^{(2 - p)} 
\bigg(\frac{F_{\rm pk}}{\rm Jy}\bigg)^{(6 + p)} \\
& \times \bigg(
\epsilon_e
(p - 2)
\bigg(\frac{f}{0.5}\bigg)
\bigg)^{-1}
\bigg]^{1/(13 + 2 p)}
\end{split}
\end{equation}

\begin{minipage}{.5\textwidth}
\begin{equation}
    \label{eq:U}
    \begin{split}
U & = 
 c_1
  (3.33\times10^{-11}  \,\rm{erg})
   \epsilon_B^{-1}
10^{\frac{75 (6 + p)}{13 + 2 p}} 
\bigg(\frac{f}{0.5}\bigg)
\bigg(\frac{\nu_{\rm{pk}}}{\rm{5\,GHz}}\bigg)^{-1} \\
&\times \Bigg[
 3.086^{6(6+p)}
 4.411\times10^{-96} 
\bigg(\frac{D}{\rm{Mpc}}\bigg)^{28 + 6 p} \\
& \times \bigg(\frac{F_{\rm{pk}}}{\rm{Jy}}\bigg)^{14 + 3 p}
 \pi^{-3 (1 + p)}
 \sin{\theta}^{-4 (1 + p)} 
 c_5^{-(14 + 3 p)}  c_6^{3 (1 + p)} \\
 & \times \bigg(
  2
   \epsilon_B
 \epsilon_e^{-1}
  (p - 2)^{-1} 
\bigg(\frac{E_l}{\rm{erg}}\bigg)^{2 - p} \times \bigg(\frac{f}{0.5}\bigg)^{-1}
 \bigg)^{11}
 \bigg]^{1/(13 + 2p)}
\end{split}
\end{equation}
\end{minipage}
\begin{minipage}{.5\textwidth}
\begin{equation}
\label{eq:ne}
n_{\rm e} = \bigg(\frac{i+2}{2\epsilon_{\rm B}}\bigg)\bigg(\frac{B^2}{8\pi}\bigg)(\mu_{e}m_{p})^{-1}(qv)^{-2}
\end{equation}
\end{minipage}

\noindent
where $f$ is a volume filling factor that represents the fraction of a sphere of radius $R$ that is emitting synchrotron radiation; $c_1$, $c_5(p)$ and $c_6(p)$ are synchrotron coefficients (see e.g., \citealt{demarchi2022}, their Appendix A); and $D$ is the distance of the TDE in Mpc. $E_l$ is the lowest energy of the electrons accelerated to the power-law. They constitute the electrons with the minimum Lorentz factor $\gamma_{\rm{min}}$. $\theta$ is  the angle between the electron velocity vector and the $B$ field. In Equation \ref{eq:ne}, $i$ is the adiabatic index of the shocked gas, $\mu_{\rm e}$ is the mean molecular weight per electron, $m_{\rm p}$ is the mass of the proton, and $q$ is the power-law index of expansion of the radius with time (i.e., $R \propto t^q$). We adopt $f=0.5$, $\sin (\theta)\approx 1$ and $\gamma_{\rm min}=1$, which gives $E_l= m_e c^2$, $i = 1$, and lastly, $q = 0.88$. This value of $q$ follows from an assumption of the contrast in the power-law of the ambient density the shock is expanding into, and the power-law of the density of the ejecta. 

\section{Derivation of Synchrotron Physical Parameters in case of \lowercase{$p < 2$}}\label{sec:pless2equations}

In this section, we derive equations for the physical parameters where $\gamma_{\rm max}$ has a finite upper bound, and is therefore valid for $p < 2$. 

The number density distribution of electrons is given by $dN(\gamma_e)/d\gamma_e = K_0 \gamma_{e}^{-p}$ for $\gamma_{e} \geq \gamma_{\rm min}$, where $N$ is the number of electrons per unit volume, $K_0$ is the normalization constant, $p$ is the power-law index of the electron population distribution. Number density distribution of electrons can also be expressed as $dN/dE = N_0 E^{-p}$ in energy space, where $N_0$ is the normalization constant, and $E = \gamma_{e} m_{e} c^2$. This gives $K_0 \gamma_{e}^{-p} d\gamma_{e} = N_0 E^{-p} dE \implies K_0 (m_e c^2)^{p-1} = N_0$. 

The relativistic electron energy density is given by:
\begin{equation}
    \label{eq:integral_Ee_pless2}
    \begin{split}
    E_{\rm e} & = \int^{E_u}_{E_{l}} E N_0 E^{-p} dE \\
    & = \int^{\gamma_{\rm max}}_{\gamma_{\rm min}}(\gamma_e m_e c^2)K_0\gamma_e^{-p} d\gamma_e\\
    & = K_0 m_e c^2 \Bigg[\frac{\gamma_{e}^{2 - p}}{2 - p}\Bigg]\Biggr|_{\gamma_{\rm min}}^{\gamma_{\rm max}}
    \end{split}
\end{equation}
If $\gamma_{\rm max} \to\infty$, Equation (\ref{eq:integral_Ee_pless2}) will revert to the form used for deriving Equation (\ref{eq:B}) -- (\ref{eq:ne}) and will require $p > 2$ to be finite. However, if we impose a finite $\gamma_{\rm max}$, then, the new $E_{\rm e}$ will take the following form:
\begin{equation}
\begin{split}
E_{\rm e} & = \frac{K_0 m_e c^2 \gamma_{\rm min}^{2-p}}{2 - p}\bigg[\bigg(\frac{\gamma_{\rm max}}{\gamma_{\rm min}}\bigg)^{2-p} - 1\bigg]\\
& = \frac{K_0 (m_e c^2)^{p-1} E_l^{2-p}}{2-p}\Lambda\\
& = \frac{N_0 E_l^{2-p}}{2-p}\Lambda
\end{split}
\end{equation}
for $p < 2$.
where $\Lambda = [(\gamma_{\rm max}/\gamma_{\rm min})^{2-p} - 1]$, and $E_l = \gamma_{\rm min} m_e c^2 $, and $p < 2$ is valid. Using this new expression of $E_{\rm e}$, we follow \cite{demarchi2022} to derive $B$, $R$ and $U$ as follows.

\begin{minipage}{.5\textwidth}
\begin{equation}
\label{eq:B_pless2}
\begin{split}
    B & = (2.50 \times10^9 \,\rm{G})
    \Big(\frac{\nu_{\rm{pk}}}{\rm{5\,GHz}} \Big) 
    \Big(\frac{1}{c_1}\Big) \\
    & \times \Bigg[
    4.69\times10^{-23}
    \bigg(\frac{E_l}{\rm{erg}}\bigg)^{2(2-p)} \Lambda^{2} \bigg(\frac{\epsilon_B}{\epsilon_e}\bigg)^2 c_5 \sin(\theta)^{\frac{1}{2}(-5-2p)} \\
    & \times (2-p)^{-2} \bigg(\frac{D}{\rm{Mpc}}\bigg)^{-2} \bigg(\frac{f}{0.5}\bigg)^{-2} \bigg(\frac{F_{\rm pk}}{\rm{Jy}}\bigg)^{-1}c_6^{-3}\Bigg]^{\frac{2}{13+2p}}
\end{split}
\end{equation}
\end{minipage}

\begin{minipage}{.5\textwidth}
\begin{equation}
\label{eq:R_pless2}
\begin{split}
R & = 
(2.50 \times10^9)^{-1}\,\rm{cm} \times 
c_1
\bigg(\frac{\nu_{\rm{pk}}}{\rm{5\,GHz}}\bigg)^{-1} \\
& \times \bigg[
12  
\epsilon_B
c_5^{-(6 + p)}
c_6^{(5 + p)} 
(9.52\times10^{25})^{(6+p)}
\sin^2{\theta}  \\
& \times \pi^{-(5 + p)} \bigg(\frac{D}{\rm{Mpc}}\bigg)^{2 (6 + p)} \bigg(\frac{E_l}{\rm{erg}}\bigg)^{(2 - p)} \Lambda 
\bigg(\frac{F_{\rm{pk}}}{\rm{Jy}}\bigg)^{(6 + p)} \\
& \times \bigg(
\epsilon_e
(2 - p)
\bigg(\frac{f}{0.5}\bigg)
\bigg)^{-1}
\bigg]^{1/(13 + 2 p)}
\end{split}
\end{equation}
\end{minipage}

\begin{minipage}{.5\textwidth}
\begin{equation}
\label{eq:U_pless2}
\begin{split}
U & = 
 c_1
  (3.33\times10^{-11}  \,\rm{erg})
   \epsilon_B^{-1}
10^{\frac{75 (6 + p)}{13 + 2 p}} 
\bigg(\frac{f}{0.5}\bigg)
\bigg(\frac{\nu_{\rm{pk}}}{\rm{5\,GHz}}\bigg)^{-1} \\
&\times \Bigg[
 3.086^{6(6+p)}
 4.411\times10^{-96} 
\bigg(\frac{D}{\rm{Mpc}}\bigg)^{28 + 6 p} \\
& \times \bigg(\frac{F_{\rm{pk}}}{\rm{Jy}}\bigg)^{14 + 3 p}
 \pi^{-3 (1 + p)}
 \sin{\theta}^{-4 (1 + p)} 
 c_5^{-(14 + 3 p)}  c_6^{3 (1 + p)} \\
 & \times \bigg(
  2
   \epsilon_B
 \epsilon_e^{-1}
  (2 - p)^{-1} 
\bigg(\frac{E_l}{\rm{erg}}\bigg)^{2 - p} \Lambda \bigg(\frac{f}{0.5}\bigg)^{-1}
 \bigg)^{11}
 \bigg]^{1/(13 + 2p)}
\end{split}
\end{equation}
\end{minipage}

\section{Mass Estimates for the SMBH in ASASSN-15\lowercase{oi}}\label{sec:smbhmass}
There have been a wide range of host SMBH mass estimates in the literature: $1.3\times10^7\,\rm{M_{\odot}}$ (\citealt{Holoien16} using the galactic scaling relationship between M$_{\rm{bulge}}$ and M$_{\rm{BH}}$); $0.5^{+1.4}_{-0.4}\times 10^6\,\rm{M_{\odot}}$ (\citealt{Wevers19, Wevers20Erratum} using M$_{\rm BH}$-$\sigma$ relationship in \citealt{Ferrarese2005} ); $(4\pm1)\times 10^6\,\rm{M_{\odot}}$ (\citealt{Mockler19}); $(5.4 \pm 4.6)\times 10^6\,\rm{M_{\odot}}$ (\citealt{Wen2020} using their slim-disc model); $9 \times 10^6\,\rm{M_{\odot}}$ (\citealt{Mummery2021_arXiv} using their evolving relativistic accretion disc model fit to the existing X-ray observations); $4.7^{+15}_{-3.4}\times 10^6\,\rm{M_{\odot}}$ (\citealt{Mummery2023} derived with their mean X-ray radius --- $M_{\rm BH}$ relation), $7^{+5}_{-5} \times 10^{5} \,\rm{M_{\odot}}$ (\citealt{Mummery2024_galscaling} using their scaling relationships between late-time UV plateau luminosity and M$_{\rm BH}$). Here, we adopt $2.5\times10^6\,\rm{M_{\odot}}$ as the SMBH mass, following \cite{Gezari17}.

\newpage
\bibliography{BibliographyFile}{}
\bibliographystyle{aasjournal}

\end{document}